\documentclass[aps,nofootinbib,noshowpacs,preprintnumbers,amsmath,amssymb]{revtex4}

\usepackage{graphicx,epsfig}
\usepackage{slashed}
\usepackage{color}
\usepackage{setspace}
\usepackage{float}

\def\be{\begin{equation}}
\def\ee{\end{equation}}
\def\bea{\begin{eqnarray}}
\def\eea{\end{eqnarray}}
\def\bes{\begin{subequations}}
\def\ees{\end{subequations}}
\newcommand{\A}{{\mathcal{A}}}
\newcommand{\tA}{{\widetilde {\mathcal{A}}}}
\newcommand{\ta}{{\widetilde a}}

\newcommand{\tk}{{\widetilde k}}
\newcommand{\td}{{\widetilde d}}
\newcommand{\tf}{{\widetilde f}}

\newcommand{\MSbar}{\overline{\rm MS}}

\newcommand{\bL}{\overline{\Lambda}}
\usepackage{graphics}
\usepackage{graphicx}
\usepackage{dcolumn}
\usepackage{bm}
\usepackage{epsfig}
\usepackage{graphicx,color}
\usepackage{multirow}
\usepackage{epstopdf}

   \graphicspath{{./figs/}}
\usepackage{color}

\def\1{\hbox{{1}\kern-.25em\hbox{l}}}
 \date{\today}
\def\be{\begin{equation}}
\def\ee{\end{equation}}
\def\bea{\begin{eqnarray}}
\def\eea{\end{eqnarray}}
\def\bear{\begin{array}}
\def\eear{\end{array}}
\def\bes{\begin{subequations}}
\def\ees{\end{subequations}}

\begin{document}
\preprint{USM-TH-367}

\title{Evaluation of neutrinoless double beta decay: QCD running to sub-GeV scales}
\author{C\'esar Ayala, Gorazd Cveti\v{c} and Lorena Gonz\'alez}
\affiliation{Department of Physics, Universidad T{\'e}cnica Federico
Santa Mar{\'\i}a (UTFSM),  
Casilla 110-V, Avenida Espa\~na 1680, 2390123 Valpara{\'\i}so, Chile}
\date{\today}

\begin{abstract}
  We evaluate QCD effects in the neutrinoless double beta ($0\nu\beta\beta$) decay, originating from new physics short-range mechanism in the form of five dimension-9 operators. For this, we employ the one-loop and two-loop renormalization group equations (RGEs) for the corresponding Wilson coefficients, performing the RGE-evolution from the new physics scales (estimated as $\Lambda ~ \sim 10^2$ GeV) to the typical spacelike $0\nu\beta\beta$-scale $Q \sim 0.1$ GeV. Since the latter scale is clearly nonperturbative, we apply various infrared-safe (IR-safe) variants of QCD where the running coupling has no Landau singularities at low spacelike $Q$. We point out that the correct treatment of the IR-safe analogs of the (noninteger) powers of the couplings is important. It turns out that in most cases of the considered operators the resulting QCD effects can be significant in this process, i.e., can be stronger than the effects of the present uncertainties in the nuclear matrix elements.
\end{abstract}

\maketitle


\section{Introduction}
\label{sec:intro}

\begin{figure}[htb] 
  \centering\includegraphics[width=40mm]{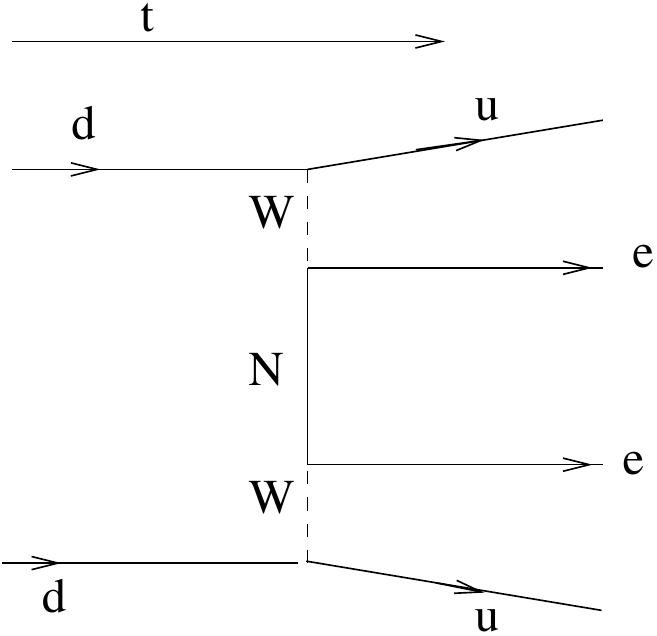}
  \vspace{-0.4cm}
 \caption{\footnotesize The decay $0\nu\beta\beta$ with the exchange of a Majorana neutrino $N$. If $M_N \gg 0.1$ GeV, then the propagators of $N$ and the off-shell $W$'s reduce together to an effective point.}
\label{Fig0nu}
\end{figure}

One of the basic questions of high energy physics is whether the neutrinos, and/or their more exotic fermionic relatives if they exist, are Majorana or Dirac particles. The question of the existence of Majorana neutrinos is closely related with the question of whether the lepton number violating (LNV) processes exist. At present, the most powerful probe of LNV processes is the neutrinoless double beta ($0\nu\beta\beta$) decay (cf.~\cite{0nubb1,DHP} for recent reviews), i.e., the process where two $d$ quarks of a nucleus transform into two $u$ quarks with the simultaneous production of two low-energy electrons. Such processes have not (yet) been observed, and one of the best lower bounds on the half-life for $0\nu\beta\beta$ is from the KamLAND-Zen experiment \cite{KamLAND} for the decay of $^{136}{\rm Xe}$
\be
T^{0\nu}_{1/2}(^{136}{\rm Xe}) > 1.07 \times 10^{26} \ {\rm yr} \; (90\% \; {\rm CL}).
\label{T12}
\ee
This decay could originate in an exchange of a Majorana neutrino in the $t$-channel topology as presented in Fig.~\ref{Fig0nu}.
Since the typical energy scales $\mu$ in the nucleus are of the order of the Fermi motion scale, $\mu \sim \mu_{\rm f} \sim 0.1$ GeV, the decay process can be regarded as a low-energy spacelike process. This means that the half-life $T^{0\nu}_{1/2} \equiv {\cal D}(Q^2)$ can be regarded as a spacelike observable with positive $Q^2 (\equiv - q^2) \sim \mu_{\rm f}^2 \sim 10^{-2} \ {\rm GeV}^2$. Furthermore, if the mass of the exchanged neutrino $N$ is heavy ($M_N \gg 0.1$ GeV), the process can be regarded as an effective pointlike process $dd \to uuee$. It can be called a short-range process, due to the high masses of the exchanged particles involved. On the other hand, such short-range (pointlike) process $dd \to uuee$ can originate also from an exotic physics \cite{Bonnet} which can be described effectively in terms of dimension-9 operators
\be
{\mathcal O}_{D=9} \sim \frac{1}{\Lambda^2_{\rm LNV}} {\bar u} {\bar u} d d {\bar e} {\bar e},
\label{OD9}   
\ee
where the scale of the new LNV-physics is expected to be $\Lambda_{\rm LNV} \gtrsim 10^2$ GeV. There are five classes of such effective (pointlike) operators (see the next Section). Since there is a very large difference between the new physics scale $\Lambda_{\rm LNV}$ and the $0\nu\beta\beta$ decay scale $\mu_{\rm f}$, the effects of the QCD corrections on the corresponding Wilson coefficients (which appear in the half-life quantity $T^{0\nu}_{1/2}$) for the process can be large. The other reason why these effects can be large lies in the color-mismatch contributions of the operators, this mismatch leads to the mixing of the operators at lower scales where the corresponding Wilson coefficients are multiplied with nuclear matrix elements (NMEs) which can have very different sizes. These short-range QCD effects can be explored by considering solutions of the renormalization group equations (RGEs) for the Wilson coefficients, and evolving them from the scales $\Lambda^2_{\rm LNV}$ of the new physics down to the Fermi motion scales $Q^2 \sim 0.01 \ {\rm GeV}^2$.

One important practical problem in such a calculation is that the mentioned RGEs, being (one- or two-loop) perturbative, are considered to involve the usual perturbative QCD coupling $a(Q^2)$ [$\equiv \alpha_s(Q^2)/\pi$] which, in turn, has the so called Landau singularities at low positive $Q^2 \lesssim 0.1 \ {\rm GeV^2}$. These singularities do not reflect the holomorphic behavior of the QCD spacelike observables ${\cal D}(Q^2)$ which must be holomorphic (i.e., analytic) functions of $Q^2$ in the $Q^2$-complex plane with the exception of a part of the negative axis: $Q^2 \in  \mathbb{C} \backslash (-\infty, -M_{\rm thr}^2]$ (where $M_{\rm thr} \sim 0.1$ GeV is a threshold scale) \cite{BS,Oehme}. The Landau singularities of perturbative QCD (pQCD) can therefore be considered as artificial, and they have their origin formally in the fact that the beta-function $\beta(a) \equiv d a(Q^2)/d \ln Q^2$ is assumed, as a function of $a$, to be a Taylor-expandable function around $a=0$ (such as a polynomial function, in the case of the $\MSbar$ scheme). This problem was addressed systematically, via Dispersion Relations (DR) for the coupling, for the first time by Shirkov and others in the nineties \cite{ShS,MS,Sh1Sh2,BMS} where a minimal analytic coupling [(F)APT: (fractional) analytic perturbation theory coupling] was constructed, $a(Q^2) \mapsto \A^{\rm (APT)}(Q^2)$, whose spectral function $\rho_{\A}(\sigma) \equiv {\rm Im} \A(Q^2=-\sigma - i \epsilon)$ was equal to the pQCD coupling spectral function for all positive $\sigma$ (i.e., negative $Q^2$), but without the Landau cut along the positive $Q^2$-axis, $\rho_{\A}(\sigma)=0$ for $\sigma < 0$. Several other holomorphic couplings $a(Q^2)$ have been constructed since then, in general modifying the discontinuity function $\rho_{\A}(\sigma)$ in the unknown nonperturbative regime of low positive $\sigma \lesssim 1 \ {\rm GeV}^2$. The couplings $\A(Q^2)$ in these approaches are in general expressed as a dispersive integral along its cut and involving the spectral function $\rho_{\A}(\sigma)$. Some of such couplings $\A(Q^2)$ attain a positive finite value at $Q^2 \to 0$ \cite{Nest2,Webber,Alekseev,CV12,1dAQCD,2dAQCD,2dCPC,Brod,Shirkovmass,KKS,Luna,Luna2,DSEdecoupFreez,PTBMF,Pelaez,Siringo,NestBook},\footnote{The holomorphic coupling of Refs.~\cite{Nest1} is infinite at $Q^2=0$.} and others attain the zero value $\A(0)=0$ \cite{Luna2,ArbZaits,Boucaud,mes2,FRGBraun,3dAQCD,Pelaez2}. All such holomorphic couplings,\footnote{For reviews of (F)APT, cf.~Refs.~\cite{Bakulev,reviews}; for additional applications of (F)APT and other $\A$QCD variants in the QCD phenomenology, see \cite{APTappl1,APTappl2,APTappl3}. Further, there exist related approaches where the dispersive method is applied directly to spacelike QCD quantities \cite{MSS1,MSS2,MagrGl,mes2,DeRafael,MagrTau,Nest3a,Nest3b,NestBook}.} i.e., couplings which are holomorphic functions of $Q^2$ for  $Q^2 \in  \mathbb{C} \backslash (-\infty, -M_{\rm thr}^2]$, are thus IR-safe and can be used in the mentioned solutions of the RGEs for the Wilson coefficients $C_j(Q^2)$, where $0 < Q^2 \sim {\mu}_{\rm f}^2 \sim 0.01 \ {\rm GeV}^2$.

In this endeavor, it is important to take into account that the analogs $\A_{\nu}(Q^2)$ of the powers $a(Q^2)^{\nu}$ (where $\nu$ is a power index with a real number value, $-1 < \nu$) are not simple powers $\A(Q^2)^{\nu}$, as already pointed out in \cite{CV12} for integer $\nu$ and later in \cite{GCAK} for general real $\nu$; in those references, the power analogs $\A_{\nu}$ were constructed in the general QCD framework ($\A$QCD) with holomorphic coupling\footnote{In the case of the minimal analytic QCD (FAPT), the existence of such analogs $\A_{\nu}$ ($\not= \A^{\nu}$) was pointed out and their construction presented in \cite{Sh1Sh2}, and an explicit FAPT construction was performed in \cite{BMS,Bakulev}.} $\A(Q^2)$.

In this work, we present in Sec.~\ref{sec:EFT} the effective Lagrangian made up of short-range dimension-9 operators, and the expression for the $0\nu\beta\beta$ half-life in terms of the Wilson coefficients of these operators at (low) Fermi motion scales and in terms of the NMEs. In Sec.~\ref{sec:RGE} we then describe the RGEs governing the evolution of the mentioned Wilson operators, and we gather the hitherto known explicit expressions of the one-loop and two-loop anomalous dimensions in Appendix \ref{app:AD}. In Sec.~\ref{sec:AQCD} we then describe the general solution of the corresponding one-loop and two-loop RGEs in the IR-safe $\A$QCD frameworks. In Appendix \ref{app:AQCD} we provide more details of the $\A$QCD formalism and a brief description of the specific $\A$QCD frameworks used in this work. In Appendix \ref{app:RGEWils} we write down the solution of the coupled system of RGEs in the case of mixing of operators. In particular, we present there the solution for the case of the degenerate mixing which, to our knowledge, has not been considered in the literature and appears in the case of the operator mixing of ${\mathcal O}_3^{LR}$-${\mathcal O}_1^{LR}$ at two-loops at low scales ($n_f=3$). In Sec.~\ref{sec:num} we present our numerical results for the RGE evolution matrices at low (sub-GeV) scales in various $\A$QCD frameworks. In addition, we present there the resulting upper bounds on the various ``bare'' LNV Wilson coefficients $C_j(\Lambda_{\rm LNV})$  (we took $\Lambda_{\rm LNV}=M_W$), where these bounds originate from the experimental lower bound on the half-life (\ref{T12}). Section \ref{sec:conc} is a summary of our conclusions.

\section{Effective Lagrangian in $0\nu\beta\beta$ decay}
\label{sec:EFT}
The effective Lagrangian within the Operator Product Expansion (OPE) formalism for the dimension-9 operators, which originate from short-range new physics and contribute to $0\nu\beta\beta$ decay, have the generic structure \cite{Pas:2000vn}
\begin{equation}
\mathcal{L}_{\mathrm{eff}}^{0, \nu \beta \beta}=\frac{G_{F}^{2}}{2 m_{p}} \sum_{i=1}^5 \sum_{X Y} C_{i}^{X Y}(\mu) \ \mathcal{O}_{i}^{X Y}(\mu)\ ,
\label{Lagr}
\end{equation}
where $G_{F}=1.166 \times 10^{-5} \ {\rm GeV}^{-2}$ is the Fermi constant, $m_{p}$ is the proton mass. The expansion (\ref{Lagr}) contains five types of dimension-9 operators; the indices ${X,Y}={L,R}$ indicate the chirality. The dimension-9 operators $\mathcal{O}_{i}^{X Y}$ can be shown in the compact notation \cite{GHK2016,GHK2018}
\bes
\label{ops}
\begin{eqnarray}
\mathcal{O}_{1}^{X Y}&=&4\left(\bar{u} P_{X} d\right)\left(\bar{u} P_{Y} d\right) j\ ,
\label{ssOp}
\\
\mathcal{O}_{2}^{X X}&=&4\left(\bar{u} \sigma^{\mu \nu} P_{X} d\right)\left(\bar{u} \sigma_{\mu \nu} P_{X} d\right) j\ ,
\label{ttOp}
\\
\mathcal{O}_{3}^{X Y}&=&4\left(\bar{u} \gamma^{\mu} P_{X} d\right)\left(\bar{u} \gamma_{\mu} P_{Y} d\right) j\ ,
\label{vvOp}
\\
\mathcal{O}_{4}^{X Y}&=&4\left(\bar{u} \gamma_{\nu} P_{X} d\right)\left(\bar{u} \sigma^{\nu \mu} P_{Y} d\right) j_{\mu} = \mathcal{O}_{4}^{X Y \mu}  j_{\mu},
\label{vtOp}
\\ \mathcal{O}_{5}^{X Y}&=&4\left(\bar{u} \gamma^{\mu} P_{X} d\right)\left(\bar{u} P_{Y} d\right) j_{\mu}\ = \mathcal{O}_{5}^{X Y \mu}  j_{\mu},
\label{vsOp}
\end{eqnarray}
\ees
where $j=\bar{e}\left(1 \pm \gamma_{5}\right) e^{c}, j_{\mu}=\bar{e} \gamma_{\mu} \gamma_{5} e^{c}$ are the lepton currents. In Eqs.~(\ref{ttOp}) and (\ref{vtOp}) we use the convention $\sigma^{\mu \nu} = (i/2) [\gamma^{\mu},\gamma^{\nu}]$.
In general, these operators mix under renormalization through QCD when we express them in terms of a color singlet structure. In this procedure, the following property is used:
\begin{equation}
\left(\lambda^{a}\right)_{\alpha \beta} \left(\lambda^{a}\right)_{\eta \xi} =-\frac{2}{N} \delta_{\alpha \beta} \delta_{\eta \xi} +2 \delta_{\alpha \xi} \delta_{\eta \beta},
\label{FierzColor}
\end{equation}
where $\lambda^a = 2 t^a$ are the Gell-Mann matrices.
This leads to the original operator [the first term on the RHS of Eq.(\ref{FierzColor})] plus a color mismatch part [the second term on the RHS of Eq.(\ref{FierzColor})].
Note that $\mathcal{O}_{2}^{X Y}=0$ for $X\neq Y$.

The effective Lagrangian (\ref{Lagr}) at high physics scales $\mu=\Lambda_{\rm LNV}$ ($\sim 10^2$-$10^3$ GeV) represents the new short-range physics. When these contributions are evolved to lower scales $\mu$, the QCD effects are the dominant contributions to the RGE evolution. The effective Lagrangian (\ref{Lagr}) must be evaluated down to a spacelike scale $\mu^2=Q^2 (\equiv -q^2)$ that enters in the $0\nu\beta\beta$ process (and in the corresponding NMEs), typically of the order of the Fermi motion scale $\mu \sim \mu_{\rm f} \sim 0.1$ GeV. In practice, when we use the Lagrangian (\ref{Lagr}) in the perturbation theory within pQCD, it is applicable only down to $\mu\sim 1$ GeV in the best scenario. This RGE-running for $0\nu\beta\beta$ decay was performed in pQCD, at one-loop level of anomalous dimensions, in \cite{Mahajan} for the set of operators ${\mathcal O}_1$-${\mathcal O}_3$, and in \cite{GHK2016} for the set ${\mathcal O}_1$-${\mathcal O}_5$. The restriction $\mu \gtrsim 1$ GeV is due to unphysical singularities, known as Landau singularities, in the pQCD running coupling at $\mu^2\sim\Lambda_{QCD}^2\approx 10^{-1}$ GeV$^2$ for $n_f=3$ active flavors. In the vicinity of these singularities our physical predictions are jeopardized. The experience shows that if we do not include some nonperturbative effects, the applicability of this series extends only down to $\mu \approx 1-2$ GeV.
 
Based on the Lagrangian (\ref{Lagr}), we can calculate the amplitude and then the $0\nu\beta\beta$ half-life as \cite{Doi:1985dx}
\begin{equation}
\left[T_{1 / 2}^{0 \nu \beta \beta}\right]^{-1}=G_1 \left| \sum_{j=1}^{3} C_{j}(Q^2_{\rm f}) \mathcal{M}_{j} \right|^{2} + G_4 \left| \sum_{j=4}^{5} C_{j}(Q^2_{\rm f}) \mathcal{M}_{j} \right|^{2} 
\label{halflife}
\end{equation}
Here, $Q^2_{\rm f} \sim 0.01 \ {\rm GeV}^2$ is the squared energy of the (spacelike) process of $0 \nu \beta \beta$ decay, $G_j$ are the phase space factors ($G_1=G_2=G_3$, and $G_4=G_5$) \cite{GDIK}, and $\mathcal{M}_{j}$ is the Nuclear Matrix element (NME) of the operator ${\mathcal O}_j$ Eqs.~(\ref{ssOp})-(\ref{vsOp}) at an effective squared energy $Q^2_{\rm f}$, $\mathcal{M}_j = \langle A_{\rm fin} | \mathcal{O}_j(Q^2_{\rm f}) | A_{\rm in} \rangle$. These constant parameters depend on the isotopes we are considering. For the considered isotope $^{136}{\rm Xe}$, the values of NMEs are given in Table \ref{tabNMEs}.
\begin{table}
  \caption{The values of the nuclear matrix elements for $^{136}{\rm Xe}$, from Ref.~\cite{DHP} (cf.~also \cite{GHK2016}), at effective Fermi motion scales $Q^2_{\rm f} = \mu^2_{\rm f} \sim 0.01 \ {\rm GeV}^2$.}
\label{tabNMEs}
\begin{ruledtabular}
\begin{tabular}{llllll}
 ${\cal M}_1$ &  ${\cal M}_2$ &     ${\cal M}_3^{(+)}$ &     ${\cal M}_3^{(-)}$ &     $|{\cal M}_4|$ &     $|{\cal M}_5|$
  \\
  \hline
  $4.5$ & $-8.5 \times 10^2$ & $6.9 \times 10^1$ & $1.1 \times 10^2$ & $9.6 \times 10^1$ & $9.3$
  \end{tabular}
\end{ruledtabular}
\end{table}
The Wilson coefficient $C_{j}(Q^2_{\rm f})$ depends on the typical scale of the $0\nu\beta\beta$ process, and as we mentioned above, this spacelike scale is quite low and some extension of the usual QCD should be taken into account. In Ref.\cite{GHK2018} the authors considered a freezing of the QCD running coupling based on the inclusion of an effective glueball mass $M$, where  $M^2 \in (0.4, 5.0) \ {\rm GeV}^2$. This inclusion was made by the shift $Q^2 \rightarrow Q^2+M^2$ in the one-loop pQCD coupling, cf.~Eq.~(\ref{AMPT}). In Ref.~\cite{Ciretal} the authors cut the effective theory of QCD at a reasonable scale (see the argumentation above) $Q=2$ GeV, and below it down to $\mu=0.1$ GeV they considered a new effective theory without quarks called Chiral Perturbation Theory (ChiPT) and many-body methods (cf.~also \cite{Prez,Grae}). Finally, in Ref.\cite{AGK} the pion mechanism is considered, where the hadronization of quarks and gluons is produced within the effective vertices given by operators (\ref{ssOp})-(\ref{vsOp}).

In the present work, we propose an alternative method to deal with this low-energy problem. We propose to extend the applicability of QCD through the Dispersion Relations, which are integrals in the complex $Q^2$-plane, which allow us to avoid the appearance of the Landau singularities in a natural way. For details on the construction of such models, we refer to Appendix \ref{app:AQCD}.

\section{Renormalization Group Equations within QCD}
\label{sec:RGE}

The renormalized effective operators (\ref{ssOp})-(\ref{vsOp}) are scale independent. Then the Renormalization Group Equation (RGE) will define the anomalous dimension matrix ${\hat \gamma}$ in the form

\begin{equation}
\frac{d \overrightarrow{\mathcal{O}}(Q^2)}{d \ln Q^2}= \frac{d \ln \boldsymbol{Z}(Q^2)}{d \ln Q^2}\equiv-\frac{1}{2} {\hat \gamma}(Q^2) \overrightarrow{\mathcal{O}}(Q^2),
\label{RGEO}
\end{equation}
where the renormalization constant matrices $\boldsymbol{Z}$ of the effective operators imply that we will have in general some mixing between them. The scale $Q^2$ is considered to be spacelike, i.e., $Q^2 \equiv - q^2$ is regarded to be nonnegative.
Now, the RGE for the Wilson coefficient follows from the fact that the Lagrangian in (\ref{Lagr}) is independent of the (spacelike) renormalization scale $\mu^2 \equiv Q^2$. As a consequence, we obtain the RGE in the matrix form
\begin{equation}
\frac{d \vec{C}(Q^2)_{\rm pt}}{d \ln Q^2}=\frac{1}{2} {\hat \gamma}^{T}(Q)_{\rm pt} \vec{C}(Q^2)_{\rm pt}.
\label{RGEw}
\end{equation}
The anomalous dimension matrix ${\hat \gamma}(Q^2)$ is extracted from the renormalization of the composite operators (\ref{ssOp})-(\ref{vsOp}). The corresponding available anomalous dimension factors and matrices are collected in Appendix \ref{app:AD}: for the operators $\mathcal{O}_1$-$\mathcal{O}_3$ from Ref.~\cite{Buras:2000if} (at the one-loop and two-loop level); for the operators $\mathcal{O}_4$-$\mathcal{O}_5$ from Refs.~\cite{LMW,Ciretal} (at the one-loop level).

If we rewrite Eq.~(\ref{RGEw}) in terms of the pQCD running coupling $a(Q^2) \equiv \alpha_s(Q^2)/\pi$, the RGE can be solved at the two-loop level explicitly, and it is given in the form (for the case of no mixing, i.e., ignoring the problems of diagonalization)
\bes
\label{RGEwCoeff}
\bea
C(a)_{\rm pt} &=& \left( \frac{a}{a_0} \right)^{\nu} \left( \frac{1 + c_1 a}{1 + c_1 a_0} \right)^{k^{(1)}/c_1} C(a_0),
\label{RGEwCoeffa}
\\
& = & \frac{\left[ a^{\nu} + k^{(1)} a^{\nu+1} + {\cal O}(a^{\nu+2}) \right]}{\left[ a_0^{\nu} + k^{(1)} a_0^{\nu+1} + {\cal O}(a_0^{\nu+2}) \right]} C(a_0),
\label{RGEwCoeffb}
\eea
\ees
where $a \equiv a(Q^2)$ and $a_0 \equiv a(Q^2_0)$; for $\beta(a)$ which appears in the renormalization group equation (RGE) Eq.~(\ref{RGE}) for the running coupling $a(Q^2)$, we took the two-loop truncated form $\beta(a) = - \beta_0 a^2 (1 + c_1 a)$. The constants $\nu$ and $k^{(1)}$ appearing in Eqs.~(\ref{RGEwCoeff}) are
\be
\nu= - \frac{1}{8 \beta_0} \gamma^{(0)}, \quad
k^{(1)} = - \frac{1}{32 \beta_0} \gamma^{(1)} - c_1 \nu,
\label{nuk1}
\ee
where $\gamma^{(j)}$ ($j=0,1$) are the one-loop and two-loop coefficients, respectively, in the anomalous dimension matrix ${\hat \gamma}$ 
\be
{\hat \gamma}(a) = {\hat \gamma}^{(0)} \frac{a(Q^2)}{4} +  {\hat \gamma}^{(1)} \left( \frac{a(Q^2)}{4} \right)^2 + \ldots
\label{hatgammaexp}
\ee
For more details and for different cases of mixing, we refer to Appendix \ref{app:RGEWils}. We note that in the expansion in Eq.~(\ref{RGEwCoeffb}) the terms ${\cal O}(a^{\nu+2})$ are not known if the three-loop anomalous dimension coefficient ${\hat \gamma}^{(2)}$ is not known.

In the case of mixing, the analogous formulas for pQCD are obtained in Appendices \ref{app:mixnondeg} and \ref{app:mixdeg} for the nondegenerate ($\nu_1 - \nu_2 \not=1$) and degenerate case ($\nu_1 - \nu_2 =1$): cf.~Eqs.~(\ref{hatgamma0}), (\ref{katk1}), and (\ref{vCres})-(\ref{hatU1}) for the nondegenerate case, and additionally Eq.~(\ref{hatU1deg}) for the degenerate case. According to our knowledge, the solution of the two-loop RGE for Wilson coefficients in the degenerate case [which occurs in the $n_f=3$ regime for the (31)$^{XY}$ mixing of operators ${\mathcal O}_3^{XY}$ and ${\mathcal O}_1^{XY}$ ($X \not= Y$)] has not been addressed in the literature hitherto.

Within the evolution procedure, the heavy quark thresholds should be taken into account. For this purpose, the evolution matrix $U(Q^2_{\rm f}, {\Lambda}^2_{\rm LNV})$, which connects the ``bare'' ${\vec C} \equiv {\vec C}({\Lambda}^2_{\rm LNV})$ at high momenta with the physical ${\vec C}(Q^2_{\rm f})$ at Fermi-motion monenta
\be
{\vec C}(Q_{\rm f}^2) = U(Q^2_{\rm f}, {\Lambda}^2_{\rm LNV}) {\vec C},
\label{Uevoldef}
\ee
can be written in the following way:
\bes
\label{Uevol}
\begin{eqnarray}
\hat{U}\left(Q^2_{\rm f}, {\Lambda}^2_{\rm LNV}=M^2_{W}\right) &=&\hat{U}^{(n_f=3)}\left(Q^2_{\rm f}, Q^2_{c}\right) \hat{U}^{(n_f=4)}\left(Q^2_{c}, Q^2_{b}\right) \hat{U}^{(n_f=5)}\left(Q^2_{b}, M^2_{W}\right), 
\label{Uevola}
\\ 
\hat{U}\left(Q^2_{\rm f}, {\Lambda}^2_{\rm LNV} > {\overline m}^2_{t}\right) &=&\hat{U}^{(n_f=3)}\left(Q^2_{\rm f}, Q^2_{c}\right) \hat{U}^{(n_f=4)}\left(Q^2_{c}, Q^2_{b}\right) \hat{U}^{(n_f=5)}\left(Q^2_{b}, Q^2_{t}\right) \hat{U}^{(n_f=6)}\left(Q^2_{t}, {\Lambda}^2_{\rm LNV} \right),
\label{Uevolb}
\end{eqnarray}
\ees
where the first equality is given for matching scale of the order of W-boson mass $M_W=80.379$ GeV \cite{PDG18}, and the second equality for large scales, where the theories beyond the standard model play a crucial role. In Eqs.~(\ref{Uevol}), the heavy quark thresholds are at $Q_t=\kappa\overline{m}_t=163\kappa$ GeV; $Q_b=\kappa\overline{m}_b=4.20\kappa$ GeV; and $Q_c=\kappa\overline{m}_c=1.27\kappa$ GeV \cite{PDG18}, where we will choose $\kappa=2$ (in general, $\kappa \sim 1$). Note that the variation of the threshold parameter $\kappa$ is numerically not important in comparison with variation of other parameters. 

We will use Eq.~(\ref{Uevola}), i.e., we will take $\Lambda_{\rm LNV}=M_W$ throughout.\footnote{If taking $\Lambda^2_{\rm LNV} =1$ TeV, the numerical results for the extracted upper bounds on the ``bare'' Wilson coefficients in general change by significantly less than 50 percent, cf.~\cite{GHK2016}.}  In the $n_f=3$ regime, we will use $\A$QCD, because the realistic Fermi motion scale $Q^2_{\rm f} \approx 0.01 \ {\rm GeV}^2$ in this regime is quite low and the deviation of the $\A$QCD couplings from the underlying pQCD couplings is significant. In the regimes $n_f \geq 4$ we use the underlying pQCD in $3 \delta$ $\A$QCD because there the $\A(Q^2)$ coupling practically coincides with the underlying pQCD coupling $a(Q^2)$ [Eq.~(\ref{diffAaN}) has ${\cal N}=5$]. In two other cases ($2 \delta$ $\A$QCD, FAPT) we use at $n_f \geq 4$ the corresponding $\A$QCD couplings out of convenience (because those coupling are available for all $n_f$). In the one-loop massive QCD (Massive Perturbation Theory: MPT) we used for $n_f \geq 4$ the underlying pQCD, for simplicity.\footnote{
  In FAPT and massive one-loop QCD (MPT), we have ${\cal N}=1$ in Eq.~(\ref{diffAaN}); in $2 \delta$ $\A$QCD we have ${\cal N}=5$, so it is practically equivalent to use $\A$QCD or the underlying pQCD couplings in the $n_f \geq 4$ regimes. Furthermore, on the basis of construction of $\A_{\nu}$ as explained in \cite{GCAK}, it is possible to show that from Eq.~(\ref{diffAaN}) we obtain $\A_{\nu}(Q^2) - a(Q^2)^{\nu} \sim (\Lambda^2/Q^2)^{\cal N}$ for all $-1 < \nu$.}

\section{Evaluation of RGE with IR-safe couplings}
\label{sec:AQCD}

As mentioned in the Introduction, in $\A$QCD the coupling $a(Q^2)=\alpha_s(Q^2)/\pi$ gets replaced by a coupling $\A(Q^2)$ where the latter reflects correctly the holomorphic (analytic) behavior of the spacelike QCD physical quantities ${\cal D}(Q^2)$. This means that $\A(Q^2)$, in contrast to $a(Q^2)$, has no Landau singularities in the complex $Q^2$-plane, or equivalently, $\A(Q^2)$ is a holomorphic function for $Q^2 \in \mathbb{C} \backslash (-\infty, -M_{\rm thr}^2]$ where $M_{\rm thr}^2$ is a positive threshold scale, $M_{\rm thr} \sim 0.1 \ {\rm GeV}^2$.\footnote{
Usually the Landau singularities of a pQCD coupling $a(Q^2)$ are cuts on the positive $Q^2$ axis, $Q^2 \in (0, \Lambda^2_{\rm Lan.})$ where $\Lambda^2_{\rm Lan.} \sim 0.1$-$1 \ {\rm GeV}^2$. The details of these singularities depend on the chosen (pQCD) renormalization scheme.} 
Here we refer to Appendix \ref{app:AQCD} for various $\A$QCD variants. Usually they are constructed with the dispersion relation approach, i.e., starting with a specific form of the discontinuity function $\rho_{\A}(\sigma) = {\rm Im} \A(Q^2= \sigma \exp(- i \pi))$ for positive $\sigma$, and the holomorphic coupling $\A(Q^2)$ is a dispersion integral involving  $\rho_{\A}(\sigma)$, cf.~Eq.~(\ref{dispA}). Due to the asymptotic freedom, $\rho_{\A}(\sigma)$ at large $\sigma > 1 \ {\rm GeV}^2$ (practically) coincides with the discontinuity function $\rho_a(\sigma)$ of the underlying pQCD coupling $a(Q^2)$ (the latter is defined in a specific chosen renormalization scheme). At low positive $\sigma \lesssim 1 \ {\rm GeV}^2$, the discontinuity function $\rho_{\A}(\sigma)$ is in principle unknown and can be parametrized with Dirac-delta functions, cf.~Eqs.~(\ref{rhoAnd}) for $n=2,3$ ($2 \delta$ and $3 \delta$ $\A$QCD), Eq.~(\ref{rhoMPT}) for one-loop ``massive'' coupling (MPT). In (Fractional) Analytic Perturbation Theory [(F)APT], the discontinuity function $\rho_{\A}(\sigma)$ is considered to coincide with its pQCD analog $\rho_a(\sigma)$ for all $\sigma$ values (all the way down to $\sigma=0$), cf.~Eq.~(\ref{dispAAPT}).   

In $\A$QCD,  the powers $a(Q^2)^{\nu +m}$ ($m=0,1,\ldots$) get replaced by their analogs as explained in Eqs.~(\ref{Anugen}), (\ref{tAn11l})
\bea
a(Q^2) & \mapsto & \A(Q^2); \; a(Q^2)^{\nu} \mapsto \A_{\nu}(Q^2) \; \left[ \not= \A(Q^2)^{\nu} \right].
\label{analyt}
\eea
Various $\A$QCD variants [$n \delta$ $\A$QCD ($n=2,3$), FAPT, and massive one-loop $\A$QCD (MPT)] are summarized in Appendix \ref{app:AQCD}. In the following we will argue that in $\A$QCD the result for the Wilson coefficient $C(Q^2)$ is really obtained from the pQCD result (\ref{RGEwCoeffb}) by the replacements (\ref{analyt}). We will show this in the case of no mixing, while the extension to the case of mixing of operators is given in Appendix \ref{app:RGEWils}.

The renormalization group equation (RGE) for a Wilson coefficient $C(Q^2)$ as a function of the effective spacelike scale $Q^2$ in pQCD has the form\footnote{
We use the conventions of \cite{Buras:2000if} (see also Appendix \ref{app:AD}), and our notations $a(Q^2) \equiv \alpha_s(Q^2)/\pi$.}
\be
\frac{d C(Q^2)_{\rm pt}}{d \ln Q^2} = \frac{1}{2} \left[ \sum_{n \geq 0} \left(\frac{a(Q^2)}{4} \right)^{n+1} \gamma^{(n)} \right] C(Q^2)_{\rm pt},
\label{pRGEWils}
\ee
where $\gamma^{(n)}$ are the ($n+1$-loop) coefficients of the anomalous dimension, and the pQCD expansion of $C(Q^2)$ in terms of $a(Q^2)$ has the form [cf.~Eq.~(\ref{RGEwCoeffb})]
\be
C(Q^2)_{\rm pt} = {\cal C} \left[ a(Q^2)^{\nu} + \sum_{j \geq 0} k^{(j)} a(Q^2)^{\nu +j} \right],
\label{Cpt}
\ee
where ${\cal C}$ is a $Q^2$-independent quantity.
Using this expansion in the RGE (\ref{pRGEWils}), and the RGE running of the pQCD coupling $a(Q^2)$ according to Eq.~(\ref{RGE}), it is straghtforward to see that the index $\nu$ and the expansion coefficients $k^{(j)}$ are [cf.~Eq.~(\ref{nuk1}) for the two-loop case]
\bes
\label{nukjpQ}
\bea
\nu & = & -\frac{1}{8 \beta_0} \gamma^{(0)},
\label{nu}
\\
k^{(1)} & = & - \frac{1}{32 \beta_0} \gamma^{(1)} - c_1 \nu,
\label{k1}
\\
k^{(2)} & = & -\frac{1}{2} \left( \frac{1}{128 \beta_0} \gamma^{(2)} + c_2 \nu \right) + \frac{1}{2} k^{(1)} (k^{(1)} - c_1),
\label{k2}
\eea
\ees
etc. The RGE in $\A$QCD is obtained by making analytic the LHS and the RHS of the RGE (\ref{pRGEWils}) where the expansion of $C(Q^2)$ has the form (\ref{Cpt}). This is performed with the replacements $a^{\nu} \mapsto \A_{\nu}$ as explained in Appendix \ref{app:AQCD}
\bea
\lefteqn {
{\cal C} \frac{d}{d \ln Q^2} \left[ \A_{\nu}(Q^2) + k^{(1)} \A_{\nu+1}(Q^2) + k^{(2)} \A_{\nu+2}(Q^2) + \cdots \right] =
}
\nonumber\\ &&
{\cal C} \frac{1}{2} {\Bigg \{} \left[ \frac{a(Q^2)}{4} \gamma^{(0)} + \left( \frac{a(Q^2)}{4} \right)^2 \gamma^{(1)} +   \left( \frac{a(Q^2)}{4} \right)^3 \gamma^{(2)} + \ldots \right] \left[ a(Q^2)^{\nu} + k^{(1)} a(Q^2)^{\nu +1} + k^{(2)} a(Q^2)^{\nu +2} + \ldots \right] {\Bigg \}}_{\rm an.} =
\nonumber\\ &&
{\cal C} \frac{1}{8} {\Bigg \{} \A_{\nu+1}(Q^2) \gamma^{(0)} +  \A_{\nu+2}(Q^2) \left( \frac{1}{4} \gamma^{(1)} + k^{(1)} \gamma^{(0)} \right) + \A_{\nu+3}(Q^2) \left( \frac{1}{4^2} \gamma^{(2)} + \frac{1}{4} k^{(1)} \gamma^{(1)} + k^{(2)} \gamma^{(0)} \right) + {\cal O}(\A_{\nu+4}) {\Bigg \}}.
\label{RGEWils1}
\eea
One may wonder whether in $\A$QCD the index $\nu$ and coefficients $k^{(j)}$ ($j=1,2,\ldots$) are the same as in pQCD Eq.~(\ref{nukjpQ}); they turn out to be the same. Namely,
the LHS of the above RGE (i.e., the first line), when using the $\A$QCD relations (\ref{AnuRGE})-(\ref{anuRGE}), can be shown be equal to
\be
{\rm LHS} \equiv {\cal C} (- \beta_0) {\Big \{} \A_{\nu+1}(Q^2) \nu + \A_{\nu+2}(Q^2) \left[ (\nu+1) k^{(1)} + \nu c_1 \right] +  \A_{\nu+3}(Q^2) \left[(\nu+2) k^{(2)} + (\nu+1) c_1 k^{(1)} + \nu c_2 \right] + {\cal O}(\A_{\nu+4}) {\Big \}}.
\label{LHSRGEWils1}
\ee
When we equate this expression with the RHS [i.e., the last line in Eq.~(\ref{RGEWils1}), we obtain for $\nu$ and $k^{(j)}$ ($j=1,2$) the same expressions Eqs.~(\ref{nukjpQ}) as obtained by the pQCD approach.

The conclusion of this exercise is that the solution of the RGE for Wilson coefficients $C(Q^2)$ in $\A$QCD is the same as in pQCD, with the replacements $a(Q^2)^{\nu+m} \mapsto \A_{\nu+m}(Q^2)$ in the pQCD expansion (\ref{Cpt}).

Therefore, the relation (\ref{RGEwCoeffb}) in $\A$QCD obtains the form
\bes
\label{C2lA}
\bea
C(Q^2)_{(\A)} & = & \left[\A_{\nu}(Q^2) + k^{(1)} \A_{\nu+1}(Q^2) + {\cal O}(\A_{\nu+2}) \right] {\cal C},
\label{C2lAa}
\\
&=& \frac{\left[ \A_{\nu}(Q^2) + k^{(1)} \A_{\nu+1}(Q^2) + {\cal O}(\A_{\nu+2}) \right]}{\left[ \A_{\nu}(Q_0^2) + k^{(1)} \A_{\nu+1}(Q_0^2) + {\cal O}(\A_{\nu+2})  \right]} C(Q_0^2)_{(\A)} \equiv U(Q^2;Q_0^2)_{(\A)}  C(Q_0^2)_{(\A)},
\label{C2lAb}
\eea
\ees
where the above expression $U(Q^2; Q_0^2)_{(\A)}$ is the RGE-evolution matrix in $\A$QCD for the Wilson coefficient from an effective (higher) scale $Q_0^2$ to an effective (lower) scale $Q^2$.

In the case of mixing, the analogous formulas for $\A$QCD are obtained in Appendices \ref{app:mixnondeg} and \ref{app:mixdeg} for the nondegenerate ($\nu_1 - \nu_2 \not=1$) and degenerate case ($\nu_1 - \nu_2 =1$): cf.~Eqs.~(\ref{hatgamma0}), (\ref{katk1}), and (\ref{AQCDresmix})-(\ref{AvCres}) for the nondegenerate case, and additionally Eqs.~(\ref{AQCDhatU1deg})-(\ref{AQCDvCresdeg2}) for the degenerate case.

In the general approach, applied in $n \delta$ $\A$QCD ($n=2,3$) and in one-loop ``massive'' $\A$QCD (MPT), where the general power analogs $\A_{\nu}$ are constructed via the generalized logarithmic-derivative analogs $\tA_{\nu+m}$, Eqs.~(\ref{Anugen}), it is important to apply the truncations in the evaluation of $\A_{\nu}$ in Eq.~(\ref{AnutAnu}) consistent with the loop-level in the expression for the Wilson coefficients. When the anomalous dimension $\gamma(a)$ is known only at one-loop level, then we have
\be
C(Q^2)_{(\A)} = \A_{\nu}(Q^2) \; {\cal C} = \frac{\A_{\nu}(Q^2)}{\A_{\nu}(Q_0^2)} C(Q_0^2)_{(\A)},
\label{C1lS}
\ee
and the expression in Eq.~(\ref{AnutAnu}) has only one term
\be
\A_{\nu}(Q^2) = \tA_{\nu}(Q^2).
\label{Anu1l}
\ee
On the other hand, when the anomalous dimension is known at the two-loop level, Eq.~(\ref{C2lA}), then the expression in Eq.~(\ref{AnutAnu}) has two terms\footnote{According to Ref.~\cite{GCAK}, $\tk_1(\nu)=-k_1(\nu)= - c_1 \nu ( H(\nu)-1)$ where  $H(\nu)$ is the Harmonic Number function.}
\be
\A_{\nu}(Q^2) = \tA_{\nu}(Q^2) + \tk_1(\nu) \tA_{\nu+1}(Q^2).
\label{Anu2l}
\ee
In practice, this implies that the (two-loop) expression (\ref{C2lA}) obtains the form
\bes
\label{C2ltA}
\bea
C(Q^2)_{(\A)} & = & \left[ \tA_{\nu}(Q^2) + \left( k^{(1)} + \tk_1(\nu) \right) \tA_{\nu+1}(Q^2) + {\cal O}(\tA_{\nu+2}) \right] \; {\cal C}
\label{C2ltAa}
\\
&=& \frac{\left[ \tA_{\nu}(Q^2) + \left( k^{(1)} + \tk_1(\nu) \right) \tA_{\nu+1}(Q^2) + {\cal O}(\tA_{\nu+2}) \right]}{\left[ \tA_{\nu}(Q_0^2) + \left( k^{(1)} + \tk_1(\nu) \right) \tA_{\nu+1}(Q_0^2) + {\cal O}(\tA_{\nu+2}) \right]} \; C(Q_0^2)_{(\A)} ,
\label{C2ltAb}
\eea
\ees
where we consistently ignore the terms  $~\sim \tA_{\nu+2}$.
It turns out that with such evaluation we get, even at low $|Q^2| < 1 \ {\rm GeV}^2$, reasonable convergence behavor for the Wilson coefficient $C(Q^2)$ when going from the one-loop to the two-loop case, see Sec.~\ref{sec:num}. This is probably related with the fact that in $\A$QCD the numerical hierarchy $|\tA_{\nu}(Q^2)| \gtrsim  |\tA_{\nu+1}(Q^2)| \gtrsim  |\tA_{\nu+2}(Q^2)| \gtrsim \ldots$ is valid in general not just for high $|Q^2|$ but even for low $|Q^2| \lesssim 1 \ {\rm GeV}^2$. There is no such hierarchy in pQCD, because of the Landau singularities at or close to $|Q^2| \lesssim 1 \ {\rm GeV}^2$. In the case of mixing, analogous approach is applied for $\nu=\nu_1$ and $\nu=\nu_2$, and we refer to Appendix \ref{app:RGEWils} for more details.

In the case of FAPT, although the use of Eqs.~(\ref{Anu1l})-(\ref{Anu2l}) is an entirely acceptable option, we will follow the more special FAPT-type approach as described in Eqs.~(\ref{FAPT1})-(\ref{FAPT2}). In the case of FAPT, this is equivalent to the evaluation of $\A_{\nu}^{\rm (FAPT)}$ as a nontruncated (resummed) sum of $\tA_{\nu+m}^{\rm (FAPT)}$ ($m=0,1,\ldots$), i.e., Eq.~(\ref{AnutAnu}) with $N \to \infty$. 

As mentioned at the end of Sec.~\ref{sec:RGE}, $\A$QCD will be applied here always in the $n_f=3$ (low-$Q^2$) regime. In the regimes $n_f \geq 4$ in general the underlying pQCD approach will be applied; in FAPT and $2 \delta$ $\A$QCD, the $\A$QCD approach will be applied also in the regimes $n_f \geq 4$ for the aformentioned reasons of conveniency.

In the cases where the anomalous dimension ${\hat \gamma}$ is known up to two-loop level, we have the mixing '(12)$^{XX}$' (of ${\mathcal O}_1^{XX}$ and ${\mathcal O}_2^{XX}$) and '(31)$^{XY}$' (of ${\mathcal O}_3^{XY}$ and ${\mathcal O}_1^{XY}$; $X \not= Y$). It turns out that in the $n_f=3$ regime we have the degeneracy $\nu_1-\nu_2=1$ in the case of mixing '(31)$^{XY}$', i.e., formulas of Appendix \ref{app:mixdeg} apply ($\nu_1=8/9$ and $\nu_2=-1/9$). The case of the mixing '(12)$^{XX}$', on the other hand, is nondegenerate, and the formulas of Appendix \ref{app:mixnondeg} apply ($\nu_1=-0.6120$ and $\nu_2=0.5379$, when $n_f=3$).

\section{Numerical results}
\label{sec:num}

\subsection{Evolution matrix elements for Wilson coefficients}
\label{subsec:evol}

For evaluation of QCD correction to the $0\nu\beta\beta$-decay, we need the physical observable, i.e., the half-life quantity based on OPE. The first question is how the evolution factors or matrices $U(Q^2_{\rm f};\Lambda^2_{\rm LNV})$ [cf.~Eqs.~(\ref{Uevoldef}) and (\ref{Uevola})] behave when the Fermi motion scale $Q^2_{\rm f}$ varies downwards towards the realistic values $Q^2_{\rm f} \sim 0.01 \ {\rm GeV}^2$. We will apply a variety of $\A$QCD frameworks: $3 \delta$ $\A$QCD \cite{3dAQCD,MathPrgs} which has the zero limit in deep IR regime, $\A(0)=0$; $2 \delta$ $\A$QCD \cite{2dAQCD,2dCPC,MathPrgs} and the one-loop ``massive'' $\A$QCD (MPT) Eq.~(\ref{AMPT}), all these having finite positive IR limit $\A(0) > 0$; and FAPT in the $\MSbar$ scheme, cf.~Eq.~(\ref{FAPT2}), which gives a nonholomorphic $\A(Q^2)$ in the point $Q^2=0$ [but has also $\A(0) >0$].

The MPT coupling (\ref{AMPT}) used is taken in two variants:
(I) The first one is with $M=1.5$ GeV (for $n_f=3$) and with the scale $\Lambda_{3}$ (i.e., at $n_f=3$) fixed in such a way that the underlying one-loop pQCD coupling achieves at $Q^2=M_Z^2$ the value $a(M_Z^2;n_f=5)=0.1181/\pi$, resulting in $\Lambda_{3}=0.1588$ GeV.\footnote{As explained in the previous Section, the quark thresholds are taken at $Q^2 = (\kappa \overline{m}_q)^2$ with $\kappa=2$. The threshold condition for the one-loop pQCD coupling is simply continuity.}. This variant will be denoted as MPT(1.5).
(II) The second variant is with $M=0.3$ GeV (for $n_f=3$) and $\Lambda_{3}=0.234$ GeV, which is suggested by the works of Refs.~\cite{Rayaetal}.\footnote{This coupling is parametrized so that it describes in the infrared an effective charge appearing in the DGLAP equation for the parton distribution functions in the pion, and in the ultraviolet it behaves as a (one-loop) pQCD coupling. The underlying pQCD coupling at $Q^2=M_Z^2$ is then $a(M_Z^2;n_f=5)=0.1264/\pi$.} This variant will be denoted MPT(0.3). Most of the variants of MPT used in the literature have $0.3 \ {\rm GeV} \leq M \leq 1.5 \ {\rm GeV}$ (cf.~also \cite{GHK2018}).

In all other cases, the couplings are normalized in such a way that, at the high scale $Q^2=M_Z^2$ (and $n_f=5$) their underlying pQCD coupling (when tranformed to the $\MSbar$ scheme, if needed) achieves the value $\alpha_s(M_Z^2; \MSbar)=0.1181$ which is the central value of the present world average \cite{PDG18}.

\begin{figure}[htb] 
\begin{minipage}[b]{.49\linewidth}
  \centering\includegraphics[width=85mm]{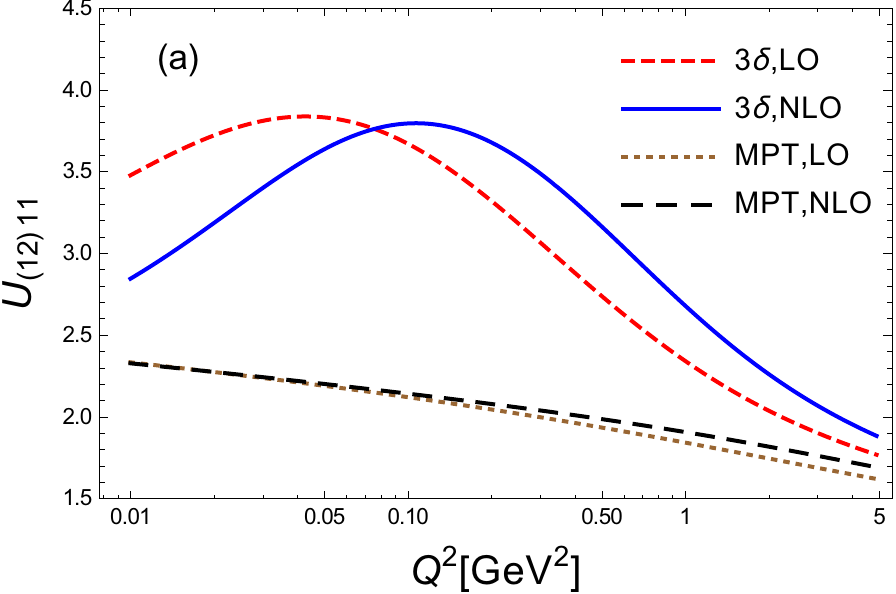}
  \end{minipage}
\begin{minipage}[b]{.49\linewidth}
  \centering\includegraphics[width=85mm]{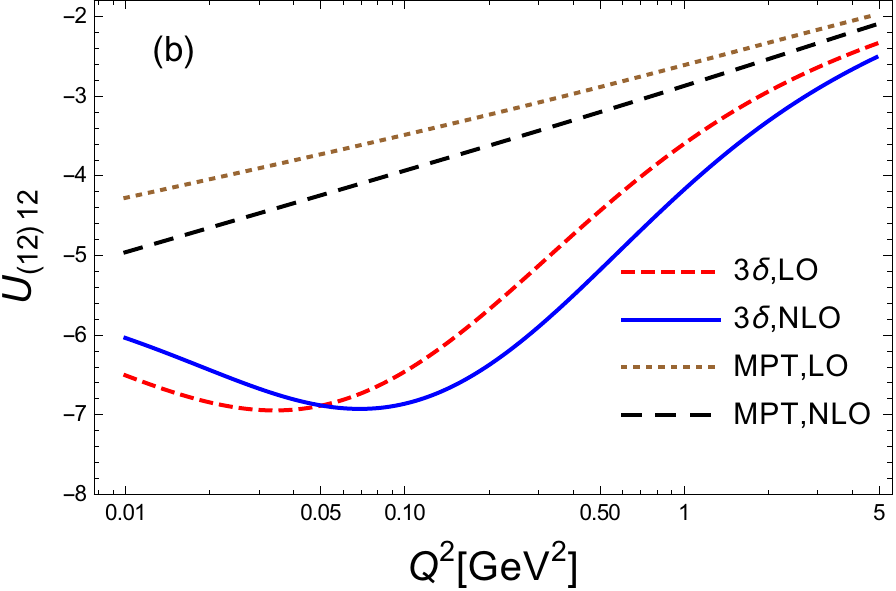}
\end{minipage}
\begin{minipage}[b]{.49\linewidth}
  \centering\includegraphics[width=85mm]{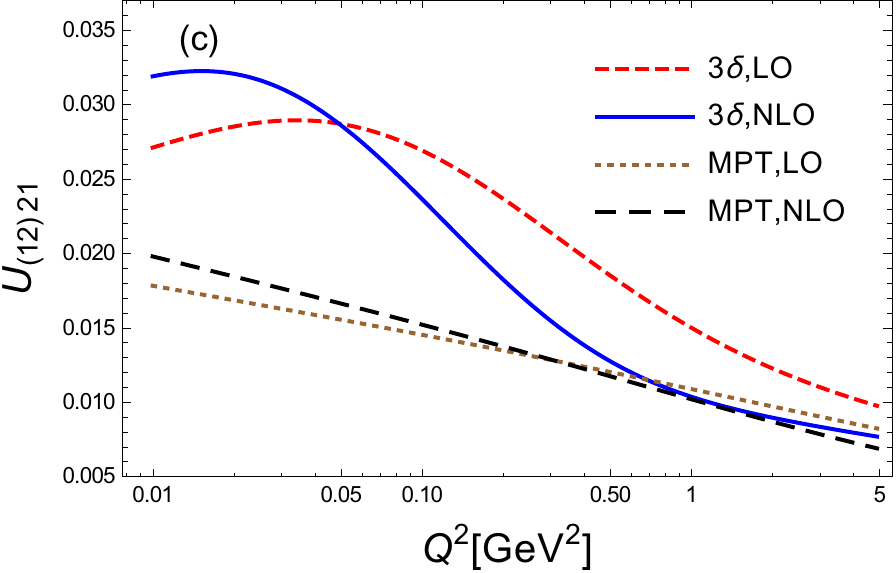}
  \end{minipage}
\begin{minipage}[b]{.49\linewidth}
  \centering\includegraphics[width=85mm]{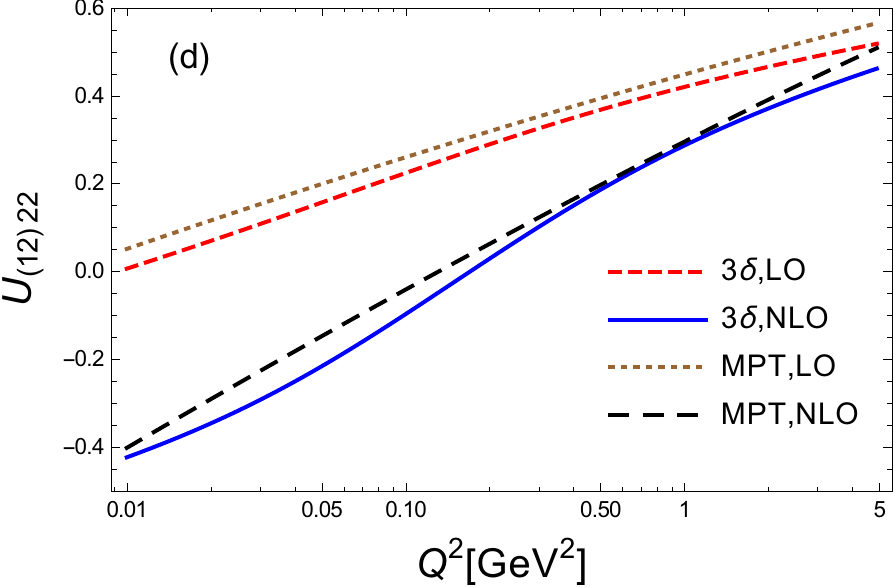}
\end{minipage}
\caption{\footnotesize Evolution matrix elements $U(Q^2; \Lambda^2_{\rm LNV})_{(12)ij}$ ($\Lambda_{\rm LNV}=M_W$) for the (12)$^{XX}$ mixing ($\Lambda_{\rm LNV}=M_W$), for $3 \delta$ $\A$QCD and MPT(1.5), at one-loop (leading order) and two-loop level (next-to-leading order) of the anomalous dimension.} 
\label{FigU12}
\end{figure}

In Fig.~\ref{FigU12} we present, for illustration, the four elements of the evolution matrix $U(Q^2; \Lambda^2_{\rm LNV})_{(12)}$ as a function of the Fermi motion scale $Q^2$ ($0 < Q^2 < 5 \ {\rm GeV}^2$), for the case of the mixing of the operators ${\mathcal O}_1^{XX}$ and ${\mathcal O}_2^{XX}$ ($X=L$, or $R$), i.e., (12)$^{XX}$. We recall that the values of the indices $\nu_j$ are in this case $\nu_1=-0.6120$ and $\nu_2=0.5379$ (when $n_f=3$). The results are given for the $3\delta$ $\A$QCD and MPT(1.5), for the cases of one-loop and two-loop anomalous dimension matrices.

\begin{figure}[htb] 
\begin{minipage}[b]{.49\linewidth}
  \centering\includegraphics[width=85mm]{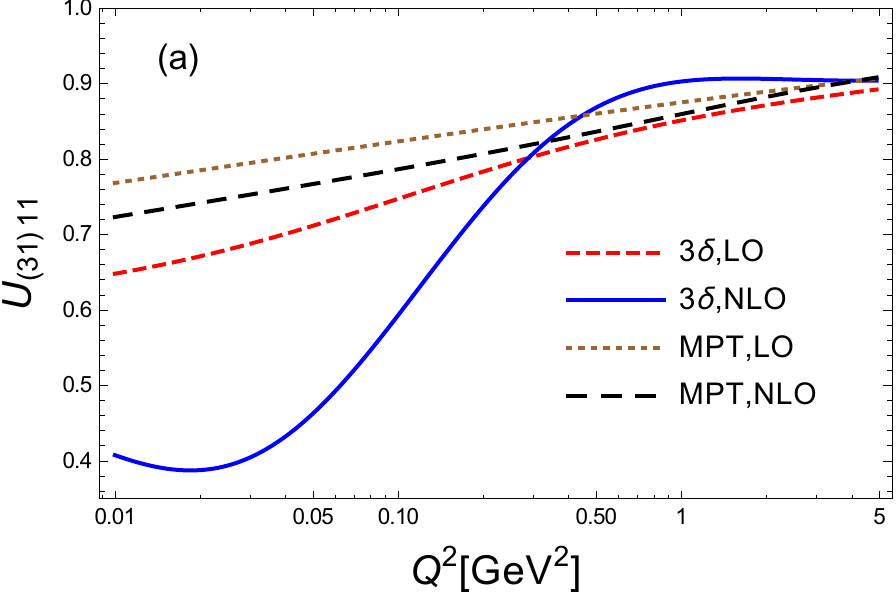}
  \end{minipage}
\begin{minipage}[b]{.49\linewidth}
    \centering\includegraphics[width=85mm]{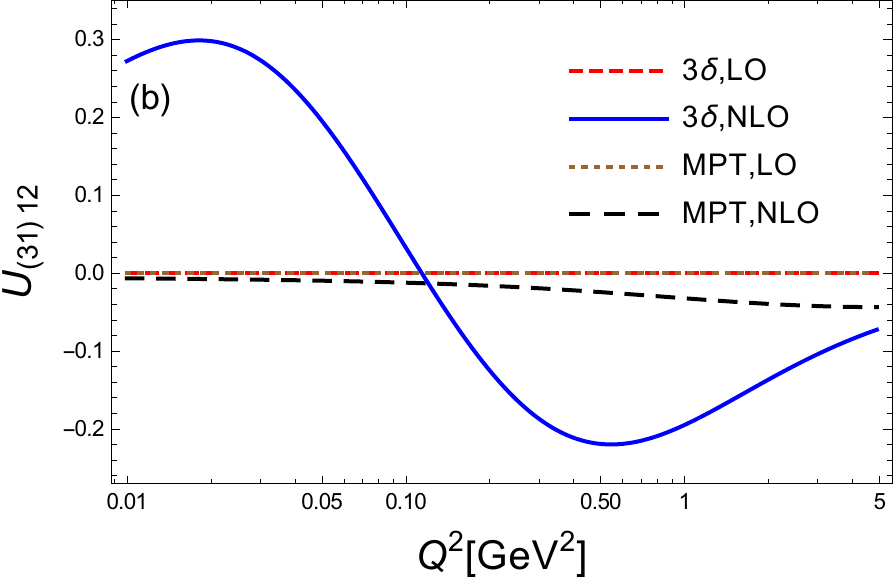}
\end{minipage}
\begin{minipage}[b]{.49\linewidth}
  \centering\includegraphics[width=85mm]{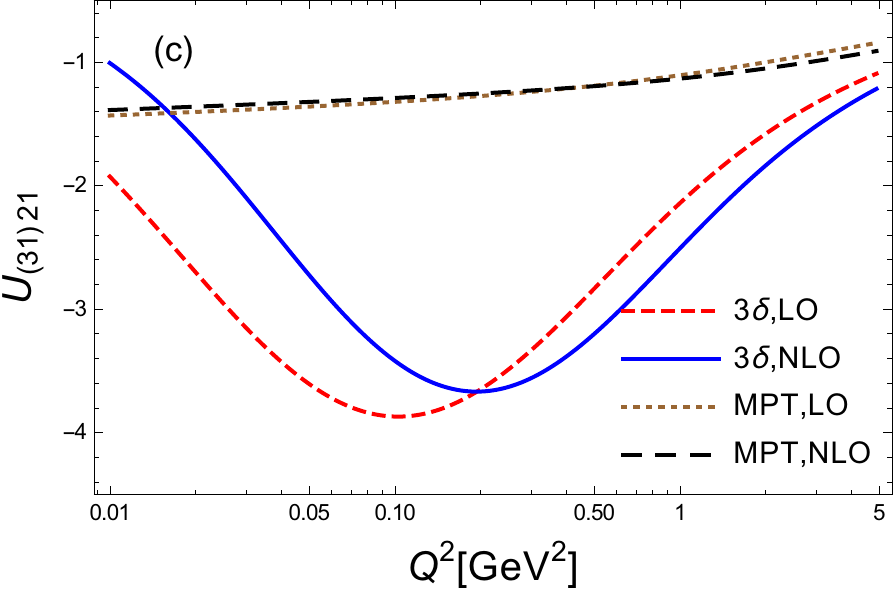}
  \end{minipage}
\begin{minipage}[b]{.49\linewidth}
  \centering\includegraphics[width=85mm]{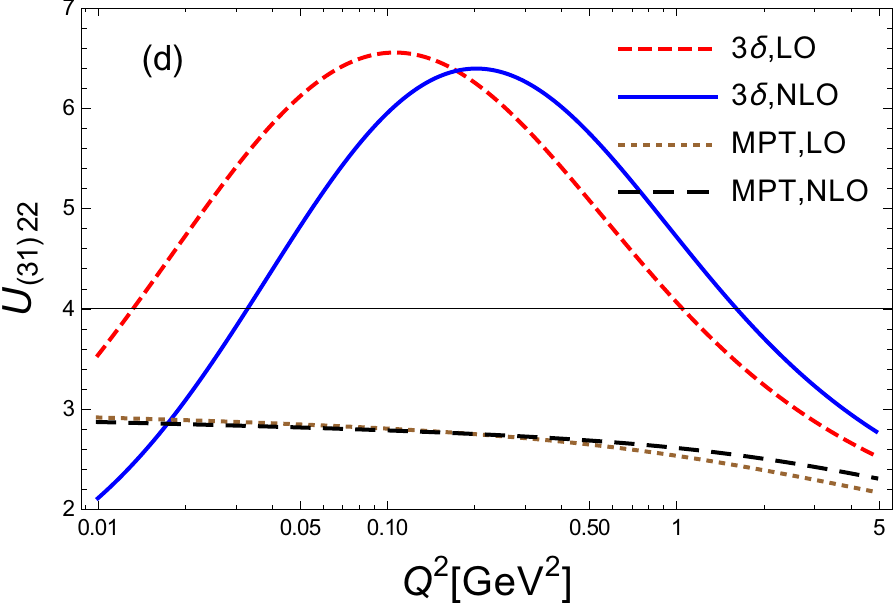}
\end{minipage}
\caption{\footnotesize   The same as in Fig.~\ref{FigU12}, but now for the evolution matrix elements  $U(Q^2; \Lambda^2_{\rm LNV})_{(31)ij}$ ($\Lambda_{\rm LNV}=M_W$) for the (31)$^{XY}$ mixing.}
\label{FigU31}
\end{figure}

\begin{figure}[htb]
  \centering\includegraphics[width=90mm]{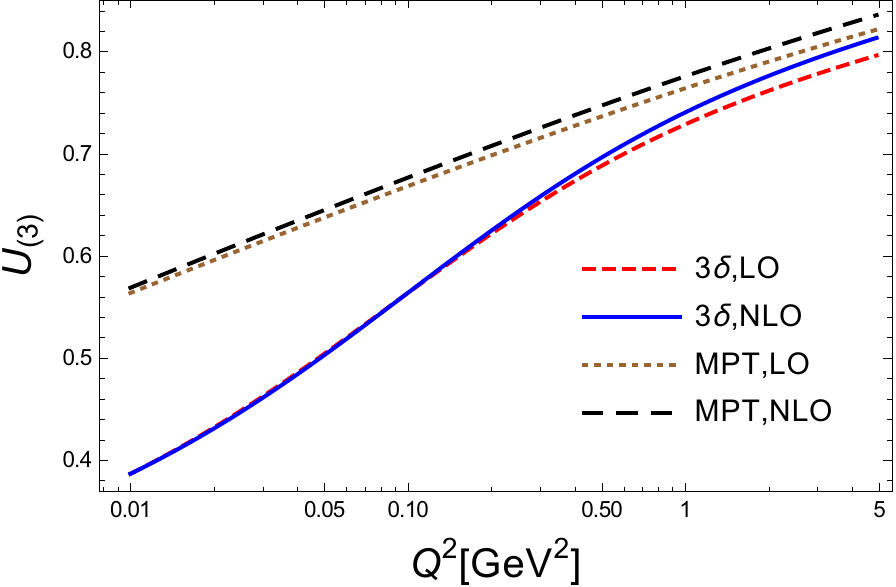}
  \caption{\footnotesize  The same as in Fig.~\ref{FigU12}, but now for the evolution factor  $U(Q^2; \Lambda^2_{\rm LNV})_{(3)}$ ($\Lambda_{\rm LNV}=M_W$) for the operator ${\mathcal O}_3^{XX}$.}
 \label{FigU3}
\end{figure}

In Figs.~\ref{FigU31} we present similarly the elements of the evolution matrix  $U(Q^2; \Lambda^2_{\rm LNV})_{(31)}$ for the case of the mixing of the operators ${\mathcal O}_3^{XY}$ and ${\mathcal O}_1^{XY}$, i.e., (31)$^{XY}$ ($X \not= Y$). Finally, in Fig.~\ref{FigU3} we present similarly the evolution factor  $U(Q^2; \Lambda^2_{\rm LNV})_{(3)}$ for the operator  ${\mathcal O}_3^{XX}$.   
At relatively large (unrealistic) values $Q^2 \approx 5 \ {\rm GeV}^2$, the matrix elements approximately coincide, as they should (asymptotic freedom). However, at more realistic Fermi motion scale values $Q^2 < 1 \ {\rm GeV}^2$, the $3\delta$ $\A$QCD and MPT(1.5) predictions in general differ significantly, especially for the two-loop (i.e., NLO) anomalous dimension case. The other interesting feature is that the predictions with two-loop (NLO) and one-loop (LO) anomalous dimension do not differ much for $3 \delta$ $\A$QCD model, and even less so for MPT(1.5) model.\footnote{
  The values of $U_{(12)22}$ may at first sight suggest otherwise, but in this case we should keep in mind that all these values are all not far from zero.} This apparent convergence suggests that the IR-safe versions of QCD ($\A$QCD), once specified, will in general give us definite quantitative predictions for the evolution matrices $U(Q^2;\Lambda^2_{\rm LNV})$ even for very low (realistic) Fermi motion scales $Q^2 \sim 0.01 \ {\rm GeV}^2$, starting at least at the two-loop level of the anomalous dimension; and at the one-loop level the predictions can be taken at least as first qualitatively correct estimates. On the other hand, the specific details of the applied $\A$QCD in the deep IR regime can affect quite significantly the values of the evolution matrices (for $Q^2 < 1 \ {\rm GeV}^2$); for example, if the $\A$QCD has zero value of the coupling $\A(Q^2)$ at $Q^2 \to 0$ (such as $3 \delta$ $\A$QCD) or a finite nonzero value [such as MPT(1.5), and even more so MPT(0.3) and $2 \delta$ $\A$QCD].   

In Table \ref{tabUall} we present the values of the elements of the evolution matrix $U(Q_{\rm f}^2;\Lambda^2_{\rm LNV})$ for $Q_{\rm f}^2=0.01 \ {\rm GeV}^2$, for the various operators (\ref{ops}) or operator mixings, for the described $\A$QCD frameworks.
We included in the Table also the values of the coupling $\pi \A(Q^2)$ at low scales $Q^2=0.01 \ {\rm GeV}^2$ and $Q^2=0$. The results for (one-loop) pQCD are not included, because the pQCD coupling has Landau singularity at $Q^2 \approx 0.025 \ {\rm GeV}^2$ which is larger than the Fermi motion scale $Q_{\rm f}^2=0.01 \ {\rm GeV}^2$.
\begin{table}
\caption{The values of the elements of the evolution matrix $U(Q_{\rm f}^2;\Lambda^2_{\rm LNV})$  ($\Lambda_{\rm LNV}=M_W$), for the Fermi motion scale $Q_{\rm f}^2=0.01 \ {\rm GeV}^2$, for various $\A$QCD frameworks. The main entries are for the case of two-loop anomalous dimension matrix; in parentheses are included the values for the cases of the one-loop anomalous dimension matrix. In the cases of operators (45)$^{XX}$, only the one-loop anomalous dimension is known. The results for (45)$^{XY}$ ($X \not= Y$) are complex conjugate of (45)$^{XX}$.}
\label{tabUall}  
\begin{ruledtabular}
\begin{tabular}{l|lllll}
  & $2\delta \A$QCD &  $3 \delta \A$QCD & FAPT  & MPT(1.5) & MPT(0.3) 
\\
\hline
$\pi \A(0)$ & 2.692 &  0.000  & 1.396 & 0.3109 & 2.810 
\\
$\pi \A(Q_{\rm f}^2)$ & 2.273 & 0.295  & 0.726 & 0.3106 & 2.318 
\\
\hline
$U^{XX}_{(12)11}$  & 9.373 (7.790) & 2.845 (3.477)  & 3.943 (3.709)&  2.327 (2.333) & 8.784 (7.432) 
\\
$U^{XX}_{(12)12}$ & -17.620 (-14.782) & -6.038 (-6.508)  & -7.569 (-7.315) & -4.961 (-4.278) & -17.954 (-14.613) 
\\
$U^{XX}_{(12)21}$  & 0.0456 (0.0616) & 0.0319 (0.0271)  & 0.0253 (0.0305) & 0.0198 (0.0178) & 0.0510 (0.0609) 
\\
$U^{XX}_{(12)22}$ & -0.846 (-0.0934) & -0.423 ( 0.00653)  & -0.331 (-0.192) & -0.402 (0.0514) & -1.931 (-0.361) 
\\
\hline
$U^{XY}_{(31)11}$ & 0.473 (0.533) & 0.407 (0.648) & 0.651 (0.665) & 0.723 (0.768) & 0.389 (0.488) 
\\
$U^{XY}_{(31)12}$ & -0.549 (0.000) & 0.272 (0.000)  &-0.122 (0.000) & -0.007 (0.000) & -0.521 (0.000) 
\\
$U^{XY}_{(31)21}$ & -12.504 (-11.714) & -1.006 (-1.926) & -3.819 (-3.700) & -1.387 (-1.432) & -11.301 (-10.575) 
\\
$U^{XY}_{(31)22}$ & 19.640 (18.103) & 2.110 (3.537)  & 6.493 (6.215) & 2.870 (2.916) & 17.764 (16.350) 
\\
\hline
$U^{XX}_{(3)}$ & 0.200 (0.242) & 0.387 (0.387) & 0.469 (0.395)  & 0.569 (0.564) & 0.102 (0.145) 
\\
\hline
$U^{XX}_{(45)11}$ & (0.0632) & (0.217)  & (0.201) & (0.411) & (-0.0763) 
\\
$U^{XX}_{(45)12}$ & ($-0.0998 i$) & ($-0.0621 i$)  & ($-0.0605 i$) & $(-0.0390 i)$ & ($-0.104 i$) 
\\
$U^{XX}_{(45)21}$ & ($-1.497 i$) & ($-0.932 i$)  & ($-0.907 i$) & $(-0.586 i)$ & ($-1.562 i$) 
\\
$U^{XX}_{(45)22}$ & (3.257) & (2.205)  & (2.136) & (1.660) & (3.256) 
\end{tabular}
\end{ruledtabular}
\end{table}
In Table \ref{tabUall} we can see that the ``strength'' of $\A$QCD in the deep infrared regime (i.e., the values of coupling $\A$ at very low $Q^2$) significantly affect the values of the evolution matrix elements $U_{ij}$. For example, the results for $U_{ij}$ in $2 \delta$ $\A$QCD and in MPT(0.3) are similar, and appear to be influenced largely by the high values of their coupling $\A(Q^2)$ in the deep IR regime.  On the other hand, however, we see that the results for $U_{i j}$ in $3 \delta$ $\A$QCD and MPT(1.5) do differ significantly (but not drastically) although the values of $\A(Q^2)$ in the deep IR regime in both of these frameworks are low.\footnote{At first sight, one notable exception are the values of $U^{XY}_{(31)12}$ which are $0.272$ and $-0.007$, respectively, representing a large relative difference. However, since the reference values of the elements of $U$ matrices are $\sim 1$, we see that both values ( $0.272$ and $-0.007$) can be considered close to zero and thus similar.} This probably has to do with the fact that $3 \delta$ $\A$QCD has significantly more complicated behavior of $\A(Q^2)$ in the deep IR than MPT(1.5) has.\footnote{
MPT(1.5), due to the high value $M=1.5$ GeV, has the coupling $\A(Q^2)$ almost ``frozen'' in a wide IR region $0 \leq Q^2 \lesssim 1 \ {\rm GeV}^2$, while $3 \delta$ $\A$QCD achieves a maximum at relatively low $Q^2 \approx 0.135 \ {\rm GeV}^2$, cf.~Fig.~\ref{FigAtA2}(a) in Appendix \ref{app:AQCD}.} 
The case of FAPT appears to be intermediate between $3 \delta$ and MPT(1.5) on one hand and $2 \delta$ and MPT(0.3) on the other hand.

Another interesting aspect which can be inferred from Table \ref{tabUall} is that all $\A$QCD frameworks give a reasonable convergence of the results when going from the one-loop to the two-loop anomalous dimension case, despite the very low (nonperturbative) Fermi motion scale $Q_{\rm f}^2=0.01 \ {\rm GeV}^2$; see also Figs.~\ref{FigU12}-\ref{FigU3}. This is connected with the holomorphic nature of $\A(Q^2)$.

\begin{figure}[htb]
  \centering\includegraphics[width=90mm]{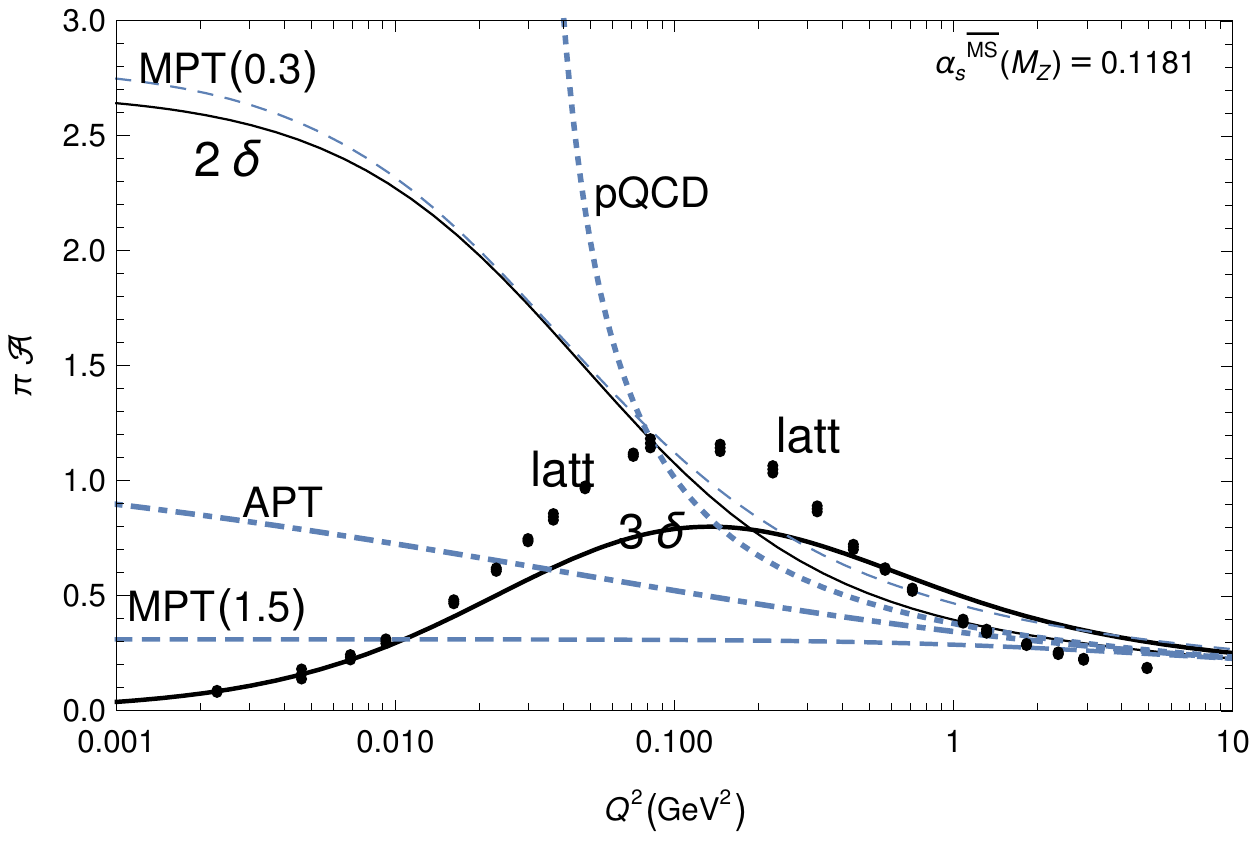}
  \caption{\footnotesize  The couplings $\pi \A(Q^2)$, i.e., the analogs of $\alpha_s(Q^2)$, in the various considered $\A$QCD frameworks, at low positive $Q^2$. Included are the lattice results \cite{LattcoupNf0} and the one-loop pQCD coupling. See the text and Appendix \ref{app:AQCD} for more details.}
 \label{FigA1various}
\end{figure}

In Fig.~\ref{FigA1various} we present the running couplings $\pi \A(Q^2)$ for low positive $Q^2$ for the mentioned $\A$QCD frameworks. In addition, the large-volume lattice results \cite{LattcoupNf0} are included (rescaled to the usual $\Lambda_{\MSbar}$-scaling convention) on which the low-$Q^2$ behaviour of the $3 \delta$ $\A$QCD is motivated (see Appendix \ref{app:AQCD} for more details). Included is also the one-loop pQCD running which is the underlying coupling for MPT(1.5). All curves are for $N_f=3$. As mentioned, the high-energy reference strength for the underlying couplings is in all cases except MPT(0.3): $\alpha_s(M_Z^2;\MSbar;N_f=5)=0.1181$. The large-volume lattice results are not reliable for $Q^2 \gtrsim 1 \ {\rm GeV}^2$. The couplings have the same scaling convention ($\Lambda_{\MSbar}$), but are in general in different renormalization schemes: lattice and $3 \delta$ $\A$QCD in the Lambert MiniMOM (LMM) scheme; $2 \delta$ $\A$QCD in the $c_2=-4.9$ Lambert scheme; (F)APT in the $\MSbar$ scheme; cf.~Appendix \ref{app:AQCD} for more details.

The results of Table \ref{tabUall}, in conjunction with the curves of Fig.~\ref{FigA1various}, reflect the fact that the conclusions on the QCD effects in $0\nu\beta\beta$ decay from short-range physics significantly depend on the specific behavior of the $\A$QCD coupling in the deep IR regime. The variation of these QCD efects, when different IR-safe $\A$QCD frameworks are used, can at the moment be regarded as an estimate of the uncertainty of these effects. On the other hand, one could adopt the view that those $\A$QCD frameworks which incorporate more physically motivated information in the IR regime than the others should be preferred in these considerations. $3 \delta$ $\A$QCD would then definitely fit into this category, because it takes into account in a natural way the additional information provided by the large-volume lattice calculations \cite{LattcoupNf0,LattcoupNf0b,LattcoupNf2,LattcoupNf4} in the deep-IR regime (see also Appendix \ref{app:AQCD}).

\begin{table}
  \caption{Comparison of the values of the elements of the evolution matrix $U(Q_{\rm f}^2;\Lambda^2_{\rm LNV})$  ($\Lambda_{\rm LNV}=M_W$), for the Fermi motion scale $Q_{\rm f}^2=0.01 \ {\rm GeV}^2$, when the power analogs $\A_{\nu}$ in MPT with $M=1.5$ GeV and $M=0.3$ GeV are treated correctly [MPT(1.5) and MPT(0.3)] on the one hand, and naively [nMPT(1.5) and nMPT(0.3)] as $\A^{\nu}$ on the other hand. Other conventions are as in Table \ref{tabUall}.}
\label{tabUMPT}  
\begin{ruledtabular}
\begin{tabular}{l|ll|ll}
 &  MPT(1.5) & nMPT(1.5) & MPT(0.3) & nMPT(0.3)
\\  \hline
$U^{XX}_{(12)11}$ & 2.327 (2.333)  &  1.943 (1.825) & 8.784 (7.432) & 8.370 (5.094)
\\
$U^{XX}_{(12)12}$  & -4.961 (-4.278) &  -2.660 (-2.501)& -17.954 (-14.613) & -13.805 (-9.392)
\\
$U^{XX}_{(12)21}$  & 0.0198 (0.0178)  & 0.00809 (0.0104) & 0.0510 (0.0609) & -0.0169 (0.0391)
\\
$U^{XX}_{(12)22}$  & -0.402 (0.0514) & 0.439 (0.491) & -1.931 (-0.361) &  -0.00693 (0.0850) 
\\
\hline
$U^{XY}_{(31)11}$  & 0.723 (0.768) & 0.901 (0.885) & 0.389 (0.488) & 2.664 (0.717)
\\
$U^{XY}_{(31)12}$  & -0.007 (0.000) & -0.078 (0.000) & -0.521 (0.000) & -3.514 (0.000)
\\
$U^{XY}_{(31)21}$ & -1.387 (-1.432) & -1.314 (-1.179) & -11.301 (-10.575) & -17.583 (-9.111) 
\\
$U^{XY}_{(31)22}$ & 2.870 (2.916) & 2.911 (2.654) & 17.764 (16.350) & 28.275 (14.383)
\\
\hline
$U^{XX}_{(3)}$ & 0.569 (0.564) & 0.806 (0.783) & 0.102 (0.145) & 0.678 (0.514)
\\
\hline
$U^{XX}_{(45)11}$  & (0.411) & (0.705) & (-0.0763) & (0.371)
\\
$U^{XX}_{(45)12}$  & $(-0.0390 i)$ & ($-0.0203 i$)  & ($-0.104 i$) & ($-0.0593 i$)
\\
$U^{XX}_{(45)21}$  & $(-0.586 i)$ & ($-0.304 i$)  & ($-1.562 i$) & ($-0.889 i$)
\\
$U^{XX}_{(45)22}$  & (1.660) & (1.353) & (3.256) & (2.269)
\end{tabular}
\end{ruledtabular}
\end{table}
In Table \ref{tabUMPT} we compare the values of the evolution factors and matrix elements $U(Q_{\rm f}^2;\Lambda^2_{\rm LNV})$  ($\Lambda_{\rm LNV}=M_W$), for the Fermi motion scale $Q_{\rm f}^2=0.01 \ {\rm GeV}^2$, in the case of $\A$QCD MPT(1.5) and MPT(0.3), with the corresponding naive versions [nMPT(1.5) and nMPT(0.3)]. The naive versions are obtained when, instead of the correct power analogs $\A_{\nu}$ we apply in the IR regime ($n_f=3$) the naive powers $\A^{\nu}$. We can see that the results change when going to the naive version. This change is especially strong in the case of nMPT(0.3), the reason for this being the strong variation of the coupling $\A$ in the deep IR regime for MPT(0.3) (cf.~the corresponding entries in the first two lines of Table \ref{tabUall}). We point out that the naive powers $\A^{\nu}$ do not treat the nonperturbative contributions correctly, in contrast to the power analogs $\A_{\nu}$, as argued in Appendix \ref{app:AQCD}. The nMPT(1.5) is close to the approach taken in Ref.~\cite{GHK2018} where one-loop anomalous dimensions were used. 
The nMPT(1.5) and nMPT(0.3) are difficult to compare with any of the $\A$QCD frameworks. Further, the values of $U_{ij}$ in nMPT(1.5) almost do not vary when $Q_{\rm f}^2$ increases from $0.01 \ {\rm GeV}^2$ upwards to $1 \ {\rm GeV}^2$, in contrast with the $\A$QCD frameworks.\footnote{This is not seen in Table \ref{tabUall} where $Q_{\rm f}^2$ is kept fixed, $Q_{\rm f}^2=0.01 \ {\rm GeV}^2$.}

\subsection{Bounds on Wilson coefficients}
\label{subsec:Wils}

The upper bounds on the Wilson coefficients can now be obtained by requiring that the expression on the RHS of Eq.~(\ref{halflife}) is larger than the lower bound on the half-life $T_{1/2}^{0\nu}(^{136}{\rm Xe})$ Eq.~(\ref{T12}). The RHS of Eq.~(\ref{halflife}) involves the NMEs (Table \ref{tabNMEs}), the space factors $G_1=  2.92 \times 10^{-14} {\rm yr}^{-1}$ and $G_4= 1.57 \times 10^{-14} {\rm yr}^{-1}$ \cite{GDIK}, and the Wilson coefficients at the Fermi motion scale
\be
C_j(Q_{\rm f}^2) = U(Q_{\rm f}^2; \Lambda^2_{\rm LNV})_{jk} C_k(\Lambda^2_{\rm LNV}).
\label{CjQf}
\ee
We recall that we use throughout this work  $\Lambda^2_{\rm LNV} = M_W^2$.
When using the expansion (\ref{CjQf}) on the RHS of Eq.~(\ref{halflife}), the following expression for the half-life in terms of the ``bare'' Wilson coefficients $C_k \equiv C_k(\Lambda^2_{\rm LNV})$ is obtained (cf.~also \cite{GHK2016})
\bea
\left[T_{1 / 2}^{0 \nu \beta \beta}\right]^{-1} &= &
G_1 {\big |} \beta_1^{XX} \left( C_1^{LL}  + C_1^{RR}  \right) +
\beta_1^{LR} 2 C_1^{LR}  + \beta_2^{XX} \left( C_2^{LL}  + C_2^{RR}  \right)
\nonumber\\ &&
+ \beta_3^{XX} \left( C_3^{LL}  + C_3^{RR}  \right) + \beta_3^{LR} 2 C_3^{LR}  {\big |}^2
\nonumber\\ &&
+ G_4 {\big |} \beta_4^{XX} \left( C_4^{LL}  + C_4^{RR}  \right) + \beta_4^{XY} \left( C_4^{LR}  + C_4^{RL}  \right) 
\nonumber\\ &&
+ \beta_5^{XX} \left( C_5^{LL}  + C_5^{RR}  \right) +
\beta_5^{XY} \left( C_5^{LR}  + C_5^{RL}  \right) {\big |}^{2} ,
\label{HLf}
\eea
where
\bes
\label{betas}
\bea
\beta_1^{XX} &= &  \mathcal{M}_{1} U^{XX}_{(12)11} + \mathcal{M}_{2} U^{XX}_{(12)21},
\qquad
\beta_1^{LR} =   \mathcal{M}_{3}^{(+)} U^{LR}_{(31)12} + \mathcal{M}_{1} U^{LR}_{(31)22},
\label{beta1}
\\
\beta_2^{XX} &= &  \mathcal{M}_{1} U^{XX}_{(12)12} + \mathcal{M}_{2} U^{XX}_{(12)22},
\label{beta2XX}
\\
\beta_3^{XX} &= &  \mathcal{M}_{3}^{(-)} U^{XX}_{(3)}, \quad
\beta_3^{LR} =   \mathcal{M}_{1} U^{LR}_{(31)21} +  \mathcal{M}_{3}^{(+)} U^{LR}_{(31)11},
\label{beta3}
\\
\beta_4^{XX} &= &  - |\mathcal{M}_{4}| U^{XX}_{(45)11} + |\mathcal{M}_{5}| U^{XX}_{(45)21},
\label{beta4XX}
\\
\beta_4^{XY} &= &  |\mathcal{M}_{4}| U^{XY}_{(45)11} + |\mathcal{M}_{5}| U^{XY}_{(45)21} \qquad (X \not= Y),
\label{beta4LR}
\\
\beta_5^{XX} &= &  - |\mathcal{M}_{4}| U^{XX}_{(45)12} + |\mathcal{M}_{5}| U^{XX}_{(45)22},
\label{beta5XX}
\\
\beta_5^{XY} &= &  |\mathcal{M}_{4}| U^{XY}_{(45)12}  + |\mathcal{M}_{5}| U^{XY}_{(45)22} \qquad (X \not= Y).
\label{beta5LR}
\eea
\ees
We used here the simplified notation $U \equiv U(Q^2_{\rm f}; \Lambda^2_{\rm LNV})$.
The mixing coefficient $U^{XY}_{(31)12}$ [appearing in Eq.~(\ref{beta1})] is zero at one-loop and nonzero at two-loop level of anomalous dimension, cf.~Table \ref{tabUall}.
We mention that factor $2$ appears in the terms with $C_j^{XY}$ ($j=1,3$); this is so because the operators ${\mathcal{O}_j^{XY}}$ ($j=1,3$) for $XY=LR$ and $RL$, Eqs.~(\ref{ops}), are symmetric under the interchange of $L$ and $R$, and hence: $C_j^{RL}=C_j^{LR}$ and $(C_j^{LR}+C_j^{RL})$ $ = 2 C_j^{XY}$ (cf.~also \cite{GHK2016}). Further, we recall that the values of NMEs are given in Table \ref{tabNMEs}.

In the two-loop running of the evolution factors or matrices $U \equiv U(Q^2_{\rm f}; \Lambda^2_{\rm LNV})$, we used the two-loop anomalous dimension ${\hat \gamma}$ of Ref.~\cite{Buras:2000if} in the naive dimensional regularization $\MSbar$ (NDR-$\MSbar$) scheme. On the other hand, the NMEs are often evaluated in a different, Regularization-Independent (RI, also named MOM) scheme. Since the values of NMEs $\mathcal{M}_j$ have large uncertainties (by about a factor of 2), we used the NDR-$\MSbar$ expressions for the anomalous dimensions, which have the attractive feature of being independent of the gauge-fixing parameter (in contrast to the case of IR scheme). 

We can now obtain the upper bounds on the values of the ``bare'' (new physics) Wilson coefficients $|C_j|$ ($\equiv |C_j(\Lambda^2_{\rm LNV})|$ by assuming that only one operator contributes dominantly to the half-life. This then gives us the upper bounds for various values of the Fermi motion scale $Q_{\rm f}^2 = 1.0, 0.1$ and $0.01 \ {\rm GeV}^2$ as given in Tables \ref{table:1llim} and \ref{table:2llim}, for the cases of one-loop and two-loop anomalous dimension matrices, respectively, for various $\A$QCD frameworks. In these Tables we included, for comparison, the results of pure pQCD approach (only for $Q_f^2 = 1 \ {\rm GeV}^2$ and $0.1 \ {\rm GeV}^2$),\footnote{
  We recall that the values $Q_f^2 < 1 \ {\rm GeV}^2$ in pQCD become very unreliable or impossible to obtain, due to the Landau singularities.} and for the ``bare'' case when there are no QCD effects (the evolutions factors or matrices are unity).
\begin{table}[htb]
\begin{center}
	\begin{tabular}{c|ccccc|ccc|c}
		\hline
		\hline
		 & $2\delta \A$QCD & $3\delta \A$QCD & FAPT & MPT(1.5) & MPT(0.3)  & nMPT(1.5) & nMPT(0.3) &pQCD & $C_i^{(0)}$ \\
        \hline
        \hline
		{$|C^{XX}_1|_{0.01}$} & $3.27$  & $7.64$ & $6.14$ & $12.2$ & $3.09$ & $87.5$ & $5.47$ & --&\multirow{3}{*}{$12.6$}\\
	    {$|C^{XX}_1|_{0.10}$} & $9.48$  & $8.89$ & $11.3$ & $20.1$ & $7.84$ & $92.8$ & $9.36$ & $9.74$&\\
	    {$|C^{XX}_1|_{1.00}$} & $57.3$  & $25.5$ & $32.0$ & $58.8$ & $28.1$ & $182$ & $29.2$ & $44.0$&\\
		\hline
		{$|C^{XY}_1|_{0.01}$} & $0.35$  & $1.78$ & $1.01$ & $2.16$ & $0.38$ & $2.37$ & $0.44$ &--&\multirow{3}{*}{$6.3$}\\
	    {$|C^{XY}_1|_{0.10}$} & $0.75$  & $0.96$ & $1.41$ & $2.24$ & $0.81$ & $2.39$ & $0.83$ & $0.87$&\\
	    {$|C^{XY}_1|_{1.00}$} & $1.88$  & $1.55$ & $2.10$ & $2.48$ & $1.81$ & $2.54$ & $1.82$ & $2.08$ &\\
		\hline
		{$|C^{XX}_2|_{0.01}$} & $4.40$  & $1.62$ & $0.43$ & $0.90$ & $0.23$ & $0.13$ & $0.49$ & --&\multirow{3}{*}{$0.07$}\\
	    {$|C^{XX}_2|_{0.10}$} & $0.20$  & $0.26$ & $0.41$ & $0.24$ & $0.43$ & $0.13$ & $0.29$ & $0.28$&\\
	    {$|C^{XX}_2|_{1.00}$} & $0.12$  & $0.15$ & $0.17$ & $0.14$ & $0.16$ & $0.13$ & $0.16$ &$0.14$&\\
		\hline
		{$|C^{XX}_3|_{0.01}$} & $2.13$  & $ 1.33$ & $1.30$ & $0.91$ & $3.55$ & $0.66$ & $1.00$ & --&\multirow{3}{*}{$0.51$}\\
	    {$|C^{XX}_3|_{0.10}$} & $0.84$  & $ 0.91$ & $0.87$ & $0.77$ & $0.94$ & $0.66$ & $0.85$ & $0.84$&\\
	    {$|C^{XX}_3|_{1.00}$} & $0.67$  & $ 0.71$ & $0.70$ & $0.67$ & $0.71$ & $0.65$ & $0.70$ &$0.68$&\\
		\hline 
		{$|C^{XY}_3|_{0.01}$} & $1.77$  & $0.78$ & $0.97$ & $0.61$ & $2.04$ & $0.51$ & $3.35$ & --&\multirow{3}{*}{$0.41$}\\
	        {$|C^{XY}_3|_{0.10}$} & $0.92$  & $0.83$ & $0.66$ & $0.56$ & $0.94$ & $0.51$ & $0.85$ & $0.82$&\\
	    {$|C^{XY}_3|_{1.00}$} & $0.54$  & $0.58$ & $0.54$ & $0.51$ & $0.55$ & $0.50$ & $0.55$ &$0.53$&\\
        \hline
		{$|C^{XX}_4|_{0.01}$} & $5.08$  & $3.42$ & $3.66$ & $1.94$ & $4.74$ & $1.14$ & $2.11$ &--&\multirow{3}{*}{$0.80$}\\
	    {$|C^{XX}_4|_{0.10}$} & $1.59$  & $1.81$ & $1.75$ & $1.45$ & $1.93$ & $1.14$ & $1.67$ & $1.64$& \\
	    {$|C^{XX}_4|_{1.00}$} & $1.17$  & $1.26$ & $1.26$ & $1.18$ & $1.26$ & $1.11$ & $1.25$ & $1.19$&\\ 
		\hline
		{$|C^{XX}_5|_{0.01}$} & $2.43$  & $3.61$ & $3.73$ & $4.86$ & $2.42$ & $6.06$ & $3.53$ &--&\multirow{3}{*}{$8.30$}\\
	    {$|C^{XX}_5|_{0.10}$} & $4.17$  & $4.00$ & $4.66$ & $5.33$ & $4.03$ & $6.07$ & $4.33$ & $4.38$&\\
	    {$|C^{XX}_5|_{1.00}$} & $5.73$  & $5.36$ & $5.66$ & $5.95$ & $5.53$ & $6.20$ & $5.56$ &$5.81$&\\
		\hline
	\end{tabular}
\caption{Upper bounds on the bare Wilson coefficients $C_j \equiv C_j(\Lambda^2_{\rm LNV})$, multiplied by $10^{8}$, for various QCD variants, where the lower (Fermi motion) scales used are $Q^2=0.01 \ \text{GeV}^2$, $0.1 \ \text{GeV}^2$ and $1 \ \text{GeV}^2$, for the isotope ${}^{136}\text{Xe}$, with one-loop anomalous dimension in the RGE, and  $\Lambda_{\rm LNV} = M_W$. The chirality superscripts are: $XX=LL$ or $RR$; $XY=LR$ or $RL$. In the case of $C_4$ and $C_5$, the results are the same for $XX$ and $XY$.}
\label{table:1llim}
\end{center}
\end{table}
\begin{table}[htb]
	\begin{tabular}{c|ccccc|ccc|c}
		\hline
		\hline
		 & $2\delta \A$QCD & $3\delta \A$QCD & FAPT & MPT(1.5) & MPT(0.3)  & nMPT(1.5) & nMPT(0.3) &pQCD & $C_i^{(0)}$ \\
        \hline
        \hline
		{$|C^{XX}_1|_{0.01}$} & $16.7$  & $3.95$ & $15.1$ & $8.90$ & $14.9$ & $30.3$ & $1.09$ & -- & \multirow{3}{*}{$12.6$}\\
	    {$|C^{XX}_1|_{0.10}$} & $6.39$  & $18.6$ & $78.6$ & $17.2$ & $14.7$ & $30.3$ & $4.06$ & $4.80$ &\\
	    {$|C^{XX}_1|_{1.00}$} & $24.6$  & $17.4$ & $68.1$ & $616$ & $41.5$ & $30.4$ & $20.5$ & $26.7$ &\\
		\hline
		{$|C^{XY}_1|_{0.01}$} & $0.56$  & $1.00$ & $1.36$ & $2.27$ & $0.64$ & $3.67$ & $0.25$ &-- &\multirow{3}{*}{$6.3$}\\
	    {$|C^{XY}_1|_{0.10}$} & $5.55$  & $0.97$ & $2.04$ & $2.42$ & $3.39$ & $3.68$ & $2.35$ & $3.83$ &\\
	    {$|C^{XY}_1|_{1.00}$} & $3.92$  & $3.64$ & $2.98$ & $2.97$ & $3.62$ & $3.75$ & $3.99$ & $3.68$ &\\
		\hline
		{$|C^{XX}_2|_{0.01}$} & $0.09$  & $0.17$ & $0.23$ & $0.18$ & $0.04$ & $0.15$ & $1.01$ &--&\multirow{3}{*}{$0.07$}\\
	    {$|C^{XX}_2|_{0.10}$} & $9.11$  & $1.10$ & $0.98$ & $3.27$ & $0.14$ & $0.15$ & $0.47$ & $0.43$ &\\
	    {$|C^{XX}_2|_{1.00}$} & $0.19$  & $0.22$ & $0.19$ & $0.21$ & $0.37$ & $0.14$ & $0.19$ &$0.17$ &\\
		\hline
		{$|C^{XX}_3|_{0.01}$} & $2.57$  & $1.33$ & $1.10$ & $0.90$ & $5.02$ & $0.64$ & $0.76$ &-- &\multirow{3}{*}{ $0.51$}\\
	    {$|C^{XX}_3|_{0.10}$} & $0.86$  & $0.91$ & $0.80$ & $0.76$ & $0.95$ & $0.64$ & $0.74$ & $0.75$ &\\
	    {$|C^{XX}_3|_{1.00}$} & $0.68$  & $0.69$ & $0.68$ & $0.66$ & $0.69$ & $0.63$ & $0.67$ & $0.65$ &\\
		\hline
		{$|C^{XY}_3|_{0.01}$} & $1.20$  & $1.20$ & $1.02$ & $0.65$ & $1.18$ & $0.50$ & $0.27$ &-- &\multirow{3}{*}{$0.41$}\\
	    {$|C^{XY}_3|_{0.10}$} & $0.72$  & $1.11$ & $0.67$ & $0.58$ & $0.82$ & $0.50$ & $0.52$ & $0.54$ &\\
	    {$|C^{XY}_3|_{1.00}$} & $0.53$  & $0.55$ & $0.54$ & $0.52$ & $0.55$ & $0.49$ & $0.53$ & $0.52$ &\\
		\hline
	\end{tabular}
\caption{Same as Table \ref{table:1llim}, but with two-loop anomalous dimension used in the RGE.}.
\label{table:2llim}
\end{table}

\begin{figure}[htb] 
\begin{minipage}[b]{.49\linewidth}
  \centering\includegraphics[width=85mm]{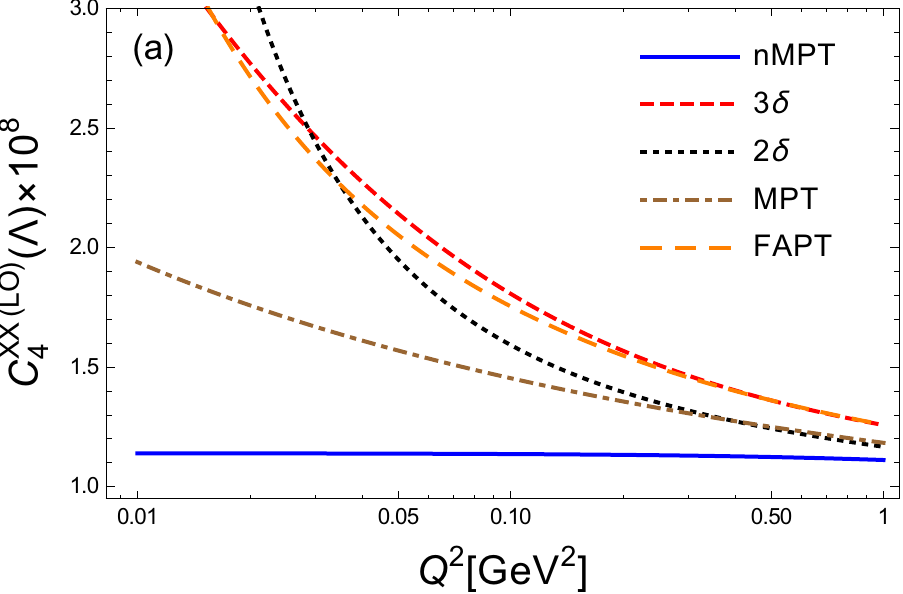}
\end{minipage}
\begin{minipage}[b]{.49\linewidth}
  \centering\includegraphics[width=85mm]{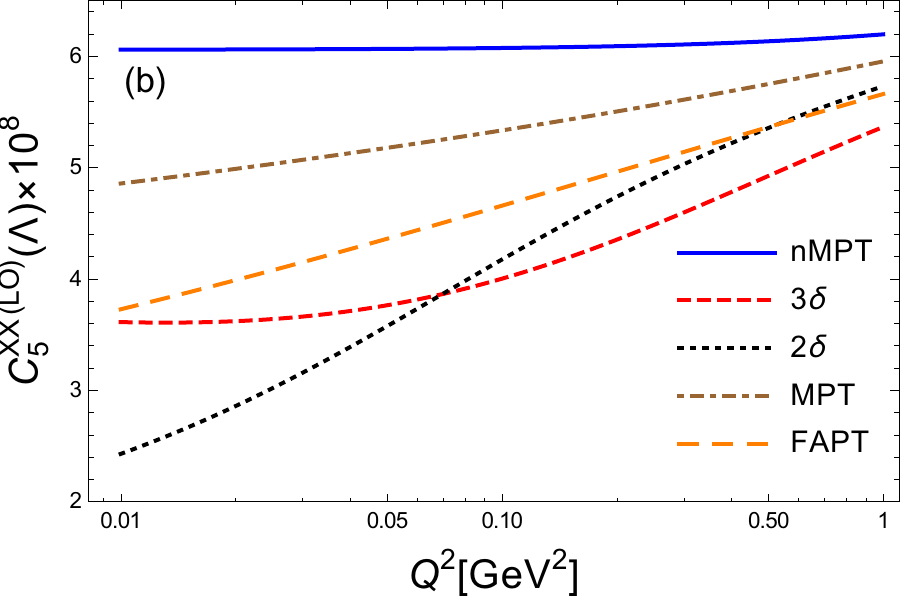}
  \end{minipage}
\caption{\footnotesize  The upper bounds for the ``bare'' Wilson coefficients $|C_j^{XX}(\Lambda^2_{\rm LNV})|$ for $j=4,5$ ($\Lambda_{\rm LNV}=M_W$), for various $\A$QCD frameworks. Note that only the one-loop ('LO') anomalous dimension is available for these calculations. MPT and naive MPT (nMPT) are for the mass $M=1.5$ GeV [MPT(1.5), nMPT(1.5)]. The results for the case $XY$ ($X \not= Y$) are the same as those for $XX$.} 
\label{FigC4C5}
\end{figure}

\begin{figure}[htb]
  \centering\includegraphics[width=90mm]{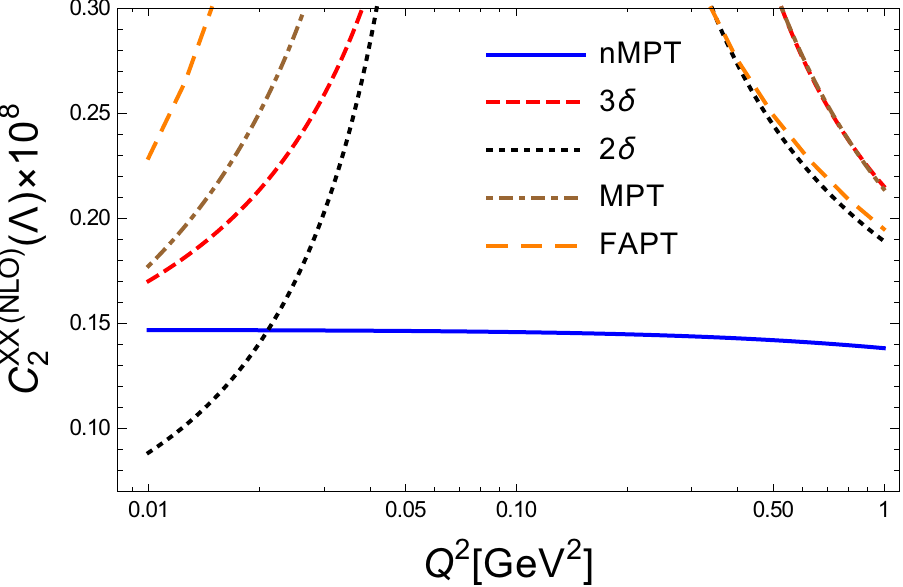}
  \caption{\footnotesize As Fig.~\ref{FigC4C5}, but for the values of $|C_2^{XX}(\Lambda^2_{\rm LNV})|$; the available two-loop (NLO) anomalous dimension was used.}
 \label{FigC2}
\end{figure}

\begin{figure}[htb] 
\begin{minipage}[b]{.49\linewidth}
  \centering\includegraphics[width=85mm]{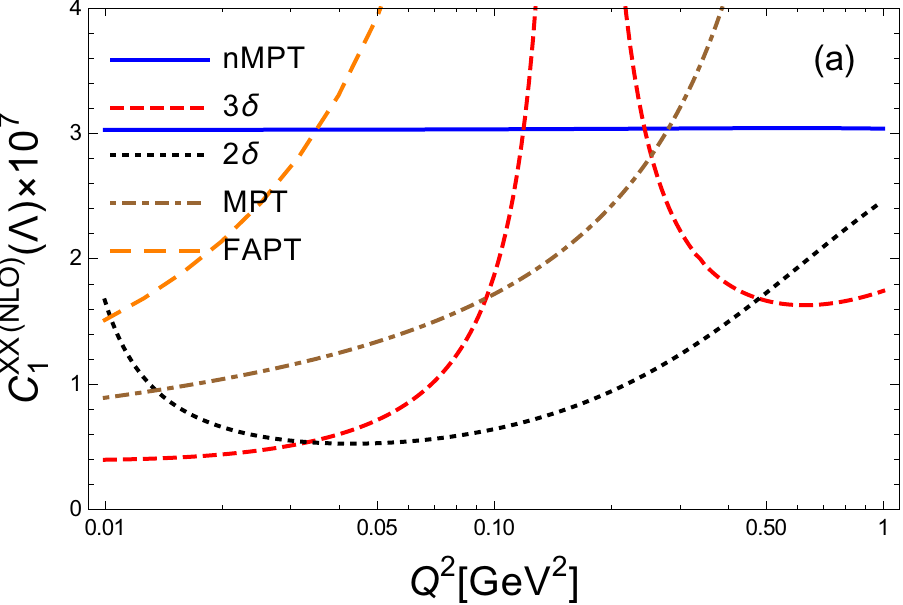}
  \end{minipage}
\begin{minipage}[b]{.49\linewidth}
  \centering\includegraphics[width=85mm]{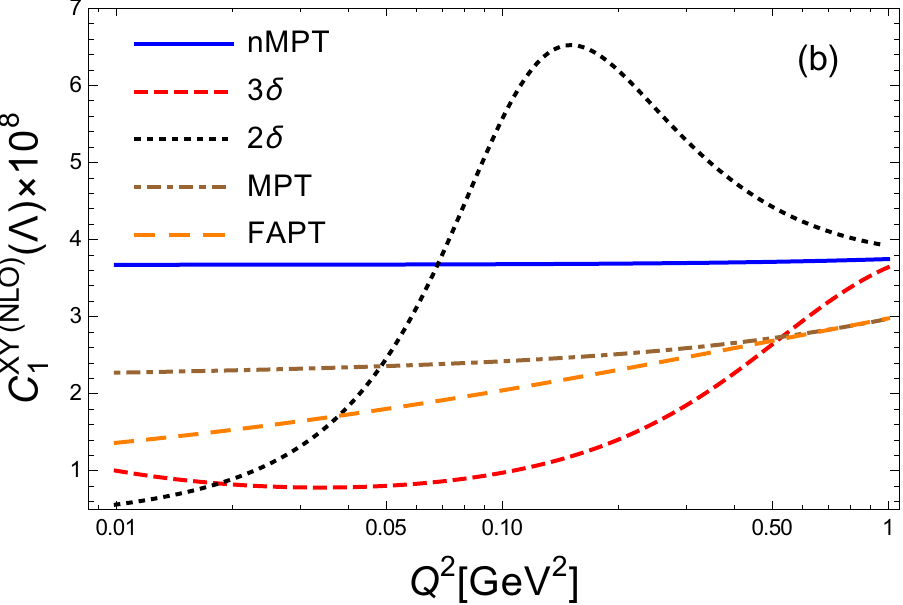}
\end{minipage}
\caption{\footnotesize  As Fig.~\ref{FigC4C5}, but for the values of $|C_1^{XX}(\Lambda^2_{\rm LNV})|$ and $|C_1^{LR}(\Lambda^2_{\rm LNV})|$; the available two-loop (NLO) anomalous dimension was used.} 
\label{FigC1}
\end{figure}

\begin{figure}[htb] 
\begin{minipage}[b]{.49\linewidth}
  \centering\includegraphics[width=85mm]{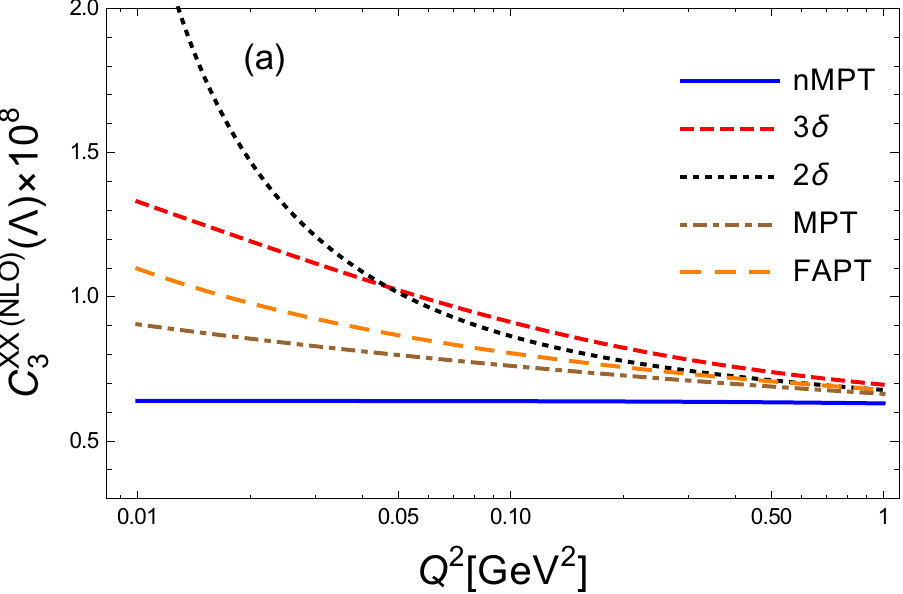}
  \end{minipage}
\begin{minipage}[b]{.49\linewidth}
  \centering\includegraphics[width=85mm]{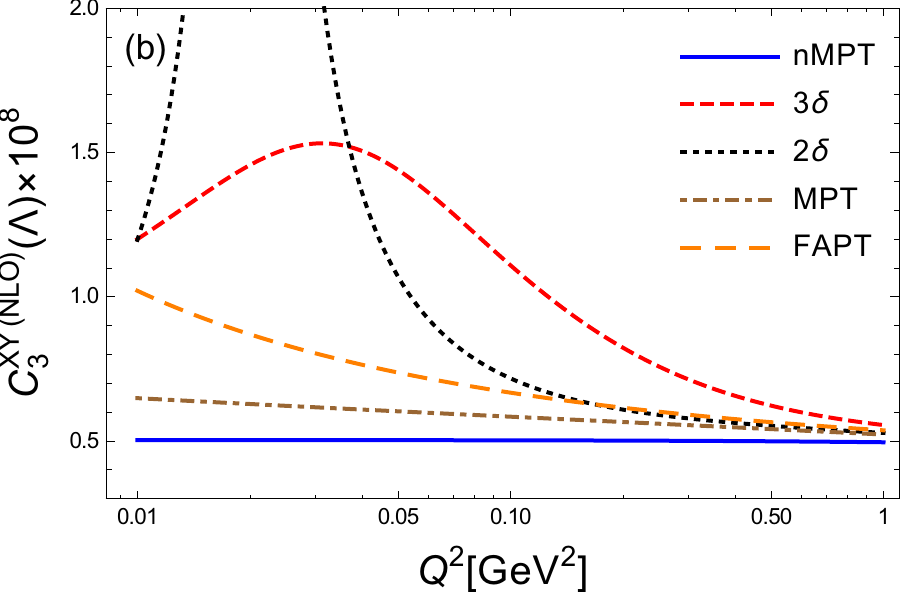}
\end{minipage}
\caption{\footnotesize   As Fig.~\ref{FigC4C5}, but for the values of $|C_3^{XX}(\Lambda^2_{\rm LNV})|$ and $|C_3^{LR}(\Lambda^2_{\rm LNV})|$; the available two-loop (NLO) anomalous dimension was used.}
\label{FigC3}
\end{figure}

In Figs.~\ref{FigC4C5}-\ref{FigC3} we present the upper bounds as a function of $Q_{\rm f}^2$ in an extended interval $0.01 \ {\rm GeV}^2 \leq Q_{\rm f}^2 < 5.0 \ {\rm GeV}^2$. At (artificially) large values of the (Fermi motion) scales $Q^2 \approx 5 \ {\rm GeV}^2$, we can see in these Figures that the upper bounds for various $\A$QCD variants approximately coincide, as it should be due to the asymtotic freedom.

We wish to point out that, when the upper bounds for the bare coefficients in the ($\A$)QCD cases are large ($10^8 |C_j^{XY}| > 10$), the upper bounds are probably not very useful (i.e., no strong restrictions). This, of course, does not imply that the ($\A$)QCD effects in such a channel are not important, but rather that these effects imply that no useful upper bound can be deduced for the corresponding Wilson coefficient.

The values of NMEs, table \ref{tabNMEs}, have relatively large uncertainties, by roughly a factor of $2$. If the coefficients are (simultaneously) multiplied by a factor of $2$, the corresponding upper bounds of the bare Wilson coefficients get reduced by this factor of $2$. Therefore, in order to discern whether the ($\A$)QCD effects for the upper bound of a Wilson coefficient $C_j(M_W^2)$ are more important than the uncertainties of the values of NMEs, we will consider that
\be
\frac{|C_j|}{|C_j^{(0)}|} < \frac{1}{2}, \qquad {\rm or} \; \frac{|C_j|}{|C_j^{(0)}|} > 2,
\label{Cjrel}
\ee
where $C_j^{(0)}(M_W^2)$ is the corresponding upper bound value when there are no QCD effects ($U$ is unity then). The first inequality in Eq.~(\ref{Cjrel}) gives a more stringent upper bound on $|C_j|$ (by at least a factor of 2) than when QCD is ignored, and the second one gives a less stringent upper bound (by at least a factor of 2).

With these naive criteriums, we can infer from Table \ref{table:2llim} our main conclusions: for the realistic Fermi motion scale value $Q_{\rm f}^2=0.01 \ {\rm GeV}^2$, the QCD effects become important for $C_1^{XY}$ in all $\A$QCD variants; for $C_2^{XX}$ and $C_3^{XX}$ in most of the $\A$QCD variants; and for $C_1^{XX}$ in $3 \delta$ $\A$QCD. In all these cases, the upper bounds become more stringent for $C_1^{XY}$ and $C_1^{XX}$, and less stringent for $C_2^{XX}$ and $C_3^{XX}$. It is interesting that most (but not all) of these qualitative conclusions for the mentioned Wilson coefficients are also valid when regarding the upper bounds obtained with the use of the one-loop anomalous dimensions, cf.~Table \ref{table:1llim}; $C_1^{XX}$ is here a notable exception.

For the coefficients $C_4$ and $C_5$, only the one-loop anomalous dimensions are available. If we regard the upper bounds for them obtained in this way as indicative, then we conclude that in all $\A$QCD variants and for almost all these coefficients the QCD effects are important, where for $C_4$ the upper bounds become less restrictive and for $C_5$ more restrictive, cf.~Table \ref{table:1llim}.

Some of the upper bounds obtained in the described analysis are quite high, at least in certain specific ranges of values of (Fermi motion scale) $Q^2$, as seen in some of Figs.~\ref{FigC4C5}-\ref{FigC3}, and in some cases in Tables \ref{table:1llim}-\ref{table:2llim}. This is so because in such cases in the corresponding coefficients $\beta_j^{XY}$ Eq.~(\ref{betas}), which consist mostly of a sum of two terms  of the type $U(Q^2)_k {\mathcal{M}_j}$, partial cancellations can occur between these two terms and thus a large value of the corresponding Wilson coefficients are allowed.

We can also note from Tables \ref{table:1llim}-\ref{table:2llim} that the incorrect, i.e., naive approaches [nMPT(1.5) and nMPT(0.3)] give different values of the upper bounds than the correct approaches [MTP(1.5) and MPT(0.3)]. The naive approach differs quite strongly in the case of nMPT(0.3) from the corresponding correct approach MPT(0.3) (when $Q^2=0.01 \ {\rm GeV}^2$); this is so because when $M=0.3$ GeV, the coupling $\A(Q^2)$ changes quite strongly in the deep IR regime, as seen also from the corresponding entries in the first two lines of Table \ref{tabUall} (cf.~also Table \ref{tabUMPT}).

\section{Nonperturbative contributions in the sub-GeV regime}
\label{sec:NPsubG}

In this work, we used for NMEs the results of the works \cite{DHP,Pas:2000vn} (cf.~also \cite{GHK2016}), which were derived by a different method than those of Ref.~\cite{Ciretal}. The NMEs are calculated in Refs.~\cite{DHP,Pas:2000vn} by using the nucleon-nucleon ($NN$) interactions in the Quasi Particle Random Phase Approximation (pn-QRPA), at effective momenta of the Fermi motion scale $Q_{\rm f} \sim 0.1$ GeV.

On the other hand, in Ref.~\cite{Ciretal}, $NN$, $\pi\pi$ and $\pi N$ interactions are used. In addition, in \cite{Ciretal} the effective Standard Model with QCD is matched to Chiral Perturbation Theory (ChiPT) already at scales $Q = \Lambda_{\chi} \sim 1$ GeV, and ChiPT couplings ($g_j^{NN}$, $g_j^{\pi \pi}$ and $g_j^{\pi N}$) are then evolved to the Fermi motion scales $Q=Q_{\rm f} \sim 0.1$ GeV by nonperturbative renormalization, where they are used in the calculation of NMEs\footnote{This approach was used also in earlier works \cite{CiretalD57}, but with operators with dimension $D < 9$ ($D=5, 7$).} by many-body methods. In a simplified notation, NMEs of Ref.~\cite{Ciretal} [without the Wilson coefficients at the scale $\Lambda^2_{\chi}$, $C_k^{XY}(\Lambda^2_{\chi})$], \textcolor{black}{here denoted as ${\mathcal M}_{(\chi)}$,}
can then be represented schematically as ${\mathcal M}_{(\chi)} = g_j^{\alpha \beta}(Q_{\rm f}^2) {\mathcal M}_{\rm red}$, where $g_j^{\alpha \beta}(Q_{\rm f}^2)$ are the RGE-evolved ChiPT couplings ($\alpha, \beta=N, \pi$) and ${\cal M}_{\rm red}$ are ``reduced'' NMEs. As argued in \cite{Ciretal}, the $NN$ ChiPT couplings $g^{NN}_{j}$ with $j=2,3,4,5$ have significant effect of the running, i.e.,
\be
g^{NN}_{j}(Q_{\rm f}^2) \approx \frac{g^{\pi \pi}_{j}}{m_{\pi}^2} \sim  \frac{g^{\pi \pi}_{j}}{Q_{\rm f}^2} \sim 10^2 \left( \sim \frac{\Lambda_{\chi}^2}{Q_{\rm f}^2} \right) \qquad (j=2,3,4,5),
\label{gNNQf} \ee
\textcolor{black}{where we took into account that $m_{\pi} \sim Q_{\rm f} \sim 0.1$ GeV, and $g^{\pi \pi}_{j} \sim 1 \ {\rm GeV}^2$ as determined by lattice calculations \cite{gjpp}. On the other hand,} the estimate $g^{NN}_{j}(\Lambda_{\chi}^2) \sim 1$ is obtained by the naive dimensional analysis (NDA).\footnote{Refs.~\cite{Prez,Grae} represent similar approaches, but without the enhancements Eq.~(\ref{gNNQf}).}
\textcolor{black}{The enhancement Eq.~(\ref{gNNQf}) comes from contributions of $g_j^{\pi\pi}$ couplings, i.e., from pion exchanges.} These important effects increase those NMEs which, in the expression for the inverse half-life $(T_{1/2}^{0\nu\beta\beta})^{-1}$, appear (in our notation) at the Wilson coefficients of the sectors (12)$^{XX}$ and (31)$^{XY}$ ($X \not= Y$). In our approach, where we evolve the Wilson coefficients down to the scales $Q_{\rm f} \sim 0.1$ GeV, the ChiPT effects Eq.~(\ref{gNNQf}) of the approach of \cite{Ciretal} can be reformulated in the (12)$^{XX}$ sector by adding to the evolution matrix ${\hat U}^{(1)}(Q^2)_{\A}$, cf.~Eq.~(\ref{AQCDhatU1}), a higher-twist ($D=2$) OPE term
\bea
{\hat U}^{(1)}(Q^2)_{\A, {\rm OPE}} & = & \left[ \A_{\hat \nu}(Q^2) + \left( {\hat k}^{(1)}_D + \frac{1}{4} {\hat J}^{(1)} \right) A_{\hat \nu + 1} + {\cal O}(\A_{\hat \nu + 2}) \right] + \frac{ {\hat \mu}_{2}}{Q^2} + {\cal O} \left( \frac{{\hat \mu}_{4}}{Q^4} \right),
\label{hatU1OPE} \eea
where ${\hat \mu}_2$ is a $D=2$ matrix condensate. \textcolor{black}{Such a nonperturbative term $\sim {\hat \mu}_2/Q^2$ would appear also in the brackets on the right-hand side of RGE (\ref{pRGEWils}). The condensate ${\hat \mu}_2$ is expected to be the expectation value of certain operators, with quark structure, in the nucleon. The ascertain the explicit structure of this operator would require a special treatment which goes beyond the scope of this work.} At $Q^2 = Q^2_{\rm f} \sim 0.01 \ {\rm GeV}^2$, the contribution $D=2$ is expected to dominate [cf.~Eq.~(\ref{AQCDvCres})]\footnote{It is expected that $D > 0$ terms are IR-regularized. A plausible scenario is that ${\hat \mu}_2/Q^2$ gets replaced by ${\hat \mu}_2/(Q^2+ M_2^2)$ where $M_2 \sim 0.1$ GeV ($\sim m_{\pi}$); \textcolor{black}{${\hat \mu}_4/Q^4$ gets replaced by ${\hat \mu}_4/(Q^2+M_4^2)^2$ where $M_4 \sim 0.01$ GeV; etc. In such case, $D=2,4,\ldots$ contributions are all suppressed at $Q \sim 1$ GeV; when $Q$ decreases, $D=2$ contribution freezes at $Q \sim 0.1$ GeV to its maximal value and $D=4$ term is still suppressed; $D=4$ contribution freezes to its maximal value at $Q \sim 0.01$ GeV; etc. In such a scenario, $D=2$ term is dominant for $Q \sim 0.1$ GeV, if we assume that the values of the dimensionless quantities ${\hat \mu}_2/M_2^2$, ${\hat \mu}_4/M_4^2$, etc., are mutually comparable.}}
\be
\frac{C_k(Q^2_{\rm f})_{(\A), {\rm OPE}}}{C_k(\Lambda^2_{\chi})_{(\A), {\rm OPE}}} \approx \frac{\Lambda^2_{\chi}}{Q^2_{\rm f}} \frac{({\hat V}^{(0)} {\hat \mu}_2 {\vec {\cal C}})_k}{({\hat V}^{(0)} {\hat \mu}_2 {\vec {\cal C}})_k} = \frac{\Lambda^2_{\chi}}{Q^2_{\rm f}} \sim 10^2.
\label{CQfCL} \ee
This implies that in this formulation, the Wilson coefficients (\ref{AQCDvCres}) get amplified by a factor of $\sim 10^2$ in the running from $Q=\Lambda_{\chi}$ down to $Q=Q_{\rm f}$. \textcolor{black}{Stated otherwise, the effect of RGE-running (enhancement) of the ChiPT coefficients $g_j^{NN}$ in the decay amplitude in the formulation of Ref.~\cite{Ciretal} is reflected in our formulation by the enhancement of the corresponding Wilson coefficients Eq.~(\ref{CQfCL}).}
Schematically, $(T_{1/2}^{0\nu\beta\beta})^{-1} \sim G_1 |C_k^{XY}(Q^2_{\rm f}) g_j^{NN}(\Lambda^2_{\chi}) {\cal M}_{\rm red}|^2 \sim  G_1 |C_k^{XY}(Q^2_{\rm f}){\cal M}_{\rm red}|^2$.
  
The leading-twist ($D=0$) term alone in the expresson (\ref{hatU1OPE}) cannot account for the behavior (\ref{CQfCL}) [$\Leftrightarrow$ (\ref{gNNQf})]. For example, the ratio of the (full evolution) matrix elements $U_{(12)11}^{XX}(Q_{\rm f}^2;M_W^2)/U_{(12)11}^{XX}(\Lambda^2_{\chi};M_W^2)$, with only $D=0$ contribution, is $\approx 1.1$ and $1.2$ for $3 \delta$ $\A$QCD and MPT(1.5), respectively; and $\approx 4.0$ and $3.6$ for $2 \delta$ $\A$QCD and MPT(0.3), respectively.

It turns out that a rough estimate can be made for the values of $D=2$ matrix condensate of the expression (\ref{hatU1OPE}). Namely, for $Q^2 > 1 \ {\rm GeV}^2$ the leading-twist ($D=0$) contribution is assumed to be dominant, and at $Q^2 \sim 1 \ {\rm GeV}^2$ the two contributions ($D=0, 2$) are assumed to start competing (i.e., become comparable). This implies
\be
{\hat \mu}_{2} \sim Q^2 {\hat U}^{(1)}(Q^2)_{\A, D=0} {\big |}_{Q=1 \ {\rm GeV}} \ .
\label{hatmu2e} \ee
This then gives, for both $3 \delta$ $\A$QCD and $2 \delta$ $\A$QCD, the values: $({\hat \mu}_2)_{11} \approx ({\hat \mu}_{2})_{21} \approx 2.4$-$2.8 \ {\rm GeV}^2$; $({\hat \mu}_{2})_{22} \approx 0.4 \ {\rm GeV}^2$; and $({\hat \mu}_{2})_{12} \sim 10^{-3} \ {\rm GeV}^2$.

These arguments were made for the sector (12)$^{XX}$. However, very similar arguments can be made for the sector (31)$^{XY}$ ($X \not= Y$).

As mentioned, the sectors (3)$^{XX}$ and (45)$^{XY}$ are apparently not affected by the possible enhancements mentioned above, i.e., $D=0$ contribution appears to be the dominant one in these sectors. This suggests that our results in these sectors, including the upper bounds on the Wilson coefficients $C_k \equiv C_k(\Lambda^2_{\rm LNV})$, are more complete in the sense that the values of NMEs in these sectors would eventually become comparable in the various approaches \cite{DHP,Prez,Grae,Ciretal}.\footnote{The contributions from $g_j^{\pi \pi}$ and $g_j^{\pi N}$ are not present in Ref.~\cite{DHP}, but are present in Refs.~\cite{Prez,Grae,Ciretal}. They give contributions to NMEs comparable with those of the ChiPT-enhanced $g_j^{NN}(Q^2_{\rm f})$. However, $g_j^{\pi \pi}$ and $g_j^{\pi N}$ are not ChiPT-enhanced, \textcolor{black}{and we do not include them in the present discussion}.}
  Therefore, the results in these sectors may allow us to decide in the future which of the $\A$QCD frameworks is better.

More detailed analysis of the consequences of different methods of the calculation of NMEs for our results goes beyond the scope of the present work. 

We wish to add a further comment on the mentioned problem of the large theoretical uncertainties of the NMEs $\mathcal{M}_j = \langle A_{\rm fin} | \mathcal{O}_j(Q^2_{\rm f}) | A_{\rm in} \rangle$. No specific regulator or scheme is involved in the approaches which calculate NMEs. On the other hand, the anomalous dimension matrices we used in RGEs for the Wilson coefficients were in the aforementioned NDR $\MSbar$ scheme \cite{Buras:2000if}. Therefore, this aspect can be regarded as part of the uncertainty of NMEs used in this work.

The present work is presented in such a way that the possible future improved results for NMEs $\mathcal{M}_j = \langle A_{\rm fin} | \mathcal{O}_j(Q^2_{\rm f}) | A_{\rm in} \rangle$ at $Q_{\rm f} \sim 0.1$ GeV (and preferably in the NDR $\MSbar$ scheme) can be readily used to generate the updated upper bounds on the Wilson coefficients, for various $\A$QCD frameworks, at least in the mentioned sectors (3)$^{XX}$ and (45)$^{XY}$ where $D=2$ contributions do not appear. Namely, the values of the evolution matrix elements, at $Q = Q_{\rm f} = 0.1$ GeV, are given in Table \ref{tabUall}, which means that they can be reused with the future new values of NMEs in Eqs.~(\ref{betas}) to obtain $\beta_j^{XY}$'s, and the new upper bounds for $|C_j(\Lambda^2_{\rm LNV})|$ can be obtained by the simple manipulation described in this Section.

\textcolor{black}{We recall that we used in our calculations the values of NMEs of Refs.~\cite{DHP,Pas:2000vn}, which were obtained in the factorization approximation and using many-body calculations with interactions between nucleons. These quantities correspond to the reduced NMEs ${\mathcal M}_{\rm red}$ (at $Q \sim 0.1$ GeV) mentioned in this Section.\footnote{\textcolor{black}{They do include, though, specific factors which can be interpreted as $g_j^{NN} \sim 1$, cf.~Eq.~(F.2) of Ref.~\cite{Ciretal}.}} On the other hand, in the discussion in this Section we referred to the NMEs of Ref.~\cite{Ciretal} which have, due to the intermediate use of ChiPT ($\chi$), the structure ${\mathcal M} = g_j^{\alpha \beta}(Q_{\rm f}^2) {\mathcal M}_{\rm red}$, and suggested that at least some of the effects of the running of the ChiPT couplings $g_j^{\alpha \beta}(Q^2)$ from $Q \sim 1$ GeV down to $Q \sim 0.1$ GeV may be contained in the $D=2$ term ${\hat \mu}_2/Q^2$ of the corresponding Wilson coefficients, Eq.~(\ref{hatU1OPE}). This then raises yet another question, namely how can ${\mathcal M}$ and ${\mathcal M}_{\rm red}$ (both at $Q \sim 0.1$ GeV) be related in principle in our approach which has no ChiPT involved. One possibility is that in our approach the use of ${\mathcal M}_{\rm red}$ is made for NMEs (i.e., ${\mathcal M} \mapsto {\mathcal M}_{\rm red}$, as we did it in this work) and that all other effects of physics from the regime $Q > 0.1$ GeV are contained in the (OPE) expansions (\ref{hatU1OPE}); in particular, that the nonperturbative physics from the regime $0.1 \ {\rm GeV} < Q < 1$ GeV is contained in some sectors dominantly (though not exclusively) in the condensate matrix\footnote{\textcolor{black}{The sectors (12)$^{XX}$ and (31)$^{XY}$ ($X \not= Y$) are expected to give two different condensate matrices ${\hat \mu}_2$.}} ${\hat \mu}_2$.  This is, of course, only a conjecture, as these difficult aspects go beyond the scope of the present work.}

\section{Conclusions}
\label{sec:conc}

In this work we investigated possible QCD effects in $0\nu\beta\beta$ decays $dd \to uuee$ within the scenarios of new LNV physics which are parametrized as short-range dimension-9 operators $\mathcal{O}_j$, cf.~Eqs.~(\ref{Lagr})-(\ref{ops}). These QCD effects are reflected in the running of the Wilson coefficients $C_j$ of such operators, from the new physics scales $\Lambda^2_{\rm LNV}$ (taken here as $M_W^2 \sim 10^4 \ {\rm GeV}^2$) to the typical $0\nu\beta\beta$-decay (sub-GeV) spacelike scales $Q^2_{\rm f} \sim 0.01 \ {\rm GeV}^2$. For some of these operators their anomalous dimension factors or matrices, which govern the RGE-evolution of the corresponding Wilson coefficients, are known up to two-loops (for $\mathcal{O}_1$-$\mathcal{O}_3$, cf.~Ref.~\cite{Buras:2000if}); for other operators they are known only up to one-loop (for $\mathcal{O}_4$-$\mathcal{O}_5$, cf.~Ref.~\cite{GHK2016,Ciretal,LMW}). The pure pQCD treatment of these RGEs is applicable only down to the (spacelike) scales $Q^2 \sim 1 \ {\rm GeV}^2$, because below such scales the pQCD coupling $a(Q^2)$ ($\equiv \alpha_s(Q^2)/\pi$) is significantly influenced by the artificial Landau singularities which are situated at $0 <Q^2 < \Lambda^2_{\rm Lan.} \sim 0.1 \ {\rm GeV}^2$. In order to achieve the running of the Wilson coefficients $C_j(Q^2)$ down to Fermi motion scales $Q^2 \sim 0.01 \ {\rm GeV}^2$, we employed various variants of QCD where the running coupling $\A(Q^2)$ [the analog of the pQCD coupling $a(Q^2)$] has no such Landau singularities, i.e., various frameworks of $\A$QCD: $3 \delta$, $2 \delta$, MPT($M$) and FAPT. We point out that in such evaluations, in order to evaluate correctly the low-momentum nonperturbative effects, it was important not to treat the analogs of the powers $a(Q^2)^{\nu}$ as naive powers $\A(Q^2)^{\nu}$, but rather as $\A_{\nu}(Q^2)$ ($\not= \A(Q^2)^{\nu}$) which are linear combinations of the (generalized) logarithmic derivatives $\tA_{\nu + m}(Q^2)$ ($m=0,1,\ldots$).\footnote{Appendix \ref{app:AQCD} is a summary of various $\A$QCD frameworks and of the evaluation of $\A_{\nu}$, all this information being available in the literature.}

The mentioned evolution of the Wilson coefficients, down to the Fermi motion scales, allowed us then in Sec.~\ref{sec:num} to evaluate the $0\nu\beta\beta$ half-life of $^{136}{\rm Xe}$ in terms of these coefficients $C_j(Q_{\rm f}^2)$ and of the corresponding nuclear matrix elements (NMEs) of the operators $\mathcal{O}_j$. Comparison of this expression with the presently available lower bound on the mentioned half-life then allowed us to extract the upper bounds for the Wilson coefficients $C_j(\Lambda^2_{\rm LNV})$ at the new physics scale.

Our main conclusions are the following. The values of the evolution factors or matrices $U(Q_{\rm f}^2,\Lambda^2_{\rm LNV})$ of Wilson coefficients, when the two-loop anomalous dimensions were used, were in all $\A$QCD frameworks not far from (and often close to) the values obtained for $U(Q^2,\Lambda^2_{\rm LNV})$ when one-loop anomalous dimensions were used. This conclusion holds even when the values of the Fermi motion scales are realistic, i.e., very low, $Q_{\rm f}^2 \sim 0.01 \ {\rm GeV}^2$. As a consequence, similar conclusion can be made for the extracted values of the upper bounds of $|C_j(\Lambda^2_{\rm LNV})|$. Further, as could be expected, the numerical results for different $\A$QCD frameworks depend largely on the behavior of the coupling $\A(Q^2)$ in the IR regime $Q^2 \lesssim 0.1 \ {\rm GeV}^2$. Therefore, for example, the results of $2 \delta$ and MPT(0.3) $\A$QCD variants were mutually comparable. The results of $3 \delta$ $\A$QCD are not easily comparable with those of other $\A$QCD frameworks, principally because the coupling $\A(Q^2)$ in $3 \delta$ $\A$QCD goes to zero in the deep IR-regime (as suggested by large-volume lattice results). Yet another conclusion of this work is that the described QCD effects are important (more than the present uncertainty of the NMEs) in most of the cases of the considered Wilson coefficients: these effects affect in such cases the upper bounds for $|C_j(Q^2)|$ (when $Q^2 = 0.01 \ {\rm GeV}^2$) by more than a factor of two.

Finally, we point out that at present there is a major uncertainty in the NMEs associated with Wilson coefficients of the operator sectors ${\mathcal O}_1^{XX}$-${\mathcal O}_2^{XX}$ and ${\mathcal O}_3^{XY}$-${\mathcal O}_1^{XY}$ ($X \not= Y$). Namely, according to Ref.~\cite{Ciretal} such NMEs may involve an enhancement effect of $\sim 10^2$ in comparison to other approaches \cite{DHP,Pas:2000vn,Prez,Grae,GDIK}. This effect may require, in our approach in the mentioned operator sectors, introduction of additional, $D=2$ contributions $\sim 1/Q^2$ in the evolution matrices.

\medskip

\begin{acknowledgments}
This work was supported in part by the Chilean FONDECYT Regular Grants No.~1200189 (C.A.) and No.~1180344 (G.C.). The work of L.G. was supported by CONICYT Chile Grant No.~21160645 and DGIIP of UTFSM.

We are grateful to M.~Gonz\'alez and S.G.~Kovalenko for helpful discussions.  
\end{acknowledgments}

\begin{appendix}
\appendix
\section{Anomalous dimension at LO and NLO}
\label{app:AD}

In this Section we write down the anomalous dimension in the full one-loop approximation and in the currently known two-loop one. The results will be expressed in terms of the number of colors $N$, and the number of active flavors $n_f$.

First, we write down the result for the mixing of operators (\ref{ssOp}) and (\ref{ttOp}), as obtained in Ref.\cite{Buras:2000if}
\begin{eqnarray}
\hat{\gamma}_{(12)}^{(0),X X}&=&\left(\begin{array}{cc}{6+\frac{6}{N}-6N} & {\frac{1}{N}-\frac{1}{2}} \\ {24+\frac{48}{N}} & {6-\frac{2}{N}+2N}\end{array}\right) \ ,
\\
\gamma_{(12)11}^{(1),X X}&=&-\frac{203}{6} N^{2}+\frac{107}{3} N+\frac{136}{3}-\frac{12}{N}-\frac{107}{2 N^{2}}+\frac{10}{3} N n_f-\frac{2}{3} n_f-\frac{10}{3 N} n_f \ ,
\nonumber\\ 
\gamma_{(12)12}^{(1),X X}&=&\frac{1}{36} N+\frac{31}{9}-\frac{9}{N}+\frac{4}{N^{2}}+\frac{1}{18} n_f-\frac{1}{9 N} n_f \ ,
\nonumber\\ 
\gamma_{(12)21}^{(1),X X}&=&\frac{364}{3} N+\frac{704}{3}+\frac{208}{N}+\frac{320}{N^{2}}-\frac{136}{3} n_f-\frac{176}{3 N} n_f \ ,
\nonumber\\ 
\gamma_{(12)22}^{(1),X X}&=&\frac{343}{18} N^{2}+21 N-\frac{188}{9}+\frac{44}{N}+\frac{21}{2 N^{2}}-\frac{26}{9} N n_f-6 n_f+\frac{2}{9 N} n_f \ .
\end{eqnarray}
We note that the off-diagonal elements here in $\gamma^{(0)}_{(12)}$ and $\gamma^{(1)}_{(12)}$  have the opposite sign to those of \cite{Buras:2000if}; this is so because in Eq.~(\ref{ttOp}) we use the convention $\sigma^{\mu \nu}=(i/2) [\gamma^{\mu}, \gamma^{\nu}]$, while  in Ref.~\cite{Buras:2000if} the convention $\sigma^{\mu \nu}=(1/2) [\gamma^{\mu}, \gamma^{\nu}]$ is used.

Then, we write down the result for the mixing of the operators (\ref{vvOp}) and (\ref{ssOp}), as obtained in Ref.\cite{Buras:2000if}
\begin{eqnarray}
\hat{\gamma}_{(31)}^{(0),X Y}&=&\left(\begin{array}{cc}{\frac{6}{N}} & {12} \\ {0} & {-6N + \frac{6}{N}}\end{array}\right)
\\
\hat{\gamma}^{(1),XY}_{(31)}&=&\left(\begin{array}{cc}{\frac{137}{6}+\frac{15}{2 N^{2}}-\frac{22}{3 N} n_f} & {\frac{200}{3} N-\frac{6}{N}-\frac{44}{3} n_f} \\ {\frac{71}{4} N+\frac{9}{N}-2 n_f} & {-\frac{203}{6} N^{2}+\frac{479}{6}+\frac{15}{2 N^{2}}+\frac{10}{3} N n_f-\frac{22}{3 N} n_f}\end{array}\right)
\end{eqnarray}

The final result that is known up to two-loop approximation corresponds to the operator (\ref{vvOp})  \cite{Buras:2000if}
\begin{eqnarray}
\gamma_{(3)}^{(0),X X}&=&6-\frac{6}{N}\ ,
\\
\gamma_{(3)}^{(1),X X}&=&-\frac{19}{6} N-\frac{22}{3}+\frac{39}{N}-\frac{57}{2 N^{2}}+\frac{2}{3} n_f-\frac{2}{3 N} n_f \ .
\end{eqnarray}

For the operators (\ref{vtOp}) and (\ref{vsOp}), only the one-loop anomalous dimension is known \cite{LMW}
\bes
\bea
\hat{\gamma}_{(45)}^{(0), XX} &=&\left(\begin{array}{cc}{3-\frac{1}{N}+N} & {\left(3+\frac{6}{N}\right) i} \\ {\left(1-\frac{2}{N}\right) i} & {3+\frac{3}{N}-3N}\end{array}\right)\ ,
\label{g45XX}
\\
\hat{\gamma}_{(45)}^{(0), XY} &=&\left(\begin{array}{cc}{3-\frac{1}{N}+N} & {\left(-3-\frac{6}{N}\right) i} \\ {\left(-1+\frac{2}{N}\right) i} & {3+\frac{3}{N}-3N}\end{array}\right)\ .
\label{g45XY}
\eea
\ees
This same result can be deduced also from the results of \cite{Ciretal}, when taking into account the following relations between the operators ${\mathcal O}_6^{\mu}$-${\mathcal O}_9^{\mu}$ of \cite{Ciretal} and the operators ${\mathcal O}_4^{XY}$- ${\mathcal O}_5^{XY}$ of Eqs.~(\ref{vtOp})-(\ref{vsOp}):
\bes
\label{CK}
\bea
{\mathcal O}_6^{\mu} & = & \frac{1}{4} {\mathcal O}_5^{LR \mu},
\label{O6} \\
{\mathcal O}_7^{\mu} & = & - \frac{i}{16} {\mathcal O}_4^{LR \mu} - \frac{(N+2)}{16 N}  {\mathcal O}_5^{LR \mu},
\label{O7} \\
{\mathcal O}_8^{\mu} & = &  \frac{1}{4} {\mathcal O}_5^{LL \mu},
\label{O8} \\      
{\mathcal O}_9^{\mu} & = & + \frac{i}{16} {\mathcal O}_4^{LL \mu} - \frac{(N+2)}{16 N}  {\mathcal O}_5^{LL \mu},
\label{O9}.
\eea \ees
For the operators ${\mathcal O}_6^{\mu \prime}$-${\mathcal O}_9^{\mu \prime}$ of \cite{Ciretal} the above relations turn out to be the same, but with $L \leftrightarrow R$ on the RHS.
These relations can be obtained by the use of Fierz transformations and of the color rearrangement  identity involving the $SU(N)$ color generators
\be
t^a_{\alpha \beta} t^a_{\eta \xi} = - \frac{1}{2 N} \delta_{\alpha \beta} \delta_{\eta \xi} + \frac{1}{2} \delta_{\alpha \xi} \delta_{\eta \beta}.
\label{tata}
\ee

\section{IR-safe couplings}
\label{app:AQCD}

This Appendix is a compendium and synthesis of several results obtained and described in various works on IR-safe holomorphic couplings, among them Refs.~\cite{Sh1Sh2,reviews,2dCPC,3dAQCD,CV12,GCAK} and \cite{renmod} (App.~B there).

The pQCD running coupling $a(Q^2) \equiv \alpha_s(Q^2)/\pi$ is defined as a function of the squared momentum $Q^2 \equiv - q^2$ in the generalized spacelike region, where $q^2 = (q^0)^2 - {\vec q}^2$ and $q$ represents a typical momentum of a considered process. When $q^2<0$ ($Q^2>0$), the momentum $q$ is considered to be spacelike in the restricted sense (e.g., appearing in deep inelastic scattering and other $t$-channel quantities, and in current correlators). When $q^2=s>0$ ($Q^2=-s<0$), the momentum is usually called timelike (e.g., appearing in the $s$-channel type decay widths and cross sections). The generalized spacelike (Euclidean) region of $Q^2$ is considered to be the entire complex plane with the exception of the timelike semiaxis: $Q^2 \in \mathbb{C} \backslash (-\infty, 0]$, and it is in this region that the running coupling $a(Q^2)$ is considered. The running coupling in this region is a solution of the (perturbative) RGE
\bes
\label{RGE}
\bea
\frac{d a(Q^2)}{d \ln Q^2} \equiv \beta(a(Q^2))
& = &
- \beta_0 a(Q^2)^2 - \beta_1 a(Q^2)^3 - \beta_2 a(Q^2)^4 - \ldots
\label{RGEa}
\\
& = &
- \beta_0 a(Q^2)^2 \left[ 1 + c_1 a(Q^2) +  c_2 a(Q^2)^2 + \ldots \right].
\label{RGEb}
\eea
\ees
Here, the first two $\beta$-coefficients,  $\beta_0 = (1/4)(11- 2 N_f/3)$ and $\beta_1=(1/16)(102 - 38 N_f/3)$, are scheme independent in mass independent schemes. The coefficients $c_j =\beta_j/\beta_0$ ($j \geq 2$) characterize the pQCD renormalization scheme \cite{Stevenson}. This means that the form of the function $\beta(a; c_2, c_3,\ldots)$ represents a definition of the renormalization scheme. Here, the momentum scale $\Lambda_{\rm QCD}$ will not be regarded as a scheme parameter, but as the momentum (re)scaling definition. This scaling change can be described equivalently as a change of the renormalization scale. Throughout this work, the $\MSbar$ scaling definition ($\Lambda^2_{\rm QCD}={\overline \Lambda}^2$) is adopted.

The pQCD coupling $a(Q^2)$, which is the solution of the perturbative RGE in a given or chosen renormalization scheme, usually has singularities on the positive axis in the $Q^2$-complex plane, $0 \leq Q^2 \lesssim \Lambda_{\rm QCD}^2$ ($\sim 0.01$-$1 \ {\rm GeV}^2$); this is in addition to the expected singularities on the negative $Q^2$-axis. However, the spacelike QCD observables ${\cal D}(Q^2)$ (such as current correlators, $t$-channel process quantities, and nucleon structure functions and their sum rules) are are holomorphic (analytic) functions in the $Q^2$-complex plane with the exception of a part of the negative semiaxis, $Q^2 \in \mathbb{C} \backslash (-\infty, -M_{\rm thr}^2]$ (with a threshold mass $M_{\rm thr} \sim 0.1$ GeV). This follows from the general principles of Quantum Field Theories \cite{BS,Oehme}. Stated otherwise, spacelike QCD observables ${\cal D}(Q^2)$ are holomorphic functions in the entire (generalized) spacelike region.

\begin{figure}[htb]
\centering\includegraphics[width=130mm]{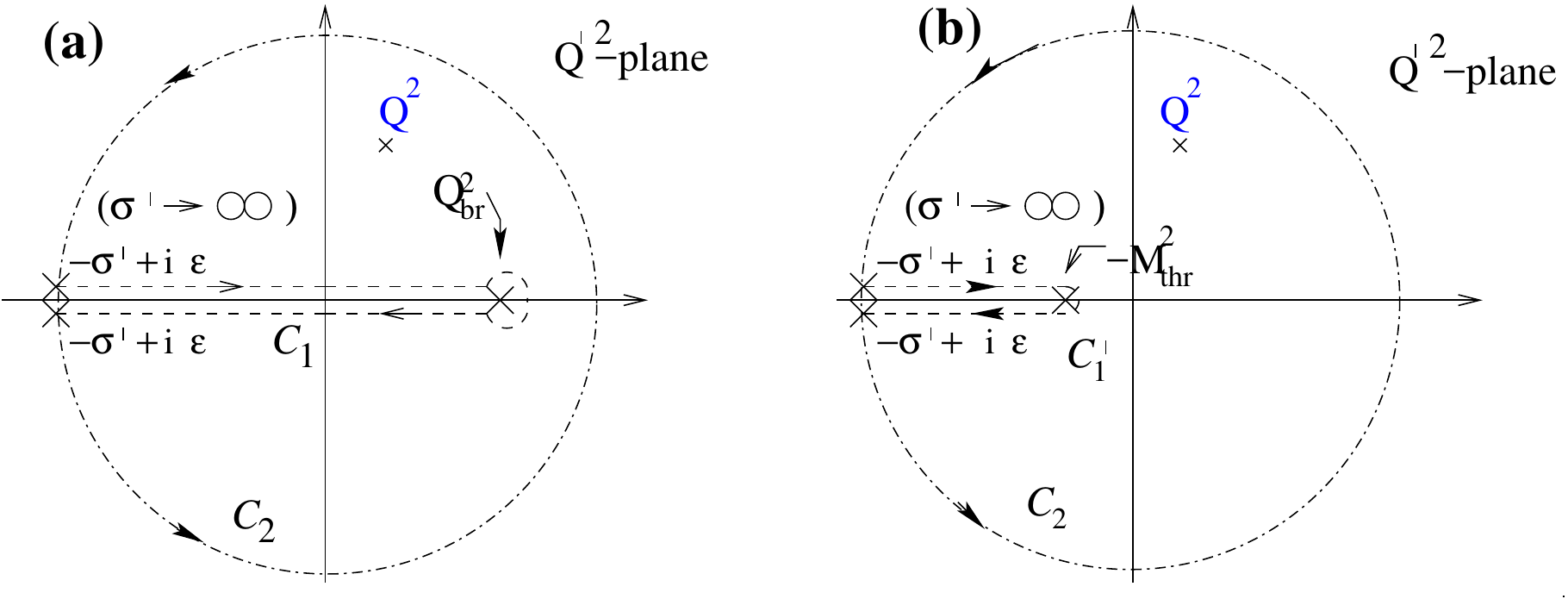}
\caption{\footnotesize (a) The integration contour for the integrand $a (Q'^2)/(Q'^2 - Q^2)$ which leads to the relation (\ref{dispa}) for  $a (Q^2)$; (b) the integration contour for the integrand  $A(Q'^2)/(Q'^2 - Q^2)$ which leads to the relation (\ref{dispA}). The radius of the circular part is $\sigma' \to \infty$.}
\label{intpath}
\end{figure}

  These properties of ${\cal D}(Q^2)$ are not reflected qualitatively in the pQCD coupling $a(Q^2)$, because the latter has the mentioned Landau singularities (cut and branching points) on the positive axis. This behavior of $a(Q^2)$ is unfortunate in the following sense: if in the evaluation of ${\cal D}(Q^2)$ at low $|Q^2| \lesssim 1 \  {\rm GeV}^2$ we use the coupling $a(Q^2)$ [or $a(\mu^2)$ with $\mu^2=\kappa Q^2 \sim Q^2$], we obtain either useless or unreliable results. These Landau singularities (also called Landau ghosts) are usually a cut on an interval $0 \leq Q^2 \leq Q_{\rm br}^2$, and the point $Q^2 =Q_{\rm br}^2$ ($\sim 0.01$-$1 \ {\rm GeV}^2$) is called the Landau branching point. When we apply the Cauchy theorem to the integrand $a(Q^{'2})/(Q^{'2}-Q^2)$ in the $Q^{'2}$-complex plane along the path in Fig.~\ref{intpath}(a), this gives the following representation of the pQCD coupling $a(Q^2)$ in the form of a dispersion integral:
\be
a(Q^2) =
\frac{1}{\pi} \int_{-Q_{\rm br}^2- \eta}^{+\infty} d \sigma \frac{\rho_a(\sigma)}{(\sigma + Q^2)}, \qquad (\eta \to +0).
\label{dispa}
\ee
Here, $\rho_a(\sigma) = {\rm Im} a(Q^{' 2} = -\sigma - i \epsilon)$ is the discontinuity (or spectral) function of the coupling $a$ along its cut.

On the other hand, the holomorphic (in the spacelike region) coupling $\A(Q^2)$ [the analog of $a(Q^2)$] has its cut only along the negative semiaxis $-\infty < Q^{'2} < -M^2_{\rm thr}$, and hence the form of its dispersion integral is [cf.~Fig.~\ref{intpath}(b)]
\be
\A(Q^2)=
\frac{1}{\pi} \int_{0}^{+\infty} d \sigma \frac{\rho_{\A}(\sigma)}{(\sigma + Q^2)} =\frac{1}{\pi} \int_{M^2_{\rm thr}-\eta}^{+\infty} d \sigma \frac{\rho_{\A}(\sigma)}{(\sigma + Q^2)}, \qquad (\eta \to +0),
\label{dispA}
\ee
where $\rho_{\A}(\sigma) = {\rm Im} \A(Q^{' 2} = -\sigma - i \epsilon)$ is the discontinuity of $\A(Q^{'2})$ along its cut in the complex $Q^{'2}$-plane.
This coupling has the cut threshold $\sigma_{\rm min} (\equiv M_{\rm thr}^2) \geq 0$.
In contrast to $a(Q^2)$, the couplings $\A(Q^2)$ represent qualitatively correctly the holomorphic properties of the QCD spacelike observables ${\cal D}(Q^2)$, and can thus be regarded as better suited for the evaluation of such quantities. However, $\A(Q^2)$ have to fulfill various physically-motivated requirements: (a) at high $|Q^2| > 1 \ {\rm GeV}^2$ they must reproduce the perturbative QCD; (b) at intermediate  $|Q^2| \sim 1 \ {\rm GeV}^2$ they must reproduce the corresponding QCD phenomenology, especially the physics of the $\tau$ lepton semihadronic decays which is well measured; (c) and at very low $|Q^2| < 1 \ {\rm GeV}^2$ we may require that they have the behavior as suggested by large-volume lattice results for the (Landau gauge) gluon and ghost dressing functions, if the running coupling there is defined in a natural way as a product of these dressing functions.

The high-momentum condition (a) can be also formulated in the following way: in a chosen renormalization scheme (i.e., for a chosen set of values of the scheme $c_j$ coefficients, $j \geq 2$), the discontinuity function $\rho_{\A}(\sigma)$ coincides at large $\sigma$ with the pQCD discontinuity function of the underlying pQCD coupling $a$
\be
\rho_{\A}(\sigma) =
\rho_a(\sigma)  \quad (\sigma \geq M_0^2 \stackrel{>}{\sim} 1 {\rm GeV}^2),
\label{rhoArhoa}
\ee
where $M_0^2$ can be called the pQCD-onset scale. Then at large $|Q^2| > 1 \ {\rm GeV}^2$ the requirement that the two running couplings practically coincide can be written as
\be
\A(Q^2) - a(Q^2) \sim
\left( \frac{\Lambda^2}{Q^2} \right)^{\cal N}
\label{diffAaN}
\ee
where $\Lambda^2 \sim 0.1 \ {\rm GeV}^2$ and index ${\cal N}$ must be relatively large, e.g. ${\cal N}=5$.

The simplest holomorphic coupling (APT) \cite{ShS} was constructed from the underlying pQCD coupling by equating $\rho_{\A}=\rho_a$ for all $\sigma \geq 0$ (and necessarily setting equal to zero the Landau cut discontinuities $\rho_{\A}(\sigma)$ at $\sigma<0$)
\be
\A^{\rm (APT)}(Q^2) =
\frac{1}{\pi} \int_{0}^{+\infty} d \sigma \frac{\rho_a(\sigma)}{(\sigma + Q^2)}.
\label{dispAAPT}
\ee
On the other hand, we constructed two types of couplings which fulfill the condition (a) [i.e., Eq.~(\ref{diffAaN}) with ${\cal N}=5$] and (b) \cite{2dAQCD,2dCPC,3dAQCD}, one type of coupling fulfilling also the deep infrared condition (c)  \cite{3dAQCD}. The discontinuity functions for these two types of couplings are parametrized in the unknown low-$\sigma$ region ($\sigma < M_0^2$) by a combination of Dirac-delta functions
\be
\rho_{\A}^{(n \delta)}(\sigma) =
\pi \sum_{j=1}^{n} {\cal R}_j \; \delta(\sigma - M_j^2)  + \Theta(\sigma - M_0^2) \rho_a(\sigma) \ ,
\label{rhoAnd}
\ee
where we expect $0 < M_1^2 < \ldots < M_n^2 < M_0^2$, and $M_0^2 \sim \ 1 \ {\rm GeV}^2$ is the pQCD-onset scale. The corresponding coupling is
\bea
\A^{(n \delta)}(Q^2) \left( \equiv \frac{1}{\pi} \int_0^{\infty} d \sigma \frac{\rho_{\A}(\sigma)}{(\sigma + Q^2)} \right) & = &
\sum_{j=1}^n \frac{{\cal R}_j}{(Q^2 + M_j^2)} + \frac{1}{\pi} \int_{M_0^2}^{\infty} d \sigma \frac{ \rho_a(\sigma) }{(Q^2 + \sigma)} \ .
\label{AQ2}
\eea
The Dirac delta functions in the spectral function give a nonperturbative contribution $\Delta \A_{\rm IR}(Q^2)$, in the form of a linear combination of simple fractions $\sim 1/(Q^2+M_j^2)$, and this can be rewritten as a near diagonal Pad\'e approximant $\Delta \A_{\rm IR}(Q^2) = [n/n-1](Q^2)$. Pad\'e approximants $[n/n-1](Q^2)$ are known to approximate the holomorphic functions in the $Q^2$-complex plane increasingly well when $n$ increases \cite{Peris}.

Such couplings $\A^{(n \delta)}$ were constructed, with two ($n=2$) and three ($n=3$) Dirac delta functions, in Refs.~\cite{2dAQCD,2dCPC} and \cite{3dAQCD}, respectively, in specific renormalization schemes.\footnote{The schemes are perturbatively defined, by the (perturbative) $\beta(a)$ function of the underlying pQCD coupling $a$.}
The values of the $(2 n +1)$ parameters [$M^2_j$, ${\cal R}_j$ ($j=1,\ldots,n$) and $M_0^2$] were then determined by several physically motivated requirements.

The requirement that $\A(Q^2)$ coupling at high $|Q^2| > 1 \ {\rm GeV}^2$ practically coincide with the underlying pQCD $a(Q^2)$ [Eq.~(\ref{diffAaN}) with ${\cal N}=5$] provides four of these requirements.

On the other hand, the fifth requirement comes from physics at moderate momenta $|Q^2| \sim m_{\tau}^2$ ($\sim 1 \ {\rm GeV}^2$): the requirement that the physics of the semihadronic $\tau$ lepton decays be reproduced, i.e., that the calculated (massless and strangeless) $\tau$ decay ratio $r_{\tau}^{(D=0)}$ gives the correct well-measured value. This is sufficient for $2 \delta$ $\A$QCD \cite{2dAQCD,2dCPC} which has five parameters.

In $3 \delta$ $\A$QCD \cite{3dAQCD}, which has seven parameters ($n=3$), two additional requirements were imposed, namely that $\A^{(3 \delta)}(Q^2) \sim Q^2$ when $Q^2$ goes to zero, and that $\A^{(3 \delta)}(Q^2)$ has for positive $Q^2$ a local maximum at $Q^2 \approx 0.135 \ {\rm GeV}^2$, in the Lambert MiniMOM (LMM) scheme. These two requirements are suggested by the large volume lattice calculations  \cite{LattcoupNf0} for $N_f=0$\footnote{Similar results were obtained also by another group  \cite{LattcoupNf0b}, for $N_f=0$. Further, similar results, but in general with lower statistics, were obtained for $N_f=2$ \cite{LattcoupNf2} and $N_f=4$ \cite{LattcoupNf4}.}
of the Landau gauge dressing functions $Z_{\rm gl}(Q^2)$ and $Z_{\rm gh}(Q^2)$ of the gluon and ghost propagators in the MiniMOM (MM) scheme \cite{MM1},\footnote{Lambert MiniMOM (LMM) scheme is the MiniMOM (MM) scheme (where the lattice calculations are performed), but with momenta rescaled to the usual $\MSbar$ scaling: $Q^2=Q^2_{\rm latt} (\Lambda_{\MSbar}/\Lambda_{\rm MM})^2 \approx Q^2_{\rm latt}/1.9^2$.}
where the lattice coupling $\A_{\rm latt.}$ was defined naturally as: $\A_{\rm latt.}(Q^2) \propto Z_{\rm gl}(Q^2) Z_{\rm gh}(Q^2)^2$. 

There is yet another, the $(2n+2)$'th ``hidden'' parameter, involved in $\A^{(n \delta)}(Q^2)$: it is the strength of the underlying pQCD coupling $a(Q^2)$ (at $N_f=3$); it can be characterized by the value of $\alpha_s(M_Z^2;\MSbar)$ (at $N_f=5$) which we take in this work in general as $\alpha_s(M_Z^2;\MSbar)=0.1181$ \cite{PDG18}.

When a specific coupling $\A(Q^2)$ has been constructed ($a \mapsto \A$), the couplings $\A_n(Q^2)$ which are the analogs of the powers $a(Q^2)^n$ of the underlying pQCD coupling ($a^n \mapsto \A_n$), can be obtained in general holomorphic $\A$QCD following the steps presented in Ref.~\cite{CV12} for integer $n$, and in Ref.~\cite{GCAK} for general (noninteger) $n$. In this construction of $\A_n(Q^2)$ from $\A(Q^2)$ for integer $n$, the logarithmic derivatives of $\A(Q^2)$ play a central role.

Here, first the construction given in Ref.~\cite{CV12} for integer $n$ will be outlined. From the linearity of the ``analytization'' $a(Q^2) \mapsto \A(Q^2)$ follows that the logarithmic derivatives $\ta_{n+1}(Q^2)$ of $a(Q^2)$
\be
\ta_{n+1}(Q^2) \equiv
\frac{(-1)^n}{\beta_0^n n!} \left( \frac{d}{d \ln Q^2} \right)^n a(Q^2) \qquad (n=0,1,2,\ldots)
\label{tan}
\ee
are replaced (i.e., ``analytized'') in $\A$QCD by the analogous logarithmic derivatives $\tA_{n+1}(Q^2)$ of $\A(Q^2)$
\bes
\label{tAn}
\bea
\left( \ta_{n+1}(Q^2) \right)_{\rm an.} &=& \tA_{n+1}(Q^2) \equiv \frac{(-1)^n}{\beta_0^n n!} \left( \frac{d}{d \ln Q^2} \right)^n \A(Q^2)
\label{tAna}
\\
& = & \frac{1}{\pi} \frac{(-1)}{\beta_0^n \Gamma(n+1)} \int_0^{\infty} \frac{d \sigma}{\sigma} \rho_{\A}(\sigma) {\rm Li}_{-n}\left( - \frac{\sigma}{Q^2} \right)
\qquad (n=0,1,2,\ldots),
\label{tAnb}
\eea
\ees
where the expression (\ref{tAnb}) is obtained by using the definition (\ref{tAna}) and the dispersion integral (\ref{dispA}).

\begin{figure}[htb] 
\begin{minipage}[b]{.49\linewidth}
  \centering\includegraphics[width=85mm]{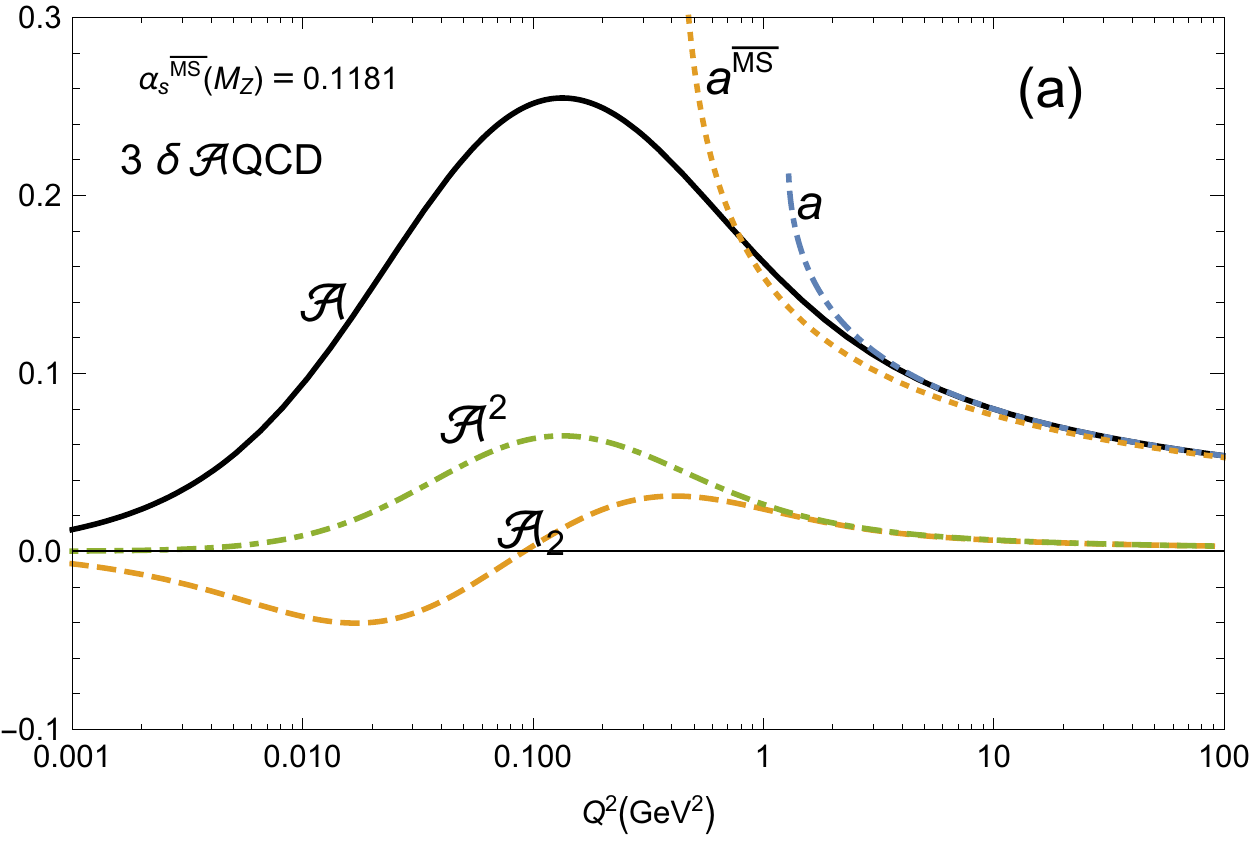}
  \end{minipage}
\begin{minipage}[b]{.49\linewidth}
  \centering\includegraphics[width=85mm]{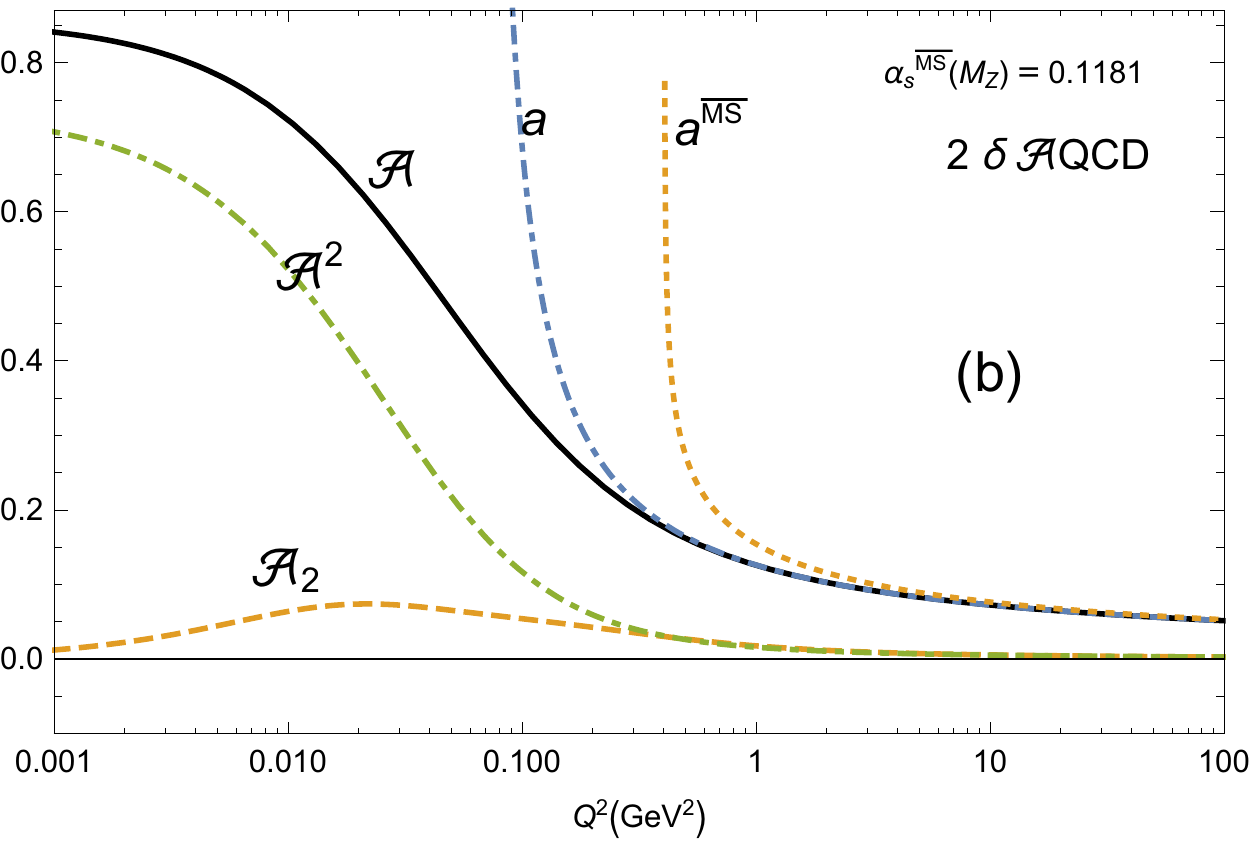}
\end{minipage}
\caption{\footnotesize  The couplings $\A(Q^2)$, $\A_2(Q^2)$, as a function of positive $Q^2$: (a) in the considered $3 \delta$ $\A$QCD case (in the LMM scheme); (b) in the considered  $2 \delta$ $\A$QCD case (in the Lambert scheme with $c_2=-4.9$). Included are the naive power $\A(Q^2)^2$ [$\not= \A_2(Q^2)$], the underlying pQCD coupling $a(Q^2)$, and the five-loop $\MSbar$ coupling $a^{\MSbar}(Q^2)$. The curves are for $N_f=3$. The corresponding ``strength'' reference value is $\alpha_s(M_Z^2,\MSbar)=0.1181$. For $\A_2$, the sum (\ref{AntAn}) with three terms was taken (and $n=1$).}
\label{FigAtA2}
\end{figure}

This construction allows us to evaluate such (truncated) $\A$QCD series whose perturbation series starts with an integer power of $a(Q^2)$, e.g., with $a(Q^2)^1$. Namely, the (leading-twist part of the) spacelike observable ${\cal D}(Q^2)$ has in such a case the power expansion
\bea
{\cal D}_{\rm pt}(Q^2) &=&
d_0 a(\mu^2) + \sum_{n \geq 1} d_n(\kappa) a(\mu^2)^{n+1},
\label{Dkappt}
\eea
where $\kappa \equiv \mu^2/Q^2$ is the renormalization scale parameter ($0 < \kappa \sim 1$). This series can be reorganized in a straightforward way as a series in the logarithmic derivatives (\ref{tan}) instead
\bea
{\cal D}_{\rm lpt}(Q^2) &=&
\td_0 a(\mu^2) + \sum_{n \geq 1} \td_n(\kappa) \ta_{n+1}(\mu^2),
\label{Dkaplpt}
\eea
where $\td_0=d_0$,  $\td_1(\kappa)=d_1(\kappa)$, $\td_2(\kappa)= d_2(\kappa) - c_1 d_1(\kappa)$, etc.\footnote{The latter relations between $\td_n$ and $d_{n-k}$'s are obtained because RGE (\ref{RGE}) implies the relations of the form $\ta_{m} = a^{m} + k_1(m) \ a^{m+1} + \ldots$.}
The resulting (truncated) series is then evaluated with the $\A$-coupling 
\bea
{\cal D}^{[N]}_{\A{\rm QCD}}(Q^2; \kappa) &=&
\td_0 \A(\kappa Q^2) + \td_1(\kappa)  \tA_2(\kappa Q^2) +
\ldots + \td_{N-1}(\kappa) \tA_{N}(\kappa Q^2).
\label{DNAkaplpt}
\eea
A weak renormalization scale dependence ($\kappa$-dependence) appears here due to the truncation effect. The analog of this expression in pQCD is the series Eq.~(\ref{Dkaplpt}) truncated at $\td_{N-1}(\kappa) \ta_{N}(\kappa Q^2)$. The truncated series (\ref{DNAkaplpt}) differs from the full sum ${\cal D}(Q^2)$ formally by a term $\sim \tA_{N+1}$ ($ \sim \ta_{N+1} \sim a^{N+1}$); this is suppressed in comparison to $\sim \tA_{N}$, because $\A$QCD frameworks in general fulfill the hierarchy $|\A(Q^2)| > |\tA_2(Q^2)| > |\tA_3(Q^2)| > \ldots$, for all (non-timelike) scales $Q^2$ (cf.~also Figs.~\ref{FigAtA2}), which appears as a consequence of the holomorphic behavior of $\A(Q^2)$.

The truncated series (\ref{DNAkaplpt}) can be reorganized (rewritten) explicitly in terms of the coefficients $d_n(\kappa)$ of the original perturbation (power) series (\ref{Dkappt})
\bea
{\cal D}^{[N]}_{\A{\rm QCD}}(Q^2; \kappa) &=&
d_0 \A(\kappa Q^2) + d_1(\kappa)  \A_2(\kappa Q^2) + \ldots + d_{N-1}(\kappa) \A_{N}(\kappa Q^2),
\label{DNAkappt}
\eea
where the power analog $\A_{n+1}$ (the $\A$-coupling analog of the power $a^{n+1}$) is a specific linear combination of the logarithmic derivatives $\tA_{n+m}$ in complete analogy with the pQCD relations
\be
\A_{n+1}  =
\tA_{n+1} + \sum_{m=1}^{N-n-1} \tk_m(n+1) \tA_{n+1+m} \qquad (n=1,\ldots,N-1).
\label{AntAn}
\ee
Here, the sums are truncated consistently at $\tA_N$; we note that $\A_N = \tA_N$ in this case. We point out that the truncated series (\ref{DNAkappt}) is equal to (\ref{DNAkaplpt}), and its pQCD analog are the series (\ref{Dkaplpt}) and (\ref{Dkappt}), both truncated at  $n+1=N$. Since $\A(Q^2)$ has in general some nonperturbative contributions in comparison to its underlying pQCD coupling $a(Q^2)$, we have $\A_n(Q^2) \not= \A(Q^2)^n$ ($n \geq 2$), and this holds even if the truncation index $N$ in the relations (\ref{AntAn}) is very high. On the other hand, at high $|Q^2| > 1 \ {\rm GeV}^2$ we have in general $\A_n(Q^2) \approx \tA_n(Q^2) \approx \A(Q^2)^n \approx a(Q^2)^n $, because ($2\delta$ and $3\delta$) $\A$QCD in the high-momentum regime practically coincides with the underlying pQCD, due to the relation (\ref{diffAaN}) with ${\cal N}=5$ there.
If in the series (\ref{DNAkappt}) the naive powers $\A(Q^2)^n$ were used instead of $\A_n(Q^2)$, this would give to the series spurious uncontrollable nonperturbative contributions at low $|Q^2| \lesssim 1 \ {\rm GeV}^2$ \cite{Techn}. Hence it is important to employ the series in logarithmic derivatives instead, i.e., Eq.~(\ref{DNAkaplpt}) [$\Leftrightarrow$ Eq.~(\ref{DNAkappt})].

In Figs.~\ref{FigAtA2}(a),(b) we present the couplings $\A(Q^2)$, $\A_2(Q^2)$, as a function of $Q^2>0$, for the considered $2\delta$ and $3\delta$ $\A$QCD, respectively, and we included the corresponding underlying pQCD coupling $a(Q^2)$ and the $\MSbar$ coupling ${\bar a}(Q^2)$ (all are for $N_f=3$). The ``strength'' reference value for the pQCD couplings is $\alpha_s(M_Z^2;\MSbar;N_f=5)=0.1181$. 
The coupling $\A_2(Q^2)$ is generated by Eq.~(\ref{AntAn}) with three terms (i.e., $N=4$ and $n=1$). The naive power $\A(Q^2)^2$ is also presented in these Figures; clearly: $\A_2(Q^2) \not\approx \A(Q^2)^2$ at low $Q^2$. The pQCD coupling $a(Q^2)$ in the LMM scheme has the branching point at $Q_{\rm br}^2 \approx 1.29 \ {\rm GeV}^2$ (not a pole). In the Lambert scheme with $c_2=-4.9$, where $2\delta$ $\A$QCD was constructed, $a(Q^2)$ has $Q^2_{\rm br} \approx 0.066 \ {\rm GeV}^2$ (a pole), and in the $\MSbar$ scheme $Q^2_{\rm br} \approx 0.36 \ {\rm GeV}^2$ (a pole). All these curves were obtained by using the programs \cite{MathPrgs}, written in {\it Mathematica}, for the evaluation of the couplings.

Until now we have described the case when $\nu=n+1$ in $\tA_{n+1}$ and $\tA_{n+1}$ is an integer. However, in many cases in physics, the physical (spacelike) quantities ${\cal F}(Q^2)$, such as here considered Wilson coefficients, have perturbation expansion in powers of $a^{\nu}=a^{\nu_0+n}$ where $\nu_0$ ($> -1$) is not integer (and $n=0,1,2,\ldots$)
\bea
{\cal F}_{\rm pt}(Q^2) & = &
f_0 a(Q^2)^{\nu_0} + \sum_{n \geq 1} f_n a(Q^2)^{\nu_0+n} \ .
\label{calFpt}
\eea
In such cases, the results for integer $\nu=1+n$ can be analytically continued to $\nu=\nu_0+n$ \cite{GCAK}, i.e., we obtain
\bes
\label{Anugen}
\bea
\lefteqn{
(a^{\nu}(Q^2))_{\rm an.} = \A_{\nu}(Q^2),
\label{anuAnu}
}  
\\
\A_{\nu} &=& \tA_{\nu} + \sum_{m=1}^{N-n-1} \tk_m({\nu}) \tA_{\nu+m} \qquad (\nu=\nu_0+n),
\label{AnutAnu}
\\
\tA_{\nu}(Q^2) &=& \tA_{\nu}(Q^2)^{\rm (FAPT,1\ell)} +
\frac{1}{\pi} \frac{(-1)}{\beta_0^{\nu-1} \Gamma(\nu)}
\int_{0}^{\infty} \ \frac{d \sigma} {\sigma}
\left[ \rho_{\A}(\sigma) - \rho_{a}(\sigma)^{\rm (1\ell)} \right]   
{\rm Li}_{-\nu+1}\left( - \frac{\sigma}{Q^2} \right) \qquad (-1 < \nu),
\label{disptAn1}
\eea
\ees
where $\nu=\nu_0+n$ ($n=0,1,\ldots,N-1$; $-1 < \nu_0$); and $\rho_{a}(\sigma)^{\rm (1\ell)}$ is the discontinuity of the one-loop pQCD coupling
\be
\rho_{a}(\sigma)^{\rm (1\ell)} = {\rm Im} a(-\sigma - i \epsilon)^{\rm (1\ell)}
= \frac{\pi}{\beta_0} \frac{1}{( \ln^2(\sigma/\bL^2) + \pi^2)},
\label{rhoa1l}
\ee
and the explicit expressions for the coefficients $\tk_m({\nu})$ appearing in the relation (\ref{AnutAnu}) are given in Ref.~\cite{GCAK}.
The unsubtracted part of the dispersive integral in Eq.~(\ref{disptAn1}) was obtained by simple continuation of the expression (\ref{tAnb}) to noninteger values ($n+1 \mapsto \nu$). The full dispersive integral in Eq.~(\ref{disptAn1}) converges in an extended regime of indices $\nu$, namely $\nu > -1$ (not just for: $\nu >0$). This is so because the basic (unsubtracted) dispersion integral was modified by subtracting and adding the one-loop (F)APT expression $\tA_{\nu}^{\rm (FAPT, 1\ell)} = \A_{\nu}^{\rm (FAPT, 1\ell)}$ which is known explicitly \cite{BMS} (when $\nu>0$, this subtraction and addition are not needed)
\be
\tA_{\nu}(Q^2)^{\rm (FAPT, 1\ell)}
= \A_{\nu}(Q^2)^{\rm (FAPT, 1\ell)} = \frac{1}{\beta_0^{\nu}}
\left(  \frac{1}{\ln^{\nu}(Q^2/\bL^2)} -
\frac{ {\rm Li}_{-\nu+1}(\bL^2/Q^2)}{\Gamma(\nu)} \right) \ .
\label{tAn11l}
\ee
Here, the scale $\bL^2 \sim 0.1 \ {\rm GeV}^2$ is arbitrary and it appears also in the (one-loop) pQCD discontinuity function $\rho_{a}(\sigma)^{\rm (1\ell)}$.

The expressions $\tA_{\nu}(Q^2)$, which are extensions of the logarithmic derivatives (\ref{tAn}) to noninteger $n+1 \mapsto \nu$, were shown to satisfy the recursive relations
\be
\frac{d}{d \ln Q^2} \tA_{\nu}(Q^2) = - \nu \beta_0 \tA_{\nu+1}(Q^2).
\label{tAnurec}
\ee
Furthermore, using the explicit expressions for the coefficients $\tk_m({\nu})$ ($m=1,2,3,4$) obtained in Ref.~\cite{GCAK}, we can check that the following RGE-type relations hold for $\A_{\nu}$:
\be
\frac{d}{d \ln Q^2} \A_{\nu}(Q^2) = - \beta_0 \nu \left[ \A_{\nu+1}(Q^2) + c_1  \A_{\nu+2}(Q^2) + c_2  \A_{\nu+3}(Q^2)  + c_3  \A_{\nu+4}(Q^2)  + c_4  \A_{\nu+5}(Q^2) +
  {\cal O}(\A_{\nu+6}) \right]
\label{AnuRGE}
\ee
This turns out to be in complete analogy with the RGE in pQCD for the power $a(Q^2)^{\nu}$
\be
\frac{d}{d \ln Q^2} a(Q^2)^{\nu} =  \nu a(Q^2)^{\nu-1} \beta(a(Q^2)) =
(-\beta_0) \nu \left[ a(Q^2)^{\nu+1} + c_1 a(Q^2)^{\nu+2} +  c_2 a(Q^2)^{\nu+3} +  \ldots \right],
\label{anuRGE}
\ee
representing thus a cross-check of consistency of our construction of the power analogs $\A_{\nu}(Q^2)$ [$\A$QCD analogs of the powers $a(Q^2)^{\nu+1}$]. 

The series (\ref{calFpt}) in (IR-safe) $\A$QCD, and its truncated version ${\cal F}^{[N]}$, are then obtained by the simple replacements (\ref{anuAnu})
\bes
\label{calFAQCD}
\bea
{\cal F}_{\A{\rm QCD}}(Q^2)  &=& 
\tf_0 \tA_{\nu_0}(Q^2) + \sum_{n \geq 1} \tf_n \tA_{\nu_0+n}(Q^2) = f_0 \A_{\nu_0}(Q^2) + \sum_{n \geq 1} f_n \A_{\nu_0+n}(Q^2)
\label{calFAQCDa}
\\
{\cal F}^{[N]}_{\A{\rm QCD}}(Q^2) &=& \tf_0 \tA_{\nu_0}(Q^2) + \sum_{n=1}^{N-1} \tf_n \tA_{\nu_0+n}(Q^2) = f_0 \A_{\nu_0}(Q^2) + \sum_{n=1}^{N-1} f_n \A_{\nu_0+n}(Q^2)
\label{calFAQCDb}
\eea
\ees
where ${\cal F}_{\A{\rm QCD}} = {\cal F}^{[N]}_{\A{\rm QCD}} + {\cal O}(\tA_{\nu_0 +N})$. In this context, we point out that the $\A$QCD frameworks in general fulfill the hierarchies $|\tA_{\nu_0}(Q^2)| > |\tA_{\nu_0+1}(Q^2)| > |\tA_{\nu_0+2}(Q^2)| > \ldots$, for all (non-timelike) scales $Q^2$, a property which appears to be a consequence of the holomorphic behavior of these quantities (and of $\A(Q^2)$). The coefficient $\tf_n$ is a linear combination of the coefficients $f_n,f_{n-1},\ldots$ due to the relations (\ref{AnutAnu}).

Sometimes, as in the degenerate case Appendix \ref{app:mixdeg}, in the perturbation expansion of physical observables we have the mixed powers $a^{\nu} \ln^k a$ (where $k=1,2,\ldots$), and they get analytized by the analogous approach \cite{Bakulev}
\be
\left[ a(Q^2)^{\nu} \ln^k a(Q^2) \right]_{\rm an.} \equiv
\left[ \left(\frac{d}{d \nu} \right)^k a(Q^2)^{\nu} \right]_{\rm an.}
= \left( \frac{d}{d \nu} \right)^k \A_{\nu}(Q^2).
\label{anulnka}
\ee

It is important to point out that the construction of the analytic analogs $\A_{\nu}(Q^2)$ of the powers $a(Q^2)^{\nu}$ [cf.~Eqs.~(\ref{tAn}) and (\ref{AntAn}) for integer $n$, and Eqs.~(\ref{Anugen}) for general $n=\nu-1$] is an operation which is linear in the (holomorphic) coupling $\A(Q^2)$, in contrast to the naive construction $(\A(Q^2))^{\nu}$. This means that, when $\A \mapsto \lambda \A$. we have: $\rho_{\A} \mapsto \lambda \rho_{\A}$, $\tA_{\nu} \mapsto \lambda \tA_{\nu}$ and $\A_{\nu} \mapsto \lambda \A_{\nu}$. Furthermore, in the case of integer $n$ it is clear from the definition (\ref{tan}) of the pQCD quantity $\ta_{n+1}(Q^2)$ that its analytic version should be $\tA_{n+1}(Q^2)$ of Eq.~(\ref{tAn}), because the transition from pQCD to $\A$QCD produces only the changes $a(Q^2) \mapsto \A(Q^2)$ and $a(Q^2 + \Delta Q^2) \mapsto \A(Q^2 + \Delta Q^2)$. More explicitly
\bea
\left( \ta_2(Q^2) \right)_{\rm an} &\equiv& \frac{(-1)}{\beta_0} Q^2 \lim_{\Delta Q^2 \to 0} \left( \frac{\left[ a(Q^2 + \Delta Q^2) - a(Q^2) \right]}{\Delta Q^2} \right)_{\rm an}
\nonumber\\
& = &  \frac{(-1)}{\beta_0} Q^2 \lim_{\Delta Q^2 \to 0} \frac{\left[ \A(Q^2 + \Delta Q^2) - \A(Q^2) \right]}{\Delta Q^2} = \tA_2(Q^2),
\label{tA2an}
\eea
and for higher $n$ analogously. One of the consequences of this construction is that $\A_{\nu}(Q^2) \not= (\A(Q^2))^{\nu}$.

The construction of the $\A$QCD analogs $\A_{\nu}(Q^2)$ of powers $a(Q^2)^{\nu}$ described here can be applied in any $\A$QCD. On the other hand, the case of APT Eq.~(\ref{dispAAPT}), where the discontinuity function $\rho_{\A}(\sigma)$ is in its {\it entirety\/} (i.e., for all $\sigma > 0$) the pQCD discontinuity function $\rho_a(\sigma)$, exceptionally allows for a more direct evaluation of $\A_{\nu}(Q^2)$, namely as
\be
\A_{\nu}^{\rm (FAPT)}(Q^2) = \frac{1}{\pi} \int_0^{+\infty} d \sigma \frac{ {\rm Im} \left[ a(-\sigma - i \epsilon)^{\nu} \right]}{(\sigma+Q^2)}  \quad (0 < \nu).
\label{FAPT1}
\ee
The extension of the convergence of this integral to the regime $-1 < \nu$ can be achieved by subtracting the one-loop (F)APT expression (\ref{tAn11l}) in the form of dispersive integral and adding it in its explicit form (\ref{tAn11l})
\bea
\A_{\nu}^{\rm (FAPT)}(Q^2) &=& \frac{1}{\pi} \int_0^{+\infty} d \sigma \frac{ \left\{ {\rm Im} \left[ a(-\sigma - i \epsilon)^{\nu} \right] -  {\rm Im} \left[ a^{(1\ell)} (-\sigma - i \epsilon)^{\nu} \right] \right\} }{(\sigma+Q^2)}
\nonumber\\
&& +  \frac{1}{\beta_0^{\nu}}
\left(  \frac{1}{\ln^{\nu}(Q^2/\bL^2)} - \frac{ {\rm Li}_{-\nu+1}(\bL^2/Q^2)}{\Gamma(\nu)} \right);
\quad (-1 < \nu).
\label{FAPT2}
\eea
It turns out that this gives the same result as the aforedescribed general method of construction of $\A_{\nu}(Q^2)$ when applied to the APT case $\rho_{\A}(\sigma) \equiv \rho_a(\sigma)$ if the truncation index in the sum on the RHS of Eq.~(\ref{AnutAnu}) is sufficiently high. We will apply the expression (\ref{FAPT2}) in the case of FAPT, using the four-loop $\MSbar$ pQCD coupling as the underlying coupling in the form given in Ref.~\cite{2dCPC} [Eq.~(6) there]. In this context, we point out that the approach (\ref{FAPT2}) can be applied only in the case of the specific $\A$QCD, namely FAPT (i.e., in the case where $\rho_{\A}(\sigma) \equiv \rho_a(\sigma)$ for all $\sigma > 0$), while the approach (\ref{Anugen}) can be applied in any $\A$QCD.

Yet another, rather popular, $\A$QCD coupling, i.e., coupling without Landau singularities, is the ``massive'' one-loop coupling (MPT)
\be
\A^{\rm (MPT)}(Q^2) = \frac{1}{\beta_0} \frac{1}{\ln \left( \frac{Q^2+M^2}{\Lambda^2} \right)},
\label{AMPT}
\ee
where $M^2 \sim 1 \ {\rm GeV}^2$ and $\Lambda^2 \sim 0.1 \ {\rm GeV}^2$. The corresponding discontinuity function is
\bea
\rho^{\rm (MPT)}_{\A}(\sigma) &=& \Theta(\sigma - M^2) \frac{\pi}{\beta_0} \frac{1}{\left[ \ln^2 \left( \frac{\sigma - M^2}{\Lambda^2} \right) + \pi^2 \right]} +
\frac{ \pi \Lambda^2}{\beta_0} \delta \left( \sigma - (M^2 - \Lambda^2) \right).
\label{rhoMPT}
\eea

In FAPT and in MPT, the deviation from the underlying pQCD at high $|Q^2|$ remains strong because it has ${\cal N}=1$ in the relation (\ref{diffAaN}). On the other hand, $2 \delta$ and $3 \delta$ $\A$QCD described before have ${\cal N}=5$, i.e., they practically coincide with the underlying pQCD at high $|Q^2| > 1 \ {\rm GeV}^2$.  

Here we summarized the evaluation of the spacelike physical quantities ${\cal D}(Q^2)$. The timelike physical quantities can, in principle, be expressed as contour integrals involving the corresponding spacelike quantities, and can thus also be evaluated in $\A$QCD (for example, cf.~\cite{3dAQCD}).

\section{RGE for Wilson coefficients: $\A$QCD}
\label{app:RGEWils}

\subsection{RGE for Wilson coefficients with mixing - nondegenerate case}
\label{app:mixnondeg}

Here we summarize the solution of the RGE for Wilson coefficients at the two-loop level in the case of ($2 \times 2$) mixing. First this will be done for the case of pQCD, and then, in accordance with the conclusions of the previous Appendix \ref{app:AQCD}, the corresponding version of the solution for $\A$QCD will be presented.

The RGE in pQCD in such a case has the form, cf.~Eqs.~(\ref{RGEw}) and (\ref{hatgammaexp})
\bea
\frac{d}{d \ln Q^2} {\vec C}(Q^2) & = & \frac{1}{2} \left[ \left( \frac{a(Q^2)}{4} \right) {\hat \gamma}^{(0)T} + \left( \frac{a(Q^2)}{4} \right)^2 {\hat \gamma}^{(1)T} + {\cal O}(a^3) \right] {\vec C}(Q^2)
\label{RGEW2la}
\eea
where ${\vec C}(Q^2)$ is the two-component vector (column) of Wilson coefficients, and ${\hat \gamma}^{(0)}$  and ${\hat \gamma}^{(1)}$ are the one-loop and two-loop $2 \times 2$ matrices which, in the cases of some operators, have been obtained in the literature, see Appendix \ref{app:AD} (cf.~\cite{Buras:2000if,GHK2016}). When changing the variable $Q^2$ to $a(Q^2)$, and taking into account the definition (\ref{RGE}), the above RGE (\ref{RGEW2la}) can be rewritten as
\bea
\frac{d}{d a} {\vec C}(a) & = & \frac{1}{2 \beta(a)} \left[ \left( \frac{a}{4} \right) {\hat \gamma}^{(0)T} + \left( \frac{a}{4} \right)^2 {\hat \gamma}^{(1)T} + {\cal O}(a^3) \right] {\vec C}(a),
\label{RGEW2lb}
\eea
where $a \equiv a(Q^2)$. Let ${\hat V}^{(0)}$ be the ``rotation'' matrix which diagonalizes the one-loop matrix ${\hat \gamma}^{(0)T}$
\be
({\hat V}^{(0)})^{-1} {\hat \gamma}^{(0)T} {\hat V}^{(0)} =  {\hat \gamma}^{(0)}_D \equiv - 8 \beta_0 {\hat \nu},
\label{hatgamma0}
\ee
where ${\hat \nu}$ is, by this definition, a diagonal matrix
\be
   {\hat \nu} = \left[ {\begin{array}{cc}
         \nu_1 & 0 \\
         0 & \nu_2 \\
  \end{array} } \right]
\label{hatnu}
\ee
When defining
\be
{\vec C}^{(0)}(Q^2) \equiv ({\hat V}^{(0)})^{-1} {\vec C}(Q^2),
\label{vC0}
\ee
the RGE (\ref{RGEW2lb}) can be rewritten in the form
\be
\frac{d}{d a} {\vec C}^{(0)}(a) = \left[ {\hat \nu} \frac{1}{a} + {\hat k}^{(1)} + {\cal O}(a) \right] {\vec C}^{(0)}(a),
\label{RGEW2lc}
\ee
where the matrix ${\hat k}^{(1)}$ incorporates the two-loop effects
\be
{\hat k}^{(1)} = - \frac{1}{32 \beta_0} ({\hat V}^{(0)})^{-1} {\hat \gamma}^{(1)T} {\hat V}^{(0)} - c_1 {\hat \nu}.
\label{katk1}   
\ee
We recall that $c_1 = \beta_1/\beta_0$ is the (universal) two-loop beta coefficient, cf.~Eqs.~(\ref{RGE}). Since the first matrix on the RHS of Eq.~(\ref{katk1}) is in general nondiagonal, an additional, two-loop, ``rotation'' is needed to obtain fully decoupled system. This is achieved by a matrix ${\hat J}^{(1)}$ which acts in the following way:
\bea
{\vec C}^{(1)}(a) & \equiv & \left[ 1 - \frac{a}{4} {\hat J}^{(1)} + {\cal O}(a^2) \right] {\vec C}^{(0)}(a)  \quad {\bigg\{} =  \left[ 1 - \frac{a}{4} {\hat J}^{(1)} + {\cal O}(a^2) \right]  ({\hat V}^{(0)})^{-1} {\vec C}(a) {\bigg\}},
\label{vC1}    
\eea
such that the RGE for ${\vec C}^{(1)}(a)$ is a decoupled system
\be
\frac{d}{d a} {\vec C}^{(1)}(a) = \left\{ {\hat \nu} \frac{1}{a} + \left[ {\hat k}^{(1)} + \frac{1}{4} [{\hat \nu}, {\hat J}^{(1)}] - \frac{1}{4} {\hat J}^{(1)} \right] + {\cal O}(a) \right\} {\vec C}^{(1)}(a),
\label{RGEW2ld}
\ee
i.e., the total  expression in brackets on the RHS of Eq.~(\ref{RGEW2ld}) is a diagonal matrix ${\hat k}^{(1)}_D$. This can be achieved by the following matrix ${\hat J}^{(1)}$:
\bes
\label{J1k1D}
\bea
{\hat J}^{(1)} & = &
\left[ {\begin{array}{cc}
         0 & \frac{4}{(1 - \nu_1 + \nu_2)} {\hat k}^{(1)}_{12} \\
         \frac{4}{(1 + \nu_1 - \nu_2)} {\hat k}^{(1)}_{21} & 0 \\
      \end{array} } \right] \; \Rightarrow
\label{hatJ1}
\\
\left[ {\hat k}^{(1)} + \frac{1}{4} [{\hat \nu}, {\hat J}^{(1)}] - \frac{1}{4} {\hat J}^{(1)} \right] & = &
\left[ {\begin{array}{cc}
         {\hat k}^{(1)}_{11} & 0 \\
         0 &  {\hat k}^{(1)}_{22} \\
   \end{array} } \right] \equiv {\hat k}^{(1)}_D.
\label{hatk1D}
\eea
\ees
The decoupled system of RGEs (\ref{RGEW2ld}) can then be integrated, resulting in
\be
{\vec C}^{(1)}(a(Q^2)) = \left[ a(Q^2)^{\hat \nu} + {\hat k}^{(1)}_D a(Q^2)^{\hat \nu +1} + {\cal O}(a^{\hat \nu +2}) \right] {\vec {\cal C}},
\label{vC1res}   
\ee
where ${\vec {\cal C}}$ is a two-component (column) vector independent of $Q^2$ scale, and $a(Q^2)^{\hat \nu}$ is a diagonal matrix according to Eq.~(\ref{hatnu})
\bea
a(Q^2)^{\nu} & = & \exp[{\hat \nu} \ln a(Q^2)] =
\left[ {\begin{array}{cc}
      a(Q^2)^{\nu_1} & 0 \\
      0 & a(Q^2)^{\nu_2}
\end{array} } \right].
\label{ahatnu}        
\eea
Using the relation (\ref{vC1}), the solution for the original vector ${\vec C}(Q^2)$ of Wilson coefficients is
\bea
{\vec C}(Q^2) & = & 
{\hat V}^{(0)} \left[ 1 + \frac{a(Q^2)}{4} {\hat J}^{(1)} + {\cal O}(a^2) \right]  {\vec C}^{(1)}(Q^2) 
\nonumber\\
& = & {\hat V}^{(0)} {\hat U}^{(1)}(a(Q^2)) {\vec {\cal C}}
\label{vCres}
\eea
where the matrix ${\hat U}^{(1)}(a)$ is
\bes
\label{hatU1}
\bea
{\hat U}^{(1)}(a) &=& a^{\hat \nu} + \left( {\hat k}^{(1)}_D + \frac{1}{4} {\hat J}^{(1)} \right) a^{\hat \nu + 1} + {\cal O}(a^{\hat \nu + 2})
\label{hatU1a}
\\
& = & 
\left[ {\begin{array}{ll}
a^{\nu_1} + {\hat k}^{(1)}_{11} a^{\nu_1+1}, & \frac{{\hat k}^{(1)}_{12}}{(1 - \nu_1 + \nu_2)} a^{\nu_2+1} \\
\frac{{\hat k}^{(1)}_{21}}{(1 + \nu_1 - \nu_2)} a^{\nu_1+1}, &  a^{\nu_2} + {\hat k}^{(1)}_{22} a^{\nu_2+1} \\      
\end{array} } \right] +  {\cal O}(a^{\hat \nu + 2}),
\label{hatU1b}
\eea
\ees
where $a \equiv a(Q^2)$, and the other parameters are $Q^2$-independent.

According to conclusions presented in Appendix \ref{app:AQCD}, in $\A$QCD the same relations are valid, but under the consistent replacements $a(Q^2)^{\nu + m} \mapsto \A_{\nu+m}(Q^2)$
\bes
\label{AQCDresmix}
\bea
{\vec C}(Q^2)_{(\A)} & = &  {\hat V}^{(0)} {\hat U}^{(1)}(Q^2)_{(\A)} {\vec {\cal C}},
\label{AQCDvCres}
\\
{\hat U}^{(1)}(Q^2)_{(\A)} & = &
\A_{\hat \nu}(Q^2) + \left( {\hat k}^{(1)}_D + \frac{1}{4} {\hat J}^{(1)} \right) A_{\hat \nu + 1} + {\cal O}(\A_{\hat \nu + 2})
\nonumber\\
& = &  
\left[ {\begin{array}{ll}
\A_{\nu_1}(Q^2) + {\hat k}^{(1)}_{11} \A_{\nu_1+1}(Q^2), & \frac{{\hat k}^{(1)}_{12}}{(1 - \nu_1 + \nu_2)} \A_{\nu_2+1}(Q^2) \\
\frac{{\hat k}^{(1)}_{21}}{(1 + \nu_1 - \nu_2)} \A_{\nu_1+1}(Q^2), &  \A_{\nu_2}(Q^2) + {\hat k}^{(1)}_{22} \A_{\nu_2+1}(Q^2) \\      
 \end{array} } \right] +  {\cal O}(\A_{\hat \nu + 2}), 
\label{AQCDhatU1}
\eea
\ees
The evaluation of $\A_{\nu_j}$ and $\A_{\nu_j+1}$ ($j=1,2$) in terms of $\tA_{\nu_j}$ and $\tA_{\nu_j+1}$ is performed along the same lines as explained in Sec.~\ref{sec:AQCD} Eqs.~(\ref{Anu2l})-(\ref{C2ltA}), but now separately for $\nu_1$ and $\nu_2$: $\A_{\nu_j} = \tA_{\nu_j} + {\widetilde k}_1(\nu_j)  \tA_{\nu_j+1}$, and $\A_{\nu_j+1}=\tA_{\nu_j+1}$ ($j=1,2$).

We recall that in Eq.~(\ref{AQCDvCres}) the vector ${\vec {\cal C}}$ is $Q^2$-independent. This allows us, equally as in pQCD [Eq.~(\ref{vCres})], to rewrite the solution in terms of the initial condition values ${\hat U}^{(1)}(Q_0^2)_{(\A)}$
\bes
\label{AvCres}
\bea
{\vec {\cal C}} &=&  \left({\hat U}^{(1)}(Q_0^2)_{(\A)} \right)^{-1} ({\hat V}^{(0)})^{-1} {\vec C}(Q_0^2)_{(\A)} \; \Rightarrow
\label{vCincon}
\\
{\vec C}(Q^2)_{(\A)} &=&  {\hat V}^{(0)} {\hat U}^{(1)}(Q^2)_{(\A)} \left({\hat U}^{(1)}(Q_0^2)_{(\A)} \right)^{-1} ({\hat V}^{(0)})^{-1} {\vec C}(Q_0^2)_{(\A)}
\equiv {\hat U}(Q^2;Q_0^2)_{(\A)}  {\vec C}(Q_0^2)_{(\A)}.
\label{AQCDvCres2}
\eea
\ees
The matrix $ {\hat U}(Q^2;Q_0^2)_{(\A)}$ is the two-loop (RGE-)evolution matrix for the Wilson coefficients ${\vec C}$ from a (higher) scale $Q_0^2$ to a (lower) scale $Q^2$, in the case of ($2 \times 2$) mixing, in $\A$QCD with IR-safe and holomorphic coupling $\A(Q^2)$.

\subsection{RGE for Wilson coefficients with mixing - degenerate case}
\label{app:mixdeg}

In some exceptional cases, the eigenvalues of the matrix $\hat \nu$ [Eq.~(\ref{hatnu})] can satisfy the relation
\be
\nu_1 - \nu_2=1.
\label{degrel}
\ee
This happens in the specific case of the (31)$^{XY}$-mixing, i.e, the mixing of operators ${\mathcal O}_3^{XY}$ and ${\mathcal O}_1^{XY}$ ($X \not= Y$) in Eqs.~(\ref{ops}) for $n_f=3$, where the anomalous dimension matrix is known at the two-loop level. We recall that $\A$QCD should be applied in the $n_f=3$ regime.

In such a case, the two-loop matrix ${\hat J}^{(1)}$ Eq.~(\ref{hatJ1}), which is needed for the decoupling of the two RGEs, does not exist because one term there has zero in the denominator. In such a case, we have to proceed in a modified way. At the two-loop level, the matrix ${\hat J}^{(1)}$ now has the limited form
\bea
{\hat J}^{(1)} & = &
\left[ {\begin{array}{cc}
         0 & 0  \\
         \frac{4}{(1 + \nu_1 - \nu_2)} {\hat k}^{(1)}_{21} & 0 \\
  \end{array} } \right]  =
\left[ {\begin{array}{cc}
         0 & 0  \\
         2 {\hat k}^{(1)}_{21} & 0 \\
  \end{array} } \right] 
\label{hatJ1deg}
\eea
With this matrix, the transformation (\ref{vC1}) leads to the partially coupled RGEs for the two components of ${\vec C}^{(1)}(a)$
\bes
\label{RGEdeg1}
\bea
\frac{d}{d a} C^{(1)}_1(a) & = & \left( \frac{1}{a} \nu_1 + {\hat k}^{(1)}_{11} \right) C^{(1)}_1(a) + {\hat k}^{(1)}_{12}  C^{(1)}_2(a),
\label{RGEdeg1a}
\\
\frac{d}{d a} C^{(1)}_2(a) & = & \left( \frac{1}{a} \nu_2 + {\hat k}^{(1)}_{22} \right) C^{(1)}_2(a),
\label{RGEdeg1b}
\eea
\ees
where the (unknown) three-loop contributions ${\cal O}(a C^{(1)}_j(a))$ on the RHS are excluded. Equation (\ref{RGEdeg1b}) can be integrated and gives
\bes
\label{C12tU2}
\bea
C^{(1)}_2(a) &=&  {\tilde U}_2(a) {\cal C}_2,
\label{C12}
\\
{\tilde U}_2(a) &=& a^{\nu_2} + {\hat k}^{(1)}_{22} a^{\nu_2+1} + {\cal O}(a^{\nu_2+2}).
\label{tU2}
\eea
\ees
Here as earlier, $a \equiv a(Q^2)$, and ${\cal C}_2$ is a $Q^2$-independent constant. The terms ${\cal O}(a^{\nu_2+2})$ are not specified in Eq.~(\ref{tU2}) because they are affected by the (unknown) three-loop contributions. Inserting the solution (\ref{C12tU2}) into the first RGE (\ref{RGEdeg1a}) gives us a nonhomogeneous differential equation for $C^{(1)}_1(a)$
\be
\frac{d}{d a} C^{(1)}_1(a) - \left( \frac{1}{a} \nu_1 + {\hat k}^{(1)}_{11} \right) C^{(1)}_1(a) =  {\hat k}^{(1)}_{12} {\tilde U}_2(a) {\cal C}_2.
\label{RGEdeg2a}
\ee
This equation can be solved by the usual mathematical methods (e.g., by the Green function approach) and gives
\be
C^{(1)}_1(a) = {\tilde U}_1(a) \frac{C^{(1)}_1(a_0)}{{\tilde U}_1(a_0)} + {\hat k}^{(1)}_{12} ({\cal G}(a) - {\cal G}(a_0)) {\tilde U}_1(a) {\cal C}_2,
\label{C11a}
\ee
where $a \equiv a(Q^2)$ and $a_0 \equiv a(Q_0^2)$. The first term on the RHS of Eq.~(\ref{C11a}) represents a solution to the homogeneous version of Eq.~(\ref{RGEdeg2a}), and the second term a particular solution to the full (nonhomogeneous) Eq.~(\ref{RGEdeg2a}); ${\tilde U}_1(a)$ is the evolution function
\bea
{\tilde U}_1(a) &=& a^{\nu_1} + {\hat k}^{(1)}_{11} a^{\nu_1+1} + {\cal O}(a^{\nu_1+2}),
\label{tU1}
\eea
and ${\cal G}(a)$ is the function
\be
{\cal G}(a) = \ln a + (- {\hat k}^{(1)}_{11} +{\hat k}^{(1)}_{22}) a + {\cal O}(a^2).
\label{calG}
\ee
The solution (\ref{C11a}) implies that the expression
\be
\frac{C^{(1)}_1(a)}{{\tilde U}_1(a)} - {\hat k}^{(1)}_{12} {\cal G}(a) {\cal C}_2
\; \left( \equiv {\cal C}_1 \right)
\label{calC1}
\ee
is a $Q^2$-independent constant (${\cal C}_1$). This, and the relation (\ref{C12}), imply that the solution for ${\vec C}^{(1)}(a)$ can be written in the form
\bes
\label{C1res}
\bea
C^{(1)}_1(a) & = & {\tilde U}_1(a) \left[ {\cal C}_1 + {\hat k}^{(1)}_{12} {\cal G}(a) {\cal C}_2 \right],
\label{C11res}
\\
C^{(1)}_2(a) & = & {\tilde U}_2(a) {\cal C}_2.
\label{C12res}
\eea
\ees
Using this solution, we can ``rotate'' back to the original basis of the Wilson coefficients using  the relation (\ref{vC1}) and the explicit form (\ref{hatJ1deg}) of ${\hat J}^{(1)}$ in the considered degenerate case. In analogy with the algebra performed in the previous Subsection \ref{app:mixnondeg}, we obtain now
\be
{\vec C}(a) = V^{(0)} {\hat U}^{(1)}(a) {\vec C},
\label{vCresdeg1}
\ee
where ${\vec C}^T = ({\cal C}_1,{\cal C}_2)$ is the vector with the two $Q^2$-independent constants, and the matrix ${\hat U}^{(1)}(a)$ is now (the considered degenerate case $\nu_1-\nu_2=1$)
\bea
{\hat U}^{(1)}(a) & = &
\left[ {\begin{array}{ll}
\left[ a^{\nu_1} + {\hat k}^{(1)}_{11} a^{\nu_1+1} \right], & {\hat k}^{(1)}_{12} \left[ a^{\nu_1} \ln a + {\hat k}^{(1)}_{11} a^{\nu_1+1} \ln a + (-{\hat k}^{(1)}_{11}+{\hat k}^{(1)}_{22}) a^{\nu_1+1} \right] \\
\frac{1}{2} {\hat k}^{(1)}_{21}  a^{\nu_1+1}, &  \left[ \frac{1}{2} {\hat k}^{(1)}_{21}{\hat k}^{(1)}_{12} a^{\nu_1+1} \ln a  + (a^{\nu_2} + {\hat k}^{(1)}_{22} a^{\nu_2+1})\right] \\      
\end{array} } \right],
\label{hatU1deg}
\eea
where terms of higher order, which are affected by (unknown) three-loop contributions, were neglected. We recall that $a \equiv a(Q^2)$.

It can be shown that the result of the nondegenerate case considered in the previous Appendix \ref{app:mixnondeg}, Eq.~(\ref{hatU1b}), is in the case of $\nu_1 - \nu_2 = 1 - \varepsilon$ (with $\epsilon \to 0$) the limiting case of the above result Eq.~(\ref{hatU1deg}), as it should be. Namely, when $\nu_1-\nu_2=1 - \epsilon$, we have
\be
\frac{a^{1 - \nu_1 + \nu_2}}{1 - \nu_1 + \nu_2} = \frac{a^{\epsilon}}{\epsilon} =
\frac{1}{\epsilon} + \ln a + (- {\hat k}^{(1)}_{11} +{\hat k}^{(1)}_{22}) a + {\cal O}(\epsilon).
\label{calGnew}
\ee
This coincides with the expression (\ref{calG}) for ${\cal G}(a)$, except for the corrections ${\cal O}(\epsilon)$ ($\to 0$) and a (large) constant $1/\epsilon$. However, this large constant is irrelevant for the final result, because only the difference ${\cal G}(a)-{\cal G}(a_0)$ matters, cf.~Eq.~(\ref{C11a}); furthermore, changing ${\cal G}$ by a constant only redefines the new constant ${\cal C}_1 \mapsto {\cal C}_1^{\rm new}$ [cf.~Eq.~(\ref{C11res})]. From here, it it straightforward to check that the limit $\epsilon \to 0$ of the nondegenerate case Eq.~(\ref{hatU1b}) is the degenerate result Eq.~(\ref{hatU1deg}).

As in the previous Subsection \ref{app:mixnondeg}, the transition to the $\A$QCD is obtained by the replacements $a^{\nu+m} \mapsto \A_{\nu +m}$ and by $a^{\nu} \ln a$ [$\equiv (d/d \nu) a^{\nu}$] $\mapsto (d/d \nu) \A_{\nu}$ in Eqs.~(\ref{vCresdeg1})-(\ref{hatU1deg})
\be
{\vec C}(Q^2)_{(\A)} =   {\hat V}^{(0)} {\hat U}^{(1)}(Q^2)_{(\A)} {\vec {\cal C}},
\label{AQCDvCresdeg}
\ee
where the matrix $ {\hat U}^{(1)}(Q^2)_{(\A)}$ in $\A$QCD is
\bea
\label{AQCDhatU1deg}
\lefteqn{
  {\hat U}^{(1)}(Q^2)_{(\A)} 
}
\\ &=&
\left[ {\begin{array}{ll}
\left[ \A_{\nu_1}(Q^2) + {\hat k}^{(1)}_{11} \A_{\nu_1+1}(Q^2) \right],  & {\hat k}^{(1)}_{12} \left[ \left(\frac{d}{d \nu} \right) \A_{\nu}(Q^2)|_{\nu=\nu_1} + {\hat k}^{(1)}_{11} \left(\frac{d}{d \nu} \right) \A_{\nu}(Q^2)|_{\nu=\nu_1+1} + (-{\hat k}^{(1)}_{11}+{\hat k}^{(1)}_{22}) \A_{\nu_1+1}(Q^2) \right] \\
\frac{1}{2} {\hat k}^{(1)}_{21}  \A_{\nu_1+1}(Q^2), &  \left[ \frac{1}{2} {\hat k}^{(1)}_{21} {\hat k}^{(1)}_{12}  \left(\frac{d}{d \nu} \right) \A_{\nu}(Q^2)|_{\nu=\nu_1+1}  + (\A_{\nu_2}(Q^2) + {\hat k}^{(1)}_{22} \A_{\nu_2+1}(Q^2)) \right] \\      
 \end{array} } \right] .
\nonumber
\eea
As in the previous Appendix \ref{app:mixnondeg}, the relation (\ref{AQCDvCresdeg}) can be written in the form
\bea
{\vec C}(Q^2)_{(\A)} &=&  {\hat V}^{(0)} {\hat U}^{(1)}(Q^2)_{(\A)} \left({\hat U}^{(1)}(Q_0^2)_{(\A)} \right)^{-1} ({\hat V}^{(0)})^{-1} {\vec C}(Q_0^2)_{(\A)}
\equiv {\hat U}(Q^2;Q_0^2)_{(\A)}  {\vec C}(Q_0^2)_{(\A)},
\label{AQCDvCresdeg2}
\eea
where the matrix
\be
{\hat U}(Q^2;Q_0^2)_{(\A)} = {\hat V}^{(0)} {\hat U}^{(1)}(Q^2)_{(\A)} \left({\hat U}^{(1)}(Q_0^2)_{(\A)} \right)^{-1} ({\hat V}^{(0)})^{-1}
\label{UQ2Q02}
\ee
is the evolution matrix for the Wilson coefficients from the (upper) scale $Q_0^2$ to the (lower) scale $Q^2$.    

Analogously as in the nondegenerate case in Appendix \ref{app:mixnondeg} [and in Sec.~\ref{sec:AQCD} Eqs.~(\ref{Anu2l})-(\ref{C2ltA}) in the case of no mixing], the evaluation of $\A_{\nu_j}$ and $\A_{\nu_j+1}$ ($j=1,2$) in Eq.~(\ref{AQCDhatU1deg}) is performed in terms of $\tA_{\nu_j}$ and $\tA_{\nu_j+1}$ as follows: $\A_{\nu_j} = \tA_{\nu_j} + {\widetilde k}_1(\nu_j)  \tA_{\nu_j+1}$, and $\A_{\nu_j+1}=\tA_{\nu_j+1}$ ($j=1,2$).

\end{appendix}

\end{document}